\documentclass[lettersize,journal]{IEEEtran}
\usepackage{amsmath,amsfonts}
\usepackage{algorithmic}
\usepackage{algorithm}
\usepackage{array}
\usepackage[caption=false,font=normalsize,labelfont=sf,textfont=sf]{subfig}
\usepackage{textcomp}
\usepackage{stfloats}
\usepackage{url}
\usepackage{verbatim}
\usepackage{graphicx}
\usepackage{cite}
\usepackage{epstopdf}
\usepackage{caption2}
\usepackage{subfig}
\usepackage{amssymb}
\usepackage{makecell}
\usepackage{xcolor}
\usepackage{bbding}
\usepackage[colorlinks,
            allcolors=blue,
            linkcolor=blue,
            anchorcolor=blue,
            citecolor=blue]{hyperref}
\newcommand{\subsubsubsection}[1]{\paragraph{#1}}
\begin{document}

\title{Silhouette Score Efficient Radio Frequency Fingerprint Feature Extraction}

\author{\IEEEauthorblockN{Xuan Yang, Dongming Li,~\IEEEmembership{Member, IEEE}, Yi Lou,~\IEEEmembership{Member, IEEE}, Xianglin Fan}
\vspace{-1.2cm}

\thanks{This work has been submitted to the IEEE for possible publication. Copyright may be transferred without notice, after which this version may no longer be accessible. This work was partially supported by the Big Data Computing Center of Southeast University and National Key Research and Development Project under Grant 2022YFB3904404. \emph{(Corresponding author: Dongming Li.)}}

\thanks{X. Yang and D. Li are with School of Cyber Science and Engineering, Southeast University, Nanjing, 211189, China (email: xuan\_yang@seu.edu.cn; lidm@seu.edu.cn).}
\thanks{Y. Lou is with School of Information Science and Engineering, Harbin Institute of Technology
 (Weihai), Weihai, 264209, China (e-mail: louyi@ieee.org).}
\thanks{X. Fan is with School of Electronic and Information Engineering, West Anhui University, Lu'an, 237012, China (email: fanxianglin@wxc.edu.cn).}
}
\maketitle

\begin{abstract}
Radio frequency fingerprint (RFF) identification technology, which exploits relatively stable hardware imperfections, is highly susceptible to constantly changing channel effects. Although various channel-robust RFF feature extraction methods have been proposed, they predominantly rely on experimental comparisons rather than theoretical analyses. This limitation hinders the progress of channel-robust RFF feature extraction and impedes the establishment of theoretical guidance for its design. In this paper, we establish a unified theoretical performance analysis framework for different RFF feature extraction methods using the silhouette score as an evaluation metric, and propose a precoding-based channel-robust RFF feature extraction method that enhances the silhouette score without requiring channel estimation. First, we employ the silhouette score as an evaluation metric and obtain the theoretical performance of various RFF feature extraction methods using the Taylor series expansion. Next, we mitigate channel effects by computing the reciprocal of the received signal in the frequency domain at the device under authentication. We then compare these methods across three different scenarios: the deterministic channel scenario, the independent and identically distributed (i.i.d.) stochastic channel scenario, and the non-i.i.d. stochastic channel scenario. Finally, simulation and experimental results demonstrate that the silhouette score is an efficient metric to evaluate classification accuracy. Furthermore, the results indicate that the proposed precoding-based channel-robust RFF feature extraction method achieves the highest silhouette score and classification accuracy under channel variations.
\end{abstract}

\begin{IEEEkeywords}
RF fingerprint, feature extraction, silhouette score, channel-robust, precoding.
\end{IEEEkeywords}

\vspace{-0.5cm}
\section{Introduction}
\IEEEPARstart{I}{n} the past several decades, the scale of wireless networks, such as WiFi, LoRa, and ZigBee, has grown exponentially. Hence, the authentication of wireless devices is critical, as successful attacks by malicious users can disrupt systems and cause significant property damage \cite{schiller2022landscape}. Traditional systems rely on digital identifiers, such as media access control (MAC) addresses, and employ cryptographic algorithms and protocols for authentication. However, digital identifiers are vulnerable to manipulation, and cryptographic methods face increasing risks of being compromised with the advent of quantum computing \cite{babun2021survey}. In addition, cryptographic algorithms and protocols are often inappropriate to implement in low-end wireless devices. As the scale of wireless networks continues to grow, the development of alternative approaches for wireless device authentication is of paramount importance.

Radio frequency fingerprint (RFF) identification is recognized as a promising technology for wireless device authentication \cite{xie2024radio}. The signals transmitted by wireless devices exhibit distinct characteristics due to unique hardware imperfections, including in-phase/quadrature (IQ) direct current (DC) offset, IQ imbalance, imperfect low-pass filters, oscillator errors, and power amplifier (PA) nonlinearities \cite{zhang2021radio}. These RFF characteristics, which are unique, unforgeable, and stable, can be extracted from wireless signals for device identification. Compared to conventional methods, RFF identification is better suited for wireless networks due to its unforgeable nature.

The implementation of RFF identification systems presents several critical challenges. First, additive noise can significantly degrade classification accuracy \cite{xing2018radio,xie2018optimized,yu2019radio,wang2024rf}. In the high signal-to-noise ratio (SNR) region, RFF feature clusters are well-separated, and samples within the same cluster are tightly grouped. However, in the low SNR region, samples within the same cluster disperse, and the boundaries between different clusters blur. As a result, classification accuracy decreases with decreased SNR. Second, channel variations can severely degrade classification performance \cite{hua2018accurate, kandel2019exploiting,jian2021radio,chen2022radio,xing2022design,chen2023channel,sun2023location,yang2023led,kong2024deepcrf,kong2024csi,dong2024robust}. This degradation arises because the statistical characteristics of the extracted RFF feature are prone to be affected by unstable channel conditions. Consequently, the decision boundaries learned by the classifier under one channel condition may become invalid under another, leading to reduced classification accuracy. Third, the receiver differences can also affect classification accuracy \cite{chen2021radio,gaskin2022tweak,bao2023receiver,shen2023towards,yu2024rf}. The reason is that different receivers introduce distinct RFF characteristics, which affect the identification process. As a result, the centers of RFF feature clusters may shift across different receivers, further degrading classification accuracy.

Many studies have attempted to address these three challenges. First, methods such as adding processes \cite{xing2018radio,yu2019radio} and wavelet-based denoising \cite{xie2018optimized,wang2024rf} have been proposed to mitigate the effects of additive noise. By adding adjacent preambles from the same device, the received SNR can be improved, thereby enhancing the classification accuracy in the low SNR region. Second, various approaches have been developed to reduce the adverse effects of channel variations, including signal characteristic extraction \cite{kandel2019exploiting}, channel estimation and compensation \cite{jian2021radio}, filtering \cite{chen2023channel,yang2023led,kong2024csi}, division \cite{chen2022radio,xing2022design}, precoding \cite{dong2024robust,sun2023location}, and data augmentation \cite{comert2022analysis}. For instance, division-based RFF extraction methods \cite{chen2022radio,xing2022design} exploit adjacent signals within the channel coherence time to eliminate channel effects through a division process, while precoding-based RFF extraction methods \cite{dong2024robust,sun2023location} leverage channel reciprocity. Third, neural network-based approaches have been employed to alleviate the impact of receiver variations. These methods include generative adversarial networks (GANs) \cite{yu2024rf}, deep adversarial neural networks \cite{bao2023receiver}, transfer learning \cite{chen2021radio}, deep metric learning \cite{gaskin2022tweak}, and adversarial training \cite{shen2023towards}. However, most of these methods require either collecting training samples from multiple receivers or performing calibration in new deployments, both of which are time-consuming.

Although existing studies have attempted to extract more robust RFF features to improve classification accuracy across various tasks, such improvements are primarily validated through experiments, while a solid theoretical foundation remains lacking. In contrast, for conventional digital modulation schemes, such as binary phase shift keying (BPSK) and quadrature phase shift keying (QPSK), the bit error rate (BER) performance can be rigorously derived and compared within a well-established communication-theoretic framework, rather than relying solely on simulations or over-the-air experiments. However, in RFF identification, the classification accuracy of RFF feature extraction methods is difficult to analyze directly from a theoretical perspective. Therefore, to theoretically analyze and compare different RFF extraction approaches, a unified metric that is strongly correlated with the final classification accuracy and is analytically tractable is required. If such a metric can be designed and a corresponding theoretical evaluation framework can be established, different RFF feature extraction methods can be directly compared on a theoretical basis, thereby reducing the reliance on extensive experimental validation. Moreover, methods that can be theoretically shown to enhance performance with respect to this unified metric can be designed under this theoretical guidance.

To achieve this goal, we adopt the silhouette score, which is commonly used as an evaluation metric in unsupervised learning \cite{rousseeuw1987silhouettes, shutaywi2021}. A higher silhouette score results from lower intra-class distance and higher inter-class distance, and indicates better clustering performance. Most existing studies focus on experimentally calculating the silhouette score \cite{shahapure2020cluster, shutaywi2021} or on determining the optimal number of clusters based on this metric \cite{bagirov2023finding, januzaj2023determining}. To the best of our knowledge, no existing work has attempted to theoretically analyze the performance of different feature extraction methods based on the silhouette score in the machine learning literature. One plausible reason is that, in most machine learning domains, such as electrocardiogram (ECG) classification and voiceprint identification, analytically tractable system models and feature extraction procedures are difficult to establish. There are two main reasons why the silhouette score can be adopted as a unified metric for RFF feature extraction: (1) The silhouette score reflects clustering quality and is strongly correlated with classification accuracy in supervised learning tasks. (2) The silhouette score is analytically tractable in a well-modeled RFF feature extraction process, as will be demonstrated in the following sections.

Some studies have also focused on the theoretical analysis of RFF identification, such as deriving the Cramer-Rao lower bound for RFF parameter estimation \cite{10158745} and analyzing the fundamental performance limits of RFF identification from an information-theoretic perspective \cite{gungor2016basic}. However, rather than deriving the theoretical limits of RFF parameters estimation and identification, developing a framework to theoretically compare different RFF feature extraction methods using a unified metric and designing more effective RFF feature extraction methods under this metric remain open research problems.

In this work, we aim to address the following questions: (1) Can a theoretical RFF feature extraction performance analysis framework be designed based on a unified and analytically tractable metric? (2) Can a novel RFF feature extraction method be developed that is theoretically proven to enhance performance under this unified metric? To answer these questions, we develop a unified theoretical performance analysis framework for RFF feature extraction methods using the silhouette score as the evaluation metric, and we propose a precoding-based, channel-robust RFF feature extraction method to enhance the silhouette score. To the best of our knowledge, this is the first study to uniformly and theoretically analyze RFF feature extraction performance using the silhouette score as the evaluation metric. The application of this work is twofold: it not only enables the theoretical analysis of the performance of existing RFF feature extraction methods but also provides guidance for the design of new methods. Future RFF feature extraction methods that are analytically tractable within the proposed framework can be directly compared with existing approaches without requiring time-consuming experiments. Furthermore, within the unified and analytically tractable metric and framework, new methods achieving higher silhouette scores can be systematically designed, thereby reducing the reliance on trial-and-error experimentation. The main contributions of this paper are summarized as follows:

1. A unified analyzing framework based on silhouette score is theoretically modeled for various RFF feature extraction methods. The silhouette scores of different methods are analytically derived using the Taylor series expansion. Theoretical comparisons of these methods are detailed under the deterministic and stochastic channel scenarios.

2. A precoding-based channel-robust RFF feature extraction method is proposed to enhance the silhouette score without requiring channel estimation. To eliminate channel effects, Bob computes the reciprocal of his received signal in the frequency domain before transmitting it to Alice for authentication. Additionally, the baseband signal is amplified at Bob's side to improve Alice's received SNR.

3. Simulation and experimental results validate the effectiveness of the silhouette score as an evaluation metric for RFF feature extraction methods, demonstrating its strong correlation with classification accuracy. Moreover, the expectation approximation using the Taylor series expansion is accurate in the high SNR region. The proposed method achieves the highest silhouette score and classification accuracy under channel variations.

The remainder of this paper is organized as follows: Section \ref{section:2} presents the RFF feature normalization process and the calculation of the silhouette score. Section \ref{section:3} obtains the theoretical expressions of the existing methods' silhouette scores under various channel conditions. Section \ref{section:4} introduces the proposed method along with its theoretical analyses based on the silhouette score.
\begin{figure}[htbp]
    \centering
    \includegraphics[width=9cm]{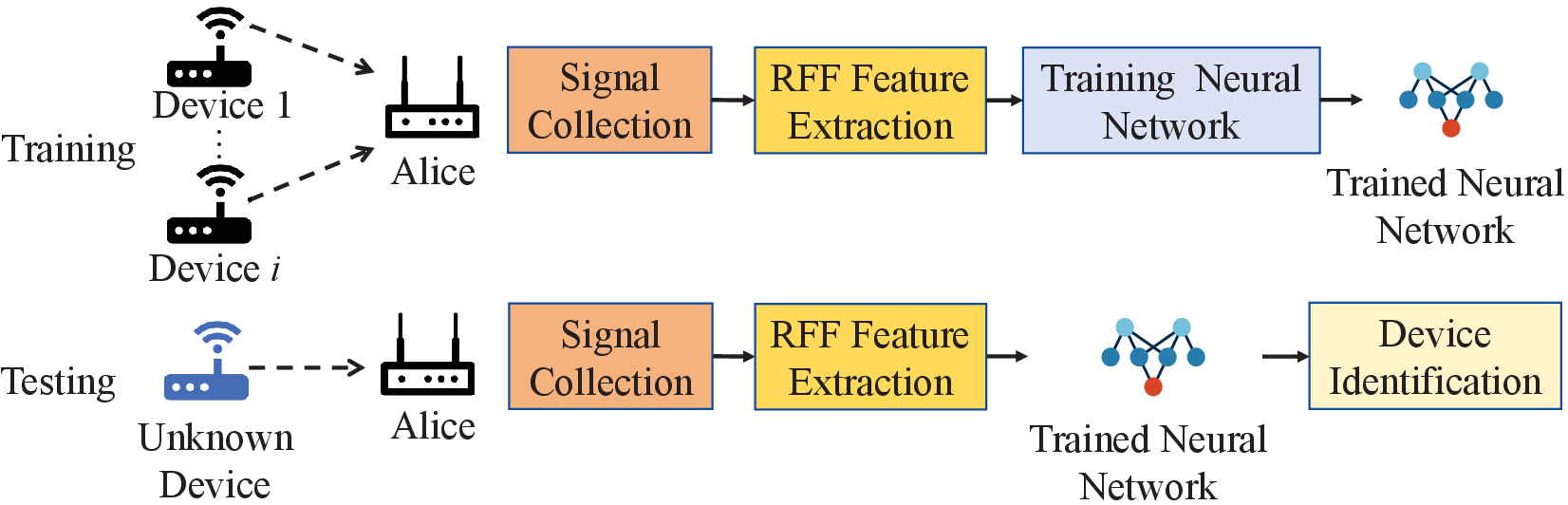}
    \vspace{-0.4cm}
    \caption{System model.}
    \label{fig1}
    \vspace{-0.3cm}
\end{figure}
In Section \ref{section:5}, the performance of RFF feature extraction methods is compared theoretically under various channel scenarios and SNR conditions. Simulation results under various channel scenarios are provided in Section \ref{section:6}, while experimental results are presented in Section \ref{section:7}. Finally, Section \ref{section:8} concludes this work.

\vspace{-0.2cm}
\section{System Model and Silhouette Score-Based Theoretical Evaluation Metric}
\label{section:2}
\vspace{-0.1cm}
\subsection{System Model}
The system model is illustrated in Fig. \ref{fig1}, where a legitimate node, i.e., Alice, is responsible for authenticating the identities of multiple devices. During the training phase, Alice collects labeled signals from the devices, extracts RFF features, and uses extracted features along with their corresponding labels to train a classification model via machine learning. In the testing phase, Alice receives unlabeled signals, extracts their RFF features, and applies the trained model to identify different devices.

\vspace{-0.5cm}
\subsection{Silhouette Score}
The silhouette score, originally proposed for comparing the performance of different clustering methods in unsupervised learning tasks \cite{rousseeuw1987silhouettes}, is utilized in this paper to evaluate the effectiveness of various RFF feature extraction methods. Before computing the silhouette score, it is necessary to determine the intra-class and inter-class distances between training and test samples. Considering two devices, i.e., device $i$ and device $j$, each has $N$ training samples and $M$ test samples. The extracted RFF feature vector of the $n$th training sample from device $i$ is denoted as ${\bf{r}}_{i,n}^{{\rm{Tr}}} = \left[ {r_{i,n,1}^{{\rm{Tr}}},r_{i,n,1}^{{\rm{Tr}}},...,r_{i,n,k}^{{\rm{Tr}}},...,r_{i,n,K}^{{\rm{Tr}}}} \right]$, where ${r_{i,n,k}^{{\rm{Tr}}}}$ represents the RFF feature on the $k$th subcarrier and $K$ denotes the number of subcarriers. The intra-class distance between the $n$th training sample and all test samples of device $i$ is given by
\begin{equation}
\label{eq1}
\begin{array}{l}
D_{i,n}^{{\rm{Intra}}} = \frac{1}{M}\sum\limits_{m = 1}^M {{{\left| {\overline {\bf{r}} _{i,n}^{{\rm{Tr}}} - \overline {\bf{r}} _{i,m}^{{\rm{Te}}}} \right|}^2}} \\
 = \frac{1}{M}\sum\limits_{m = 1}^M {{{\left| {\frac{{{\bf{r}}_{i,n}^{{\rm{Tr}}} - \mu _{i,n}^{{\rm{Tr}}}}}{{\sigma _{i,n}^{{\rm{Tr}}}}} - \frac{{{\bf{r}}_{i,m}^{{\rm{Te}}} - \mu _{i,m}^{{\rm{Te}}}}}{{\sigma _{i,m}^{{\rm{Te}}}}}} \right|}^2}} \\
 = \frac{1}{M}\sum\limits_{m = 1}^M {\sum\limits_{k = 1}^K {\left( {\frac{{r_{i,n,k}^{{\rm{Tr}}} - \mu _{i,n}^{{\rm{Tr}}}}}{{\sigma _{i,n}^{{\rm{Tr}}}}}} \right.} } {\left. { - \frac{{r_{i,m,k}^{{\rm{Te}}} - \mu _{i,m}^{{\rm{Te}}}}}{{\sigma _{i,m}^{{\rm{Te}}}}}} \right)^2},
\end{array}
\end{equation}
where ${r_{i,m,k}^{{\rm{Te}}}}$ denote the RFF feature of the $m$th samples from device $i$ on the $k$th subcarrier in the test set. ${{\bf{r}}_{i,m}^{{\rm{Te}}}}$ represent the RFF feature vector of the $m$th test samples from device $i$, extracted across all subcarriers. ${\mu _{i,n}^{{\rm{Tr}}}}$ and ${\sigma _{i,n}^{{\rm{Tr}}}}$ represent the mean and standard deviation of the feature amplitudes across subcarriers for ${{\bf{r}}_{i,n}^{{\rm{Tr}}}}$, while ${\mu _{i,m}^{{\rm{Te}}}}$ and ${\sigma _{i,m}^{{\rm{Te}}}}$ denote the corresponding statistics for ${{\bf{r}}_{i,m}^{{\rm{Te}}}}$. ${\overline {\mathbf{r}} _{i,n}^{{\rm{Tr}}}}$ and ${\overline {\mathbf{r}} _{i,m}^{{\rm{Te}}}}$ represent the normalized RFF feature vectors of device $i$'s $n$th training sample and $m$th test sample, respectively. The RFF feature normalization process, similar to min-max normalization, is adopted for its analytical tractability and its ability to achieve high classification accuracy, as demonstrated in Section \ref{section:3} and Section \ref{section:7}. Assuming that the RFF features are independently and identically distributed (i.i.d.) across different subcarriers, the intra-class distance of the $n$th training sample and all test samples of device $i$ can be expressed as
\begin{equation}
\begin{array}{l}
\label{eq2}
D_{i,n}^{{\rm{Intra}}} = \frac{K}{M}\sum\limits_{m = 1}^M {{{\left( {\frac{{{r}_{i,n,1}^{{\rm{Tr}}} - \mu _{i,n}^{{\rm{Tr}}}}}{{\sigma _{i,n}^{{\rm{Tr}}}}} - \frac{{{r}_{i,m,1}^{{\rm{Te}}} - \mu _{i,m}^{{\rm{Te}}}}}{{\sigma _{i,m}^{{\rm{Te}}}}}} \right)}^2}.}
\end{array}
\end{equation}
Meanwhile, the inter-class distance between the $n$th training sample of device $i$ and the test samples of device $j$ is defined as
\begin{equation}
\begin{array}{l}
\label{eq3}
D_{i,j,n}^{{\rm{Inter}}} = \frac{1}{M}\sum\limits_{m = 1}^M {{{\left| {\overline {{\bf{r}}} _{i,n}^{{\rm{Tr}}} - \overline {{\bf{r}}} _{j,m}^{{\rm{Te}}}} \right|}^2},}
\end{array}
\end{equation}
where ${\overline {\mathbf{r}} _{j,m}^{{\rm{Te}}}}$ represents the normalized RFF feature vector of $m$th test sample from device $j$. Furthermore, since the RFF feature is i.i.d. across different subcarriers, the inter-class distance between the $n$th training sample of device $i$ and all test samples from device $j$ can be expressed as
\begin{equation}
\label{eq4}
\begin{array}{l}
D_{i,j,n}^{{\rm{Inter}}} = \frac{K}{M}\sum\limits_{m = 1}^M {{{\left( {\frac{{r_{i,n,1}^{{\rm{Tr}}} - \mu _{i,n}^{{\rm{Tr}}}}}{{\sigma _{i,n}^{{\rm{Tr}}}}} - \frac{{r_{j,m,1}^{{\rm{Te}}} - \mu _{j,m}^{{\rm{Te}}}}}{{\sigma _{j,m}^{{\rm{Te}}}}}} \right)}^2}.}
\end{array}
\end{equation}
Then, the silhouette coefficient of the $n$th training sample from device $i$ is
\begin{equation}
\begin{array}{l}
\label{eq5}
{S_{i,n}} = \frac{{D_{i,j,n}^{{\rm{Inter}}} - D_{i,n}^{{\rm{Intra}}}}}{{\max \left( {D_{i,j,n}^{{\rm{Inter}}},D_{i,n}^{{\rm{Intra}}}} \right)}}.
\end{array}
\end{equation}
It can be observed that the silhouette coefficient satisfies the condition $ - 1 \le {S_{i,n}} \le 1$. Then, the average silhouette score can be obtained as
\begin{equation}
\begin{array}{l}
\label{eq6}
S = \frac{1}{{2N}}\sum\limits_{d \in \left\{ {i,j} \right\}} {\sum\limits_{n = 1}^N {{S_{d,n}}} }.
\end{array}
\end{equation}
It should be noted that a higher silhouette score $S$ indicates lower intra-class distance and higher inter-class distance. Both lower intra-class distance and higher inter-class distance are beneficial for enhancing classification accuracy.

\vspace{-0.3cm}
\section{The Existing Division-Based and Precoding-Based RFF Extraction Methods and Their Silhouette Scores}
\label{section:3}
\vspace{-0.1cm}
\subsection{The Existing Division-Based and Precoding-Based RFF Extraction Methods}
In this subsection, we analyze three existing RFF feature extraction methods along with a baseline method. Three different channel scenarios are considered, including the deterministic channel scenario, i.i.d. stochastic channel scenario, and non-i.i.d. stochastic channel scenario, which are denoted as $\mathrm{det}$, $\mathrm{iid}$, and $\mathrm{non}$, respectively.
\subsubsection{Division-Based RFF Extraction Method Using Adjacent Preamble}
In \cite{chen2022radio}, the RFF feature is extracted by dividing the WiFi frame's legacy short training field (L-STF) part and legacy long training field (L-LTF) part in the frequency domain, which is denoted as the short-long division (SL) method. The L-STF part occupies 12 subcarriers, excluding the DC subcarrier. Since RFF is assumed to manifest as a convolutional effect in the time domain, the signals transmitted by device $i$ and received by Alice in the L-STF and L-LTF parts can be expressed as
\begin{equation}
\begin{array}{l}
\label{eq7}
y_{\mathrm{SL},\mathrm{S},i}^{c}=f_{\mathrm{RxA}}\otimes h_c\otimes f_{\mathrm{TxU}_{\mathrm{S},i}}\otimes x_{\mathrm{T}}+n_{\mathrm{SL},\mathrm{S}},
\\
y_{\mathrm{SL},\mathrm{L},i}^{c}=f_{\mathrm{RxA}}\otimes h_c\otimes f_{\mathrm{TxU}_{\mathrm{L},i}}\otimes x_{\mathrm{T}}+n_{\mathrm{SL},\mathrm{L}},
\end{array}
\end{equation}
where $f_{\mathrm{RxA}}$ denotes the RFF introduced by Alice's receiving front end. $h_c$ represents the channel between the device and Alice under channel scenario $c$, where $c\in \left\{ \mathrm{det,iid,non} \right\}$. $f_{\mathrm{TxU}_{\mathrm{S},i}}$ and $f_{\mathrm{TxU}_{\mathrm{L},i}}$ denote the RFF of device $i$'s transmitting front end in the L-STF part and L-LTF part, respectively. $x_{\mathrm{T}}$ denotes the preamble signal, while $n_{\mathrm{SL},\mathrm{S}}$ and $n_{\mathrm{SL},\mathrm{L}}$ represent Gaussian noise with zero mean and variance $\sigma _N^2$ in the L-STF and L-LTF parts, respectively. Let the Fourier transform of $y_{\mathrm{SL},\mathrm{S},i}^{c}$ and $y_{\mathrm{SL},\mathrm{L},i}^{c}$ be noted as $Y_{\mathrm{SL},\mathrm{S},i}^{c}$ and $Y_{\mathrm{SL},\mathrm{L},i}^{c}$, respectively. By dividing $Y_{\mathrm{SL},\mathrm{S},i}^{c}$ by $Y_{\mathrm{SL},\mathrm{L},i}^{c}$, the extracted RFF feature of device $i$ using the SL method is given by
\begin{equation}
\begin{array}{l}
\label{eq8}
r_{\mathrm{SL},i}^{c}=\frac{F_{\mathrm{RA}}\cdot H_c\cdot F_{\mathrm{TU}_{\mathrm{S},i}}\cdot X+N_{\mathrm{SL},\mathrm{S}}}{F_{\mathrm{RA}}\cdot H_c\cdot F_{\mathrm{TU}_{\mathrm{L},i}}\cdot X+N_{\mathrm{SL},\mathrm{L}}},
\end{array}
\end{equation}
where $F_{\mathrm{RA}}$, $H_c$, $F_{\mathrm{TU}_{\mathrm{S},i}}$, $F_{\mathrm{TU}_{\mathrm{L},i}}$, $X$, $N_{\mathrm{SL},\mathrm{S}}$, and $N_{\mathrm{SL},\mathrm{L}}$ denote the Fourier transforms of $f_{\mathrm{RxA}}$, $h_c$, $f_{\mathrm{TxU}_{\mathrm{S},i}}$, $f_{\mathrm{TxU}_{\mathrm{L},i}}$, $x_\mathrm{T}$, $n_{\mathrm{SL},\mathrm{S}}$, and $n_{\mathrm{SL},\mathrm{L}}$, respectively.

The extracted RFF feature of the $d$th sample from device $i$ on the $k$th subcarrier in the frequency domain can be expressed as
\begin{equation}
\begin{array}{l}
\label{eq9}
r_{{\rm{SL}},i,d,k}^c = \frac{{{F_{{\rm{RA}}}} \cdot {H_{c,k}} \cdot {F_{{\rm{T}}{{\rm{U}}_{{\rm{S}},i,k}}}} \cdot X + {N_{{\rm{SL}},{\rm{S}},d,k}}}}{{{F_{{\rm{RA}}}} \cdot {H_{c,k}} \cdot {F_{{\rm{T}}{{\rm{U}}_{{\rm{L}},i,k}}}} \cdot X + {N_{{\rm{SL}},{\rm{L}},d,k}}}},
\end{array}
\end{equation}
where $H_{c,k}$ represents the channel state information (CSI) on the $k$th subcarrier under channel scenario $c$. ${{F_{\mathrm{T}{\mathrm{U}_{\mathrm{S},i,k}}}}}$ and ${{F_{\mathrm{T}{\mathrm{U}_{\mathrm{L},i,k}}}}}$ are the RFF of device $i$'s transmitting front end on the $k$th subcarrier in the L-STF part and L-LTF part, respectively. ${N_{{\rm{SL}},{\rm{S}},d,k}}$ and ${N_{{\rm{SL}},{\rm{L}},d,k}}$ represent Gaussian noise with zero mean and variance $\sigma _N^2$ in the L-STF part and L-LTF part of the $d$th sample on the $k$th subcarrier, respectively.

Since the legitimate node Alice can be equipped with high-end hardware, it is assumed that $F_\mathrm{RA}$ remains constant across different subcarriers and is denoted by $f_\mathrm{RA}$. To focus on the RFF effects in the L-STF part, we assume that $F_{\mathrm{T}{\mathrm{U}_{\mathrm{L},i,k}}}$ is constant across different subcarriers and denote it as $f_{\mathrm{T}{\mathrm{U}_{\mathrm{L},i}}}$. Moreover, the differences in RFF effects at the receivers of device $i$ and device $j$ in the L-LTF part are assumed to be negligible, i.e., $f_{\mathrm{TU}_{\mathrm{L},i}}=f_{\mathrm{TU}_{\mathrm{L},j}}=f_{\mathrm{TU_L}}$. Since RFF is influenced by multiple hardware impairments, $F_{\mathrm{T}{\mathrm{U}_{\mathrm{S},i,k}}}$ is assumed to follow the i.i.d. Gaussian distribution across different subcarriers with mean ${\mu _{{\mathrm{S}_i}}}$ and variance $\sigma _{{\mathrm{S}_i}}^2$, in accordance with the central limit theorem. Furthermore, the transmitter RFF effects of device $i$ and device $j$ in the L-STF part are assumed to follow the i.i.d. Gaussian distribution, i.e., $\mu _{\mathrm{S}_i}=\mu _{\mathrm{S}_j}=\mu _\mathrm{S}$ and $\sigma _{\mathrm{S}_i}^{2}=\sigma _{\mathrm{S}_j}^{2}=\sigma _\mathrm{S}^{2}$. The amplitude value of the preamble signal $X$ is constant across different subcarriers and can be denoted as $x$.

To derive the mean and variance of ${r}_{{\rm{SL}},i,d,k}^c$, i.e., the extracted RFF feature of the $d$th sample from device $i$ on the $k$th subcarrier in the frequency domain, and to formulate the expression for expected intra-class distance, the following claim is introduced.

\textbf{Claim 1.} Let $Z = \frac{G}{{\rho G + W}}$, where $G$ is a Gaussian random variable with mean ${\mu _G}$ and variance $\sigma _G^2$. $W$ is a Gaussian random variable with mean ${\mu _W}$ and variance $\sigma _W^2$, where ${\mu _W} = 0$. Then, it can be obtained that
\begin{equation}
\label{eq10}
\begin{array}{l}
{\mathbb E}\left[ Z \right] \approx \frac{{{\rho ^2}\mu _G^2 + \sigma _W^2}}{{{\rho ^3}\mu _G^2}},\\
{\mathbb E}\left[ {{Z^2}} \right] \approx \frac{{{\rho ^2}\mu _G^2 + 3\sigma _W^2}}{{{\rho ^4}\mu _G^2}}.
\end{array}
\end{equation}

\textit{Proof}: For further details, please refer to Appendix \ref{appendix:A}.$\hfill\blacksquare$

The mean and variance of the RFF feature ${r}_{{\rm{SL}},i,d,k}^c$ can then be obtained as
\begin{equation}	
\begin{small}
\label{eq11}
\begin{aligned}
\mu _{\mathrm{SL}_c}\approx& \frac{f_\mathrm{RA}x\mu _\mathrm{S}\left( \gamma ^2\mu _{H_c}^{2}+\sigma _{N}^{2} \right)}{\gamma ^3\mu _{H_c}^{2}},
\\
\sigma _{\mathrm{SL}_c}^{2}\approx& \{f_\mathrm{RA}^{2}x^2[\mu _\mathrm{S}^{2}\sigma _{N}^{2}\left( \gamma ^2\mu _{H_c}^{2}-\sigma _{N}^{2} \right) +\gamma ^2\mu _{H_c}^{2}\sigma _\mathrm{S}^{2}(\gamma ^2\mu _{H_c}^{2}
\\
&+3\sigma _{N}^{2})]+\gamma ^2\sigma _{N}^{2}\left( \gamma ^2\mu _{H_c}^{2}+3\gamma ^2\sigma _{H_c}^{2}+3\sigma _{N}^{2} \right\} /\left( \gamma ^6\mu _{H_c}^{4} \right) ,
\end{aligned}
\end{small}
\end{equation}
where $\gamma$ is an auxiliary variable, defined as $\gamma \triangleq f_{\mathrm{RA}}\cdot f_{\mathrm{TU}_{\mathrm{L}}}\cdot x$.

\subsubsection{Division-Based RFF Extraction Method with Additional Perfect Link}
In \cite{dong2024robust}, an additional perfect link is required, and the RFF feature is extracted by dividing the signal received by Alice and the signal received by device $i$ in the frequency domain. Since the challenge-response mode is employed in \cite{dong2024robust}, this method is referred to as the challenge-response (CR) method. In frequency domain, the extracted RFF feature of $d$th sample from device $i$ on the $k$th subcarrier, where $c\in \left\{ \mathrm{det,iid,non} \right\}$, can be expressed as
\begin{equation}
\begin{array}{l}
\label{eq12}
{r}_{{\rm{CR}},i,d,k}^c = \frac{{{F_{{\rm{RA}}}} \cdot {H_{c,k}} \cdot {F_{{\rm{T}}{{\rm{U}}_{i,k}}}} \cdot X + {N_{{\rm{CR}},{\rm{A}},d,k}}}}{{{F_{{\rm{R}}{{\rm{U}}_{i,k}}}} \cdot {H_{c,k}} \cdot {F_{{\rm{TA}}}} \cdot X + {N_{{\rm{CR}},{\rm{U}},d,k}}}},
\end{array}
\end{equation}
where ${{F_{\mathrm{R}{\mathrm{U}_{i,k}}}}}$ and ${{F_{\mathrm{T}{\mathrm{U}_{i,k}}}}}$ are the RFF of device $i$ on the $k$th subcarrier in receiving and transmitting front ends, respectively. $F_{\mathrm{TA}}$ is the Alice's RFF in the transmitting front end. ${{N_{{\rm{CR}},{\rm{A}},d,k}}}$ and ${{N_{{\rm{CR}},{\rm{U}},d,k}}}$ are Gaussian noise with zero mean and variance $\sigma _N^2$.

To isolate the transmitter RFF effect of device $i$, the receiver RFF effect $F_{\mathrm{RU}_{i,k}}$ is assumed to be constant over subcarriers and can be denoted as $f_{\mathrm{RU}_i}$. Similar to most RFF feature extraction studies \cite{zhang2021radio, xie2024radio}, the difference in receiver RFF effects between device $i$ and device $j$ is assumed to be negligible, i.e., $f_{\mathrm{RU}_i}=f_{\mathrm{RU}_j}=f_\mathrm{RU}$. The transmitter RFF effect of device $i$, i.e., $F_{\mathrm{TU}_{i,k}}$, is assumed to follow the i.i.d. Gaussian distribution across different subcarriers with mean ${\mu _{{\mathrm{U}_i}}}$ and variance $\sigma _{{\mathrm{U}_i}}^2$. Moreover, the transmitter RFF effects of device $i$ and device $j$ are assumed to follow the i.i.d. Gaussian distribution, i.e., $\mu _{\mathrm{U}_i}=\mu _{\mathrm{U}_j}=\mu _\mathrm{U}$ and $\sigma _{\mathrm{U}_i}^{2}=\sigma _{\mathrm{U}_j}^{2}=\sigma _\mathrm{U}^{2}$.

The mean and variance of ${r}_{{\rm{CR}},i,d,k}^c$, i.e., the extracted RFF feature of $d$th sample from device $i$ on the $k$th subcarrier, can then be expressed as
\begin{equation}	
\begin{small}
\label{eq13}
\begin{aligned}
{\mu _{{\rm{C}}{{\rm{R}}_c}}} \approx& \frac{{{f_\mathrm{RA}}x{\mu _{\mathrm{U}}}\left( {{\beta ^2}\mu _{{H_c}}^2 + \sigma _N^2} \right)}}{{{\beta ^3}\mu _{{H_c}}^2}},\\
\sigma _{{\rm{C}}{{\rm{R}}_c}}^2 \approx& \{ f_\mathrm{RA}^2{x^2}[\mu _{\mathrm{U}}^2\sigma _N^2\left( {{\beta ^2}\mu _{{H_c}}^2 - \sigma _N^2} \right) + {\beta ^2}\mu _{{H_c}}^2\sigma _{\mathrm{U}}^2({\beta ^2}\mu _{{H_c}}^2\\
& + 3\sigma _N^2)] + {\beta ^2}\sigma _N^2\left( {{\beta ^2}\mu _{{H_c}}^2 + 3{\beta ^2}\sigma _{{H_c}}^2 + 3\sigma _N^2} \right)\} /\left( {{\beta ^6}\mu _{{H_c}}^4} \right),
\end{aligned}
\end{small}
\end{equation}
where $\beta$ is an auxiliary variable, defined as $\beta \triangleq f_{\mathrm{RU}}\cdot f_{\mathrm{TA}}\cdot x$.

\subsubsection{Precoding-Based RFF Extraction Method with Channel Estimation}
Since the feature extraction in \cite{sun2023location} involves a precoding procedure, the corresponding method is referred to as the precoding (PC) method. In \cite{sun2023location}, Alice first transmits a preamble signal to the device to be identified. In the frequency domain, the signal received by the device can be expressed as
\begin{equation}
\label{eq14}
{Y_{{\rm{PC}},{\rm{U}}}} = {F_{{\rm{RU}}}} \cdot {H_{{\rm{AU}}}} \cdot {F_{{\rm{TA}}}} \cdot X + {N_{{\rm{PC}},{\rm{U}}}},
\end{equation}
where $H_\mathrm{AU}$ denotes the channel from Alice to the device and $N_{\mathrm{PC,U}}$ is Gaussian noise with zero mean and variance $\sigma _N^2$. Then, the device estimates the channel as
\begin{equation}
\label{eq15}
\widetilde{H}_\mathrm{AU}=\frac{Y_{\mathrm{PC,U}}}{X}=\frac{F_{\mathrm{RU}}\cdot H_\mathrm{AU}\cdot F_\mathrm{TA}\cdot X+N_{\mathrm{PC,U}}}{X}.
\end{equation}

After that, the device applies the precoding process, and the processed signal can be expressed as
\begin{equation}
\label{eq16}
Y_{\mathrm{PC,U}}^{\prime}=\frac{Y_{\mathrm{PC,U}}}{\widetilde{H}_\mathrm{AU}^{2}}=\frac{X^2}{F_{\mathrm{RU}}\cdot H_\mathrm{AU}\cdot F_\mathrm{TA}\cdot X+N_{\mathrm{PC,U}}}.
\end{equation}

After the device transmits $Y_{\mathrm{PC,U}}^{\prime}$ to Alice, the signal received by Alice can be expressed as
\begin{equation}
\label{eq17}
\begin{aligned}
Y_{\mathrm{PC,A}}=&F_\mathrm{RA}\cdot H_\mathrm{UA}\cdot F_{\mathrm{TU}}\cdot Y_{\mathrm{PC,U}}^{\prime}+N_{\mathrm{PC,A}}
\\
=&\frac{F_\mathrm{RA}\cdot H_\mathrm{UA}\cdot F_{\mathrm{TU}}\cdot X^2}{F_{\mathrm{RU}}\cdot H_\mathrm{AU}\cdot F_\mathrm{TA}\cdot X+N_{\mathrm{PC,U}}}+N_{\mathrm{PC,A}},
\end{aligned}
\end{equation}
where $H_\mathrm{UA}$ represents the channel from the device to Alice and $N_{\mathrm{PC,A}}$ is Gaussian noise with zero mean and variance $\sigma _N^2$. The extracted RFF feature of the $d$th sample from device $i$ on the $k$th subcarrier, where $c\in \left\{ \mathrm{det,iid,non} \right\}$, can be expressed as
\begin{equation}
\begin{small}
\begin{aligned}
\label{eq18}
r_{{\rm{PC}},i,d,k}^c = \frac{{{F_{{\rm{RA}}}} \cdot {H_{c,k}} \cdot {F_{{\rm{T}}{{\rm{U}}_{i,k}}}} \cdot {X^2}}}{{{F_{{\rm{R}}{{\rm{U}}_{i,k}}}} \cdot {H_{c,k}} \cdot {F_{{\rm{TA}}}} \cdot X + {N_{{\rm{PC}},{\rm{U}},d,k}}}} + {N_{{\rm{PC}},{\rm{A}},d,k}}.
\end{aligned}
\end{small}
\end{equation}
The mean and variance of $r_{{\rm{PC}},i,d,k}^c$ can be derived as
\begin{equation}	
\begin{small}
\label{eq19}
\begin{aligned}
\mu _{\mathrm{PC}_c}\approx &\frac{f_\mathrm{RA}x^2\mu _\mathrm{U}(\beta ^2\mu _{H_c}^{2}+\sigma _{N}^{2})}{\beta ^3\mu _{H_c}^{2}},
\\
\sigma _{\mathrm{PC}_c}^{2}\approx& \left\{ f_\mathrm{RA}^{2}x^4\left[ \mu _\mathrm{U}^{2}\sigma _{N}^{2}\left( \beta ^2\mu _{H_c}^{2}-\sigma _{N}^{2} \right) \right. \right.
\\
&\left. \left. +\beta ^2\mu _{H_c}^{2}\sigma _{\mathrm{U}}^{2}\left( \beta ^2\mu _{H_c}^{2}+3\sigma _{N}^{2} \right) \right] /\left( \beta ^6\mu _{H_c}^{4} \right) \right\} +\sigma _{N}^{2}.
\end{aligned}
\end{small}
\end{equation}
Meanwhile, we denote the baseline method, which uses the received raw IQ data directly without a channel removing process, as the raw (RAW) method. In the RAW method, the received signal from device $i$ to Alice can be expressed as
\vspace{-0.2cm}
\begin{equation}
\begin{array}{l}
\label{eq20}
Y_{\mathrm{RAW,A}}=F_\mathrm{RA}\cdot H_\mathrm{UA}\cdot F_{\mathrm{TU}_i}\cdot X+N_{\mathrm{RAW,A}},
\end{array}
\end{equation}
where $N_{\mathrm{RAW,A}}$ is Gaussian noise with zero mean and variance $\sigma _N^2$. The extracted RFF feature of the $d$th sample from device $i$ on the $k$th subcarrier can be expressed as
\begin{equation}
\begin{array}{l}
\label{eq21}
{r}_{{\rm{RAW}},i,d,k}^c = {F_{{\rm{RA}}}} \cdot {H_{c,k}} \cdot {F_{{\rm{T}}{{\rm{U}}_{i,k}}}} \cdot X + {N_{{\rm{RAW,A}},d,k}}.
\end{array}
\end{equation}
The mean and variance of $r_{\mathrm{RAW},i,d,k}^{c}$ can be obtained as
\begin{equation}	
\begin{small}
\label{eq22}
\begin{aligned}
&\mu _{\mathrm{RAW}c}=f_\mathrm{RA}x\mu _\mathrm{U}\mu _{H_c},
\\
&\sigma _{\mathrm{RAW}c}^{2}=f_\mathrm{RA}^{2}x^2\left( \mu _\mathrm{U}^{2}\sigma _{H_c}^{2}+\sigma _\mathrm{U}^{2}\mu _{H_c}^{2}+\sigma _\mathrm{U}^{2}\sigma _{H_c}^{2} \right) +\sigma _{N}^{2}.
\end{aligned}
\end{small}
\end{equation}
\vspace{-0.3cm}

\subsection{Silhouette Score under Various Channel Scenarios}
In this subsection, three different channel scenarios are considered, including the deterministic channel scenario, i.i.d. stochastic channel scenario, and non-i.i.d. stochastic channel scenario. In the deterministic channel scenario, both the training and test samples experience the same channel condition, i.e., $H_\mathrm{det}$, which follows a Gaussian distribution with mean ${\mu _{{H_\mathrm{det}}}}$ and variance $\sigma _{{H_\mathrm{det}}}^2$. In the i.i.d. stochastic channel scenario, the training and test samples experience i.i.d. Gaussian channels, i.e., $H_\mathrm{det}$ and $H_\mathrm{iid}$, both of which follow i.i.d. Gaussian distributions. The mean and variance of $H_\mathrm{iid}$ are ${\mu _{{H_\mathrm{iid}}}}$ and $\sigma _{H_\mathrm{iid}}^2$, respectively. Moreover, the equality $\mu _{H_\mathrm{det}}=\mu _{H_\mathrm{iid}}$ and $\sigma _{H_\mathrm{det}}^{2}=\sigma _{H_\mathrm{iid}}^{2}$ hold. For expressional simplicity, we denote $\mu _{H_\mathrm{det}}=\mu _{H_\mathrm{iid}}=\mu _H$ and $\sigma _{H_\mathrm{det}}^{2}=\sigma _{H_\mathrm{iid}}^{2}=\sigma _H^{2}$. In the non-i.i.d. stochastic channel scenario, the training and test samples experience non-i.i.d. Gaussian channels, i.e., $H_\mathrm{det}$ and $H_\mathrm{non}$, where $H_\mathrm{non}$ follows a Gaussian distribution with mean ${\mu _{{H_\mathrm{non}}}}$ and variance $\sigma _{{H_\mathrm{non}}}^2$. The channels $H_\mathrm{det}$ and $H_\mathrm{non}$ are independent, and either ${\mu _{{H_\mathrm{det}}}} \ne {\mu _{{H_\mathrm{non}}}}$ or $\sigma _{{H_\mathrm{det}}}^2 \ne \sigma _{{H_\mathrm{non}}}^2$ holds. It should be noted that the deterministic channel scenario corresponds to a static channel, for example, when the transmitter, receiver, and surrounding environment remain unchanged. The i.i.d. stochastic channel scenario corresponds to channels that vary over time while the distribution of the CSI remains invariant, such as when the transmitter moves along a circular trajectory around the receiver. The non-i.i.d. stochastic channel scenario corresponds to both time-varying channels and varying CSI distributions, for instance, when the distance between the transmitter and receiver changes or under mixed LOS and NLOS conditions.

\subsubsection{Deterministic Channel Scenario}
Under the deterministic channel scenario, we can obtain the intra-class distances, inter-class distances, and silhouette scores for different RFF feature extraction methods as follows.

\subsubsubsection{Intra-Class Distance} It should be noted that the terms ${\mu _{i,n}^{{\rm{Tr}}}}$ and ${\mu _{i,m}^{{\rm{Te}}}}$ in (\ref{eq2}), i.e., the means of the feature amplitudes across subcarriers for ${{\bf{r}}_{i,n}^{{\rm{Tr}}}}$ and ${{\bf{r}}_{i,m}^{{\rm{Te}}}}$, are identical under the deterministic channel scenario. Similarly, the terms ${\sigma _{i,n}^{{\rm{Tr}}}}$ and ${\sigma _{i,m}^{{\rm{Te}}}}$ in (\ref{eq2}), i.e., the standard deviations of the feature amplitudes across subcarriers for ${{\bf{r}}_{i,n}^{{\rm{Tr}}}}$ and ${{\bf{r}}_{i,m}^{{\rm{Te}}}}$, are also identical. Therefore, the expected intra-class distances for device $i$ under the deterministic channel scenario for the RAW, SL, CR, and PC methods can be expressed as
\vspace{-0.2cm}
\begin{equation}	
\begin{small}
\label{eq23}
\begin{array}{l}
D_{\mathrm{RAW},i}^{\mathrm{Intra},\det}=\frac{2R_{\mathrm{L}}\sigma _{N}^{2}}{\sigma _{\mathrm{RAW}_{\det}}^{2}},
\\
D_{\mathrm{SL},i}^{\mathrm{Intra},\det}\approx \frac{2R_{\mathrm{S}}\left\{ f_{\mathrm{RA}}^{2}x^2\mu _{H}^{2}\sigma _{N}^{2}\left( \mu _{\mathrm{S}}^{2}+\sigma _{\mathrm{S}}^{2} \right) +\sigma _{N}^{2}\left[ \gamma ^2\left( \mu _{H}^{2}+3\sigma _{H}^{2} \right) +3\sigma _{N}^{2} \right] \right\}}{\sigma _{\mathrm{SL}_{\det}}^{2}\gamma ^4\mu _{H}^{4}},
\\
D_{\mathrm{CR},i}^{\mathrm{Intra},\det}\approx \frac{2R_{\mathrm{L}}\left\{ f_{\mathrm{RA}}^{2}x^2\mu _{H}^{2}\sigma _{N}^{2}\left( \mu _{\mathrm{U}}^{2}+\sigma _{\mathrm{U}}^{2} \right) +\sigma _{N}^{2}\left[ \beta ^2\left( \mu _{H}^{2}+3\sigma _{H}^{2} \right) +3\sigma _{N}^{2} \right] \right\}}{\sigma _{\mathrm{CR}_{\det}}^{2}\beta ^4\mu _{H}^{4}},
\\
D_{\mathrm{PC},i}^{\mathrm{Intra},\det}\approx \frac{2R_{\mathrm{L}}}{\sigma _{\mathrm{PC}_{\det}}^{2}}\left[ \frac{f_{\mathrm{RA}}^{2}x^4\sigma _{N}^{2}\left( \mu _{\mathrm{U}}^{2}+\sigma _{\mathrm{U}}^{2} \right)}{\beta ^4\mu _{H}^{2}}+\sigma _{N}^{2} \right] ,
\end{array}
\end{small}
\end{equation}
where $R_\mathrm{S}$ and $R_\mathrm{L}$ are the number of occupied subcarriers in the L-STF part and the L-LTF part, respectively.

\subsubsubsection{Inter-Class Distance} The following claim is introduced to derive the expected inter-class distance for each training sample of device $i$ in the PC method.

\textbf{Claim 2.} Let $Z = \frac{{{G^2}}}{{\left( {\rho G + {W_1}} \right)\left( {\rho G + {W_2}} \right)}}$, where $G$ is a Gaussian random variable with mean $\mu _G$ and variance $\sigma _G^2$. $W_1$ and $W_2$ are i.i.d. Gaussian random variables with zero mean and variance $\sigma _W^2$. Then, we have
\begin{equation}
\begin{array}{l}
\label{eq24}
{\mathbb E}\left[ Z \right] \approx \frac{{{\rho ^2}\mu _G^2 + 2\sigma _W^2}}{{{\rho ^4}\mu _G^2}}.
\end{array}
\end{equation}

\textit{Proof}: By applying the Taylor series expansion to $Z$ around $(\mu_G,0,0)$ and subsequently taking the expectation, the results in (\ref{eq24}) can be derived.
$\hfill\blacksquare$

In (\ref{eq4}), the terms ${\mu _{i,n}^{{\rm{Tr}}}}$ and ${\mu _{j,m}^{{\rm{Te}}}}$ are identical under the deterministic channel scenario. Similarly, the terms ${\sigma _{i,n}^{{\rm{Tr}}}}$ and ${\sigma _{j,m}^{{\rm{Te}}}}$ are also identical. The expected inter-class distances for the RAW, SL, CR, and PC methods are then given by
\vspace{-0.1cm}
 \begin{equation}	
 \begin{small}
 \label{eq25}
 \begin{aligned}
D_{\mathrm{RAW},i,j}^{\mathrm{Inter},\det}=&2R_{\mathrm{L}}\left[ f_{\mathrm{RA}}^{2}x^2\sigma _{\mathrm{U}}^{2}(\mu _{H}^{2}+\sigma _{H}^{2})+\sigma _{N}^{2} \right] /\sigma _{\mathrm{RAW}_{\det}}^{2},
\\
D_{\mathrm{SL},i,j}^{\mathrm{Inter},\det}\approx& 2R_{\mathrm{S}}\{f_{\mathrm{RA}}^{2}x^2\mu _{H}^{2}\left[ \mu _{\mathrm{S}}^{2}\sigma _{N}^{2}+\sigma _{\mathrm{S}}^{2}(\gamma ^2\mu _{H}^{2}+3\sigma _{N}^{2}) \right]
\\
&+\sigma _{N}^{2}[\gamma ^2(\mu _{H}^{2}+3\sigma _{H}^{2})+3\sigma _{N}^{2}]\}/(\sigma _{\mathrm{SL}_{\det}}^{2}\gamma ^4\mu _{H}^{4}),
\\
D_{\mathrm{CR},i,j}^{\mathrm{Inter},\det}\approx& 2R_{\mathrm{L}}\{f_{\mathrm{RA}}^{2}x^2\mu _{H}^{2}\left[ \mu _{\mathrm{U}}^{2}\sigma _{N}^{2}+\sigma _{\mathrm{U}}^{2}(\beta ^2\mu _{H}^{2}+3\sigma _{N}^{2}) \right]
\\
&+\sigma _{N}^{2}\left[ \beta ^2(\mu _{H}^{2}+3\sigma _{H}^{2})+3\sigma _{N}^{2} \right\} /(\sigma _{\mathrm{CR}_{\det}}^{2}\beta ^4\mu _{H}^{4}),
\\
D_{\mathrm{PC},i,j}^{\mathrm{Inter},\det}\approx& 2R_{\mathrm{L}}\{f_{\mathrm{RA}}^{2}x^4\left[ \mu _{\mathrm{U}}^{2}\sigma _{N}^{2}+\sigma _{\mathrm{U}}^{2}(\beta ^2\mu _{H}^{2}+3\sigma _{N}^{2}) \right]
\\
&+\beta ^4\mu _{H}^{2}\sigma _{N}^{2}\}/(\sigma _{\mathrm{PC}_{\det}}^{2}\beta ^4\mu _{H}^{2}).
\end{aligned}
\end{small}
 \end{equation}
\vspace{-0.1cm}

\subsubsubsection{Silhouette Score} By substituting the intra-class and inter-class distances from (\ref{eq23}) and (\ref{eq25}) into (\ref{eq5}) and applying (\ref{eq6}), the expected silhouette score can be obtained. The expected silhouette scores for different methods are given as
\begin{equation}	
\begin{small}
    \label{eq26}
    \begin{array}{l}
S_{{\rm{RAW}}}^\mathrm{det} = \frac{{f_\mathrm{RA}^2{x^2}\sigma _\mathrm{U}^2\left( {\mu _H^2 + \sigma _H^2} \right)}}{{f_\mathrm{RA}^2{x^2}\sigma _\mathrm{U}^2\left( {\mu _H^2 + \sigma _H^2} \right) + \sigma _N^2}},\\
S_{{\rm{SL}}}^\mathrm{det} \approx \frac{{f_\mathrm{RA}^2{x^2}\sigma _\mathrm{S}^2\mu _H^2\left( {{\gamma ^2}\mu _H^2 + 2\sigma _N^2} \right)}}{{f_\mathrm{RA}^2{x^2}\mu _H^2\left[ {\mu _\mathrm{S}^2\sigma _N^2 + \sigma _\mathrm{S}^2\left( {{\gamma ^2}\mu _H^2 + 3\sigma _N^2} \right)} \right] + \sigma _N^2\left[ {{\gamma ^2}\left( {\mu _H^2 + 3\sigma _H^2} \right) + 3\sigma _N^2} \right]}},\\
S_{{\rm{CR}}}^\mathrm{det} \approx \frac{{f_\mathrm{RA}^2{x^2}\sigma _\mathrm{U}^2\mu _H^2\left( {{\beta ^2}\mu _H^2 + 2\sigma _N^2} \right)}}{{f_\mathrm{RA}^2{x^2}\mu _H^2\left[ {\mu _\mathrm{U}^2\sigma _N^2 + \sigma _\mathrm{U}^2\left( {{\beta ^2}\mu _H^2 + 3\sigma _N^2} \right)} \right] + \sigma _N^2\left[ {{\beta ^2}\left( {\mu _H^2 + 3\sigma _H^2} \right) + 3\sigma _N^2} \right]}},\\
S_{{\rm{PC}}}^\mathrm{det} \approx \frac{{f_\mathrm{RA}^2{x^4}\sigma _\mathrm{U}^2\left( {{\beta ^2}\mu _H^2 + 2\sigma _N^2} \right)}}{{f_\mathrm{RA}^2{x^4}\left[ {\mu _\mathrm{U}^2\sigma _N^2 + \sigma _\mathrm{U}^2\left( {{\beta ^2}\mu _H^2 + 3\sigma _N^2} \right)} \right] + {\beta ^4}\mu _H^2\sigma _N^2}}.
    \end{array}
\end{small}
\end{equation}

\subsubsection{I.i.d. Stochastic Channel Scenario}
The following derivation and results can be obtained under the i.i.d. stochastic channel scenario.

\subsubsubsection{Intra-Class Distance}
The following claim is introduced to derive the expression for device $i$'s intra-class distance in the PC method under the i.i.d. stochastic channel scenario.

\textbf{Claim 3.} Let $Z = \frac{{{G_1}{W_2} - {G_2}{W_1}}}{{\left( {\rho {G_1} + {W_1}} \right)\left( {\rho {G_2} + {W_2}} \right)}}$, where $G_1$ and $G_2$ are i.i.d. Gaussian random variables, with mean $\mu _G$ and variance $\sigma _G^2$. $W_1$ and $W_2$ are also i.i.d. Gaussian random variables with zero mean and variance $\sigma _W^2$. It can be obtained that
\begin{equation}
\label{eq27}
\begin{array}{l}
{\mathbb E}\left[ Z \right] \approx 0,\\
{\mathbb E}\left[ {{Z^2}} \right] \approx \frac{{2\sigma _W^2}}{{{\rho ^4}\mu _G^2}}.
\end{array}
\end{equation}

\textit{Proof}: Please refer to Appendix \ref{appendix:B} for further details.
$\hfill\blacksquare$

Similar to the case under the deterministic channel scenario, the terms ${\mu _{i,n}^{{\rm{Tr}}}}$ and ${\mu _{i,m}^{{\rm{Te}}}}$ in (\ref{eq2}) are identical, and the terms ${\sigma _{i,n}^{{\rm{Tr}}}}$ and ${\sigma _{i,m}^{{\rm{Te}}}}$ in (\ref{eq2}) are also the same. Then, the expected intra-class distances for the RAW, SL, CR, and PC methods can be derived as
\begin{equation}	
\begin{small}
 \label{eq28}
 \begin{aligned}
D_{\mathrm{RAW},i}^{\mathrm{Intra},\mathrm{iid}}=&\frac{2R_{\mathrm{L}}\left[ f_{\mathrm{RA}}^{2}x^2\sigma _{H}^{2}\left( \mu _{\mathrm{U}}^{2}+\sigma _{\mathrm{U}}^{2} \right) +\sigma _{N}^{2} \right]}{\sigma _{\mathrm{RAW}_{\mathrm{iid}}}^{2}},
\\
D_{\mathrm{SL},i}^{\mathrm{Intra},\mathrm{iid}}\approx& 2R_{\mathrm{S}}[f_{\mathrm{RA}}^{2}x^2\sigma _{N}^{2}\left( \mu _{\mathrm{S}}^{2}+\sigma _{\mathrm{S}}^{2} \right) \left( \gamma ^2\mu _{H}^{2}-\sigma _{N}^{2} \right)
\\
&+\gamma ^2\sigma _{N}^{2}\left( \gamma ^2\mu _{H}^{2}+3\gamma ^2\sigma _{H}^{2}+3\sigma _{N}^{2} \right) ]/\left( \sigma _{\mathrm{SL}_{\mathrm{iid}}}^{2}\gamma ^6\mu _{H}^{4} \right) ,
\\
D_{\mathrm{CR},i}^{\mathrm{Intra},\mathrm{iid}}\approx& 2R_{\mathrm{L}}[f_{\mathrm{RA}}^{2}x^2\sigma _{N}^{2}\left( \mu _{\mathrm{U}}^{2}+\sigma _{\mathrm{U}}^{2} \right) \left( \beta ^2\mu _{H}^{2}-\sigma _{N}^{2} \right)
\\
&+\beta ^2\sigma _{N}^{2}\left( \beta ^2\mu _{H}^{2}+3\beta ^2\sigma _{H}^{2}+3\sigma _{N}^{2} \right) ]/\left( \sigma _{\mathrm{CR}_{\mathrm{iid}}}^{2}\beta ^6\mu _{H}^{4} \right) ,
\\
D_{\mathrm{PC},i}^{\mathrm{Intra},\mathrm{iid}}\approx& \frac{2R_{\mathrm{L}}}{\sigma _{\mathrm{PC}_{\mathrm{iid}}}^{2}}\left[ \frac{f_{\mathrm{RA}}^{2}x^4\sigma _{N}^{2}\left( \mu _{\mathrm{U}}^{2}+\sigma _{\mathrm{U}}^{2} \right) \left( \beta ^2\mu _{H}^{2}-\sigma _{N}^{2} \right)}{\beta ^6\mu _{H}^{4}}+\sigma _{N}^{2} \right] .
 \end{aligned}
 \end{small}
\end{equation}
Since the channel conditions under training set and test set are assumed to be i.i.d., the following equalities hold: $\sigma _{{\rm{RA}}{{\rm{W}}_{{\rm{iid}}}}}^2 = \sigma _{{\rm{RA}}{{\rm{W}}_{\det }}}^2$, $\sigma _{{\rm{S}}{{\rm{L}}_{{\rm{iid}}}}}^2 = \sigma _{{\rm{S}}{{\rm{L}}_{\det }}}^2$, $\sigma _{{\rm{C}}{{\rm{R}}_{{\rm{iid}}}}}^2 = \sigma _{{\rm{C}}{{\rm{R}}_{\det }}}^2$, and $\sigma _{{\rm{P}}{{\rm{C}}_{{\rm{iid}}}}}^2 = \sigma _{{\rm{P}}{{\rm{C}}_{\det }}}^2$. These can be uniformly denoted as $\sigma _{{\rm{RAW}}}^2$, $\sigma _{{\rm{SL}}}^2$, $\sigma _{{\rm{CR}}}^2$, and $\sigma _{{\rm{PC}}}^2$, respectively.
\subsubsubsection{Inter-Class Distance}
In calculating the expected inter-class distance for device $i$ under the i.i.d. stochastic channel scenario, the terms ${\mu _{i,n}^{{\rm{Tr}}}}$ and ${\mu _{j,m}^{{\rm{Te}}}}$ in (\ref{eq4}) are equal, while the terms ${\sigma _{i,n}^{{\rm{Tr}}}}$ and ${\sigma _{j,m}^{{\rm{Te}}}}$ are also identical. By getting the expression of extracted RFF features in the training set and test set, the expected inter-class distances of different methods can be obtained using (\ref{eq4}). The expected inter-class distances for the RAW, SL, CR, and PC methods are presented in (\ref{eq65}) in Appendix \ref{appendix:C}.

\subsubsubsection{Silhouette Score}
The silhouette coefficient is obtained by substituting the intra-class and inter-class distances from (\ref{eq28}) and (\ref{eq65}) in Appendix \ref{appendix:C} into (\ref{eq5}). The expected silhouette score is then computed using (\ref{eq6}). The expected silhouette scores for the RAW, SL, CR, and PC methods are provided in (\ref{eq66}) in Appendix \ref{appendix:D}.

\subsubsection{Non-i.i.d. Stochastic Channel Scenario}
Under the non-i.i.d. stochastic channel scenario, we can get the following intra-class and inter-class distances and silhouette scores.
\subsubsubsection{Intra-Class Distance}
Based on (\ref{eq2}), the expected intra-class distances for the RAW, SL, CR, and PC methods can be obtained. These results are presented in (\ref{eq67}) in Appendix \ref{appendix:E}. Note that the terms ${\mu _{i,n}^{{\rm{Tr}}}}$ and ${\mu _{i,m}^{{\rm{Te}}}}$ in (\ref{eq2}) are not equal while the terms ${\sigma _{i,n}^{{\rm{Tr}}}}$ and ${\sigma _{i,m}^{{\rm{Te}}}}$ also differ due to the non-i.i.d. stochastic channel.

\subsubsubsection{Inter-Class Distance}
The expected inter-class distance can be derived based on (\ref{eq4}). It should be noted that the terms ${\mu _{i,n}^{{\rm{Tr}}}}$ and ${\mu _{j,m}^{{\rm{Te}}}}$ in (\ref{eq4}) are not equal, and the terms ${\sigma _{i,n}^{{\rm{Tr}}}}$ and ${\sigma_{j,m}^{{\rm{Te}}}}$ are also differ. The expected inter-class distances for different methods are provided in (\ref{eq68}) in Appendix \ref{appendix:F}.

\subsubsubsection{Silhouette Score}
After obtaining the intra-class and inter-class distances, the silhouette scores can be derived using (\ref{eq5}) and (\ref{eq6}). The expected silhouette scores for different methods are presented in (\ref{eq69}) in Appendix \ref{appendix:G}.

\vspace{-0.3cm}
\section{Silhouette Score Efficient RFF Feature Extraction}
\label{section:4}
In this section, a precoding-based channel-robust RFF feature extraction method is proposed to enhance the silhouette score without requiring channel estimation. Then, the expected silhouette scores of the proposed method under different channel scenarios are theoretically deduced.
\vspace{-0.3cm}
\subsection{Silhouette Score Efficient RFF Feature Extraction}
First, Alice transmits a preamble signal to the device that is to be authenticated. In the frequency domain, the signal received by the device can be expressed as
\begin{equation}
\label{eq29}
Y_{\mathrm{RC},\mathrm{U}}=F_{\mathrm{RU}}\cdot H_{\mathrm{AU}}\cdot F_{\mathrm{TA}}\cdot X+N_{\mathrm{RC},\mathrm{U}},
\end{equation}
where $N_{\mathrm{RC,U}}$ denotes Gaussian noise with zero mean and variance $\mathrm{\sigma}_{N}^{2}$. The proposed method is referred to as the reciprocal (RC) method, as it involves applying a reciprocal operation in the subsequent step.

After performing the reciprocal and amplification operations at baseband, the signal at the device side can be expressed as
\begin{equation}
\label{eq30}
Y_{\mathrm{RC},\mathrm{U}}^{\prime}=\frac{\alpha _{\mathrm{U}}}{F_{\mathrm{RU}}\cdot H_{\mathrm{AU}}\cdot F_{\mathrm{TA}}\cdot X+N_{\mathrm{RC},\mathrm{U}}},
\end{equation}
where $\alpha_\mathrm{U}$ is chosen to enhance the power of the baseband signal $Y_{\mathrm{RC,U}}^{\prime}$. We assume that the maximum power of the transmitted baseband signal is $\eta$ under the transmitter power amplifier's linear amplification region. Let $P$ denote the variance of the term $\frac{1}{F_{\mathrm{RU}}\cdot H_{\mathrm{AU}}\cdot F_{\mathrm{TA}}\cdot X+N_{\mathrm{RC},\mathrm{U}}}$. Then, $\alpha _\mathrm{U}$ can be calculated as
\begin{equation}
\label{eq31}
\alpha _\mathrm{U} = \sqrt {\frac{\eta }{P}}.
\end{equation}
The transmission baseband signal is constrained within the linear amplification region of the transmitter power amplifier. If the signal power exceeds the linear amplification region, signal clipping may occur, which can degrade both normal communication and the RFF identification process.

Subsequently, the device transmits the processed signal $Y_{\mathrm{RC,U}}^{\prime}$ to Alice. The signal received by Alice in the frequency domain can be expressed as
\begin{equation}
\label{eq32}
\begin{aligned}
Y_{\mathrm{RC},\mathrm{A}}=&F_{\mathrm{RA}}\cdot H_{\mathrm{UA}}\cdot F_{\mathrm{TU}}\cdot Y_{\mathrm{RC},\mathrm{U}}^{\prime}+N_{\mathrm{RC},\mathrm{A}}
\\
=&\frac{\alpha _{\mathrm{U}}\cdot F_{\mathrm{RA}}\cdot H_{\mathrm{UA}}\cdot F_{\mathrm{TU}}}{F_{\mathrm{RU}}\cdot H_{\mathrm{AU}}\cdot F_{\mathrm{TA}}\cdot X+N_{\mathrm{RC},\mathrm{U}}}+N_{\mathrm{RC},\mathrm{A}},
\end{aligned}
\end{equation}
where $N_{\mathrm{RC,A}}$ denotes Gaussian noise with zero mean and variance $\sigma _{N}^{2}$.

Within the channel coherent time, the channel reciprocity can be held, i.e., $H_\mathrm{AU} = H_\mathrm{UA}$. Additionally, the noise received by the device can be neglected in the high SNR region. Therefore, ${Y_{{\mathrm{RC,A}}}}$, i.e., the signal received by Alice in the frequency domain, can be approximated as
\begin{equation}
\label{eq33}
Y_{\mathrm{RC},\mathrm{A}}\approx \frac{\alpha _{\mathrm{U}}\cdot F_{\mathrm{RA}}\cdot F_{\mathrm{TU}}}{F_{\mathrm{RU}}\cdot F_{\mathrm{TA}}\cdot X}+N_{\mathrm{RC},\mathrm{A}}.
\end{equation}
Thus, the channel effects are eliminated accordingly. Compared to the PC method, the proposed RC method eliminates the channel estimation process, resulting in more efficient signal processing. Assume that Alice and different devices have the same number of transmitting antennas $N_t$ and receiving antennas $N_r$. The computational complexity of different methods in the frequency domain is summarized in Table \ref{tab111} in on the top of the following page, where $L$ denotes the number of occupied subcarriers. The computational complexity of the proposed RC method is $\mathcal{O} \left( 3N_rN_tL \right)$, which is lower than that of the PC method, i.e., $\mathcal{O} \left( 4N_rN_tL \right)$.

\begin{table*}[h!t]
\scriptsize
   \centering
   \caption{Computational Complexity}
   \vspace{-0.1cm}
    \begin{center}
     \begin{tabular}{||c c c c c c c||}
     \hline
     Method & {\makecell[c]{Signal Transmission\\ from Alice to Bob}} & {\makecell[c]{Channel Estimation\\by Bob}} & {\makecell[c]{Channel Compensation\\by Bob }} & {\makecell[c]{Transmission Signal\\ Generation by Bob}} & {\makecell[c]{RFF Feature\\Extraction by Alice}} & Total\\
     \hline
     RAW & {N/A}  & {N/A} & {N/A} & {$\mathcal{O} \left( N_rN_tL \right)$} & {$\mathcal{O} \left(1\right)$} & {$\mathcal{O} \left( N_rN_tL \right)$}\\
     \hline
     SL \cite{chen2022radio} & {N/A} & {N/A} & {N/A} & {$\mathcal{O} \left( N_rN_tL \right)$} & {$\mathcal{O} \left( N_rN_tL \right)$} & {$\mathcal{O} \left( 2N_rN_tL \right)$}\\
     \hline
     CR \cite{dong2024robust} & {$\mathcal{O} \left( N_rN_tL \right)$} & {N/A} & {N/A} & {$\mathcal{O} \left( 2N_rN_tL \right)$} & {$\mathcal{O} \left( N_rN_tL \right)$} & {$\mathcal{O} \left( 4N_rN_tL \right)$}\\
     \hline
     PC \cite{sun2023location} & {$\mathcal{O} \left( N_rN_tL \right)$}  & {$\mathcal{O} \left( N_rN_tL \right)$} & {$\mathcal{O} \left( N_rN_tL \right)$} & {$\mathcal{O} \left( N_rN_tL \right)$} & {$\mathcal{O} \left(1\right)$} & {$\mathcal{O} \left( 4N_rN_tL \right)$}\\
     \hline
     RC & {$\mathcal{O} \left( N_rN_tL \right)$} & {N/A}  & {N/A} & {$\mathcal{O} \left( 2N_rN_tL \right)$} & {$\mathcal{O} \left(1\right)$} & {$\mathcal{O} \left( 3N_rN_tL \right)$}\\
     \hline
    \end{tabular}
    \end{center}
\label{tab111}
\vspace{-0.8cm}
\end{table*}

The extracted RFF feature of device $i$ on the $k$th subcarrier of the $d$th sample under channel $c$, where $c\in \left\{ \mathrm{det,iid,non} \right\}$, can be expressed as
\begin{equation}
\label{eq34}
r_{{\rm{RC}},i,d,k}^c = \frac{{{\alpha _c} \cdot {F_{{\rm{RA}}}} \cdot {H_{c,k}} \cdot {F_{{\rm{T}}{{\rm{U}}_{i,k}}}}}}{{{F_{{\rm{RU}}}} \cdot {H_{c,k}} \cdot {F_{{\rm{TA}}}} \cdot X + {N_{{\rm{RC}},{\rm{U}},d,k}}}} + {N_{{\rm{RC}},{\rm{A}},d,k}}.
\end{equation}
The parameter $\alpha _c$ is determined based on the variance of the term $\frac{1}{f_{\mathrm{RU}}\cdot H_{c,k}\cdot f_\mathrm{TA}\cdot x+N_{\mathrm{RC,U},d,k}}$. The following claim is introduced to derive the variance of the term.

\textbf{Claim 4.} Let $Z = \frac{1}{{\rho G + W}}$, where $G$ is a Gaussian random variable with mean ${\mu _G}$ and variance $\sigma _G^2$, and $W$ is a Gaussian random variable with zero mean and variance $\sigma _W^2$. Then, it can be derived that
\begin{equation}
\label{eq35}
\begin{array}{l}
{\mathbb E}\left[ Z \right] \approx \frac{{{\rho ^2}\mu _G^2 + {\rho ^2}\sigma _G^2 + \sigma _W^2}}{{{\rho ^3}\mu _G^3}},\\
{\mathbb E}\left[ {{Z^2}} \right] \approx \frac{{{\rho ^2}\mu _G^2 + 3{\rho ^2}\sigma _G^2 + 3\sigma _W^2}}{{{\rho ^4}\mu _G^4}}.
\end{array}
\end{equation}

\textit{Proof}: The proof can be obtained by applying the Taylor series expansion to $Z$ and $Z^2$ around $\left( {{\mu _G},0} \right)$ and calculating the expectation, which is similar to the proof of Claim 1.
$\hfill\blacksquare$

By substituting $\rho$ with ${{f_{\mathrm{RU}}} \cdot {f_{\mathrm{TA}}} \cdot x}$ in Claim 4, the variance of the term $\frac{1}{f_{\mathrm{RU}}\cdot H_{c,k}\cdot f_\mathrm{TA}\cdot x+N_{\mathrm{RC,U},d,k}}$ can be obtained as
\begin{equation}
\begin{array}{l}
\label{eq36}
{P_c} = \frac{{{\beta ^2}\mu _{{H_c}}^2 + 3{\beta ^2}\sigma _{{H_c}}^2 + 3\sigma _N^2}}{{{\beta ^4}\mu _{{H_c}}^4}}.
\end{array}
\end{equation}
Accordingly, we have
\begin{equation}
\begin{array}{l}
\label{eq37}
{\alpha _c} = \sqrt {\frac{\eta }{{{P_{\rm{c}}}}}}  \approx \sqrt {\frac{{\eta {\beta ^4}\mu _{{H_c}}^4}}{{{\beta ^2}\mu _{{H_c}}^2 + 3{\beta ^2}\sigma _{{H_c}}^2 + 3\sigma _N^2}}} .
\end{array}
\end{equation}
According to Claim 1, the mean and variance of $r_{{\rm{RC}},i,d,k}^c$, i.e., the extracted RFF feature of device $i$ on the $k$th subcarrier of the $d$th sample under channel $c$, can be obtained as
\begin{equation}
\begin{small}
\label{eq38}
\begin{aligned}
{\mu _{{\rm{R}}{{\rm{C}}_c}}} \approx& \frac{{{\alpha _c}{f_{{\rm{RA}}}}{\mu _{\rm{U}}}\left( {{\beta ^2}\mu _{{H_c}}^2 + \sigma _N^2} \right)}}{{{\beta ^3}\mu _{{H_c}}^2}},\\
\sigma _{\mathrm{RC}_c}^{2}\approx& \left\{ \alpha _{c}^{2}f_{\mathrm{RA}}^{2}\left[ \mu _{\mathrm{U}}^{2}\sigma _{N}^{2}\left( \beta ^2\mu _{H_c}^{2}-\sigma _{N}^{2} \right) \right. \right.
\\
&\left. \left. +\sigma _{\mathrm{U}}^{2}\left( \beta ^4\mu _{H_c}^{4}+3\beta ^2\mu _{H_c}^{2}\sigma _{N}^{2} \right) \right] /\left( \beta ^6\mu _{H_c}^{4} \right) \right\} +\sigma _{N}^{2}.
\end{aligned}
\end{small}
\end{equation}
\vspace{-0.3cm}
\subsection{Silhouette Score under Various Channel Scenarios}

\subsubsection{Deterministic Channel Scenario}
Under the deterministic channel scenario, where the channel condition $c = \det $ holds in (\ref{eq38}), the expected intra-class and inter-class distances for each training sample of device $i$ can be obtained by substituting the mean and variance of $r_{{\rm{RC}},i,d,k}^{\rm{det}}$ from (\ref{eq38}) into (\ref{eq2}) and (\ref{eq4}). These distances can then be computed by applying the result of Claim 2 as
\begin{equation}
\begin{small}
\label{eq39}
\begin{aligned}
D_{\mathrm{RC},i}^{\mathrm{Intra},\det}\approx& \frac{2R_{\mathrm{L}}}{\sigma _{\mathrm{RC}_{\det}}^{2}}\left[ \frac{\alpha _{\det}^{2}f_{\mathrm{RA}}^{2}\sigma _{N}^{2}\left( \mu _{\mathrm{U}}^{2}+\sigma _{\mathrm{U}}^{2} \right)}{\beta ^4\mu _{H}^{2}}+\sigma _{N}^{2} \right] ,
\\
D_{\mathrm{RC},i,j}^{\mathrm{Inter},\det}\approx& \frac{2R_{\mathrm{L}}}{\sigma _{\mathrm{RC}_{\det}}^{2}}\left\{ \frac{\alpha _{\det}^{2}f_{\mathrm{RA}}^{2}\left[ \mu _{\mathrm{U}}^{2}\sigma _{N}^{2}+\sigma _{\mathrm{U}}^{2}\left( \beta ^2\mu _{H}^{2}+3\sigma _{N}^{2} \right) \right]}{\beta ^4\mu _{H}^{2}} \right.
\\
&\left. +\sigma _{N}^{2} \right\} .
\end{aligned}
\end{small}
\end{equation}

Since the transmitter RFF effects of device $i$ and device $j$ are assumed to follow the i.i.d. Gaussian distribution, the expected intra-class and inter-class distances for each training sample of device $j$ are identical to those of device $i$. Therefore, by substituting the intra-class and inter-class distances from (\ref{eq39}) into (\ref{eq5}), the silhouette coefficient can be obtained. Using this silhouette coefficient along with (\ref{eq6}), the expected silhouette score can be expressed as
 \begin{equation}
 \begin{small}
 \begin{array}{l}
 \label{eq40}
S_{{\rm{RC}}}^{{\rm{det}}} \approx \frac{{\alpha _{\det }^2f_{{\rm{RA}}}^2\sigma _{\rm{U}}^2\left( {{\beta ^2}\mu _H^2 + 2\sigma _N^2} \right)}}{{\alpha _{\det }^2f_{{\rm{RA}}}^2\left[ {\mu _{\rm{U}}^2\sigma _N^2 + \sigma _{\rm{U}}^2\left( {{\beta ^2}\mu _H^2 + 3\sigma _N^2} \right)} \right] + {\beta ^4}\mu _H^2\sigma _N^2}}.
\end{array}
\end{small}
 \end{equation}

\subsubsection{I.i.d. Stochastic Channel Scenario}
Under the i.i.d. stochastic channel scenario, the channel condition $c = \rm{iid}$ holds in (\ref{eq38}). By substituting the mean and variance of $r_{{\rm{RC}},i,d,k}^{\rm{iid}}$ from (\ref{eq38}) into (\ref{eq2}) and (\ref{eq4}), and applying the results of Claim 3, the expected intra-class and inter-class distances for each training sample of device $i$ can be expressed as
\begin{equation}
\begin{small}	
\label{eq41}
 \begin{aligned}
D_{\mathrm{RC},i}^{\mathrm{Intra},\mathrm{iid}}\approx& \frac{2R_{\mathrm{L}}}{\sigma _{\mathrm{RC}_{\mathrm{iid}}}^{2}}\left[ \frac{\alpha _{\mathrm{iid}}^{2}f_{\mathrm{RA}}^{2}\sigma _{N}^{2}\left( \mu _{\mathrm{U}}^{2}+\sigma _{\mathrm{U}}^{2} \right) \left( \beta ^2\mu _{H_{\mathrm{iid}}}^{2}-\sigma _{N}^{2} \right)}{\beta ^6\mu _{H_{\mathrm{iid}}}^{4}} \right.
\\
&\left. +\sigma _{N}^{2} \right],\\
D_{{\rm{RC}},i,j}^{{\rm{Inter}},{\rm{iid}}} \approx& 2{R_{\rm{L}}}\left\{ {\alpha _{{\rm{iid}}}^2f_{{\rm{RA}}}^2[\mu _{\rm{U}}^2\sigma _N^2\left( {{\beta ^2}\mu _{{H_{{\rm{iid}}}}}^2 - \sigma _N^2} \right) + \sigma _{\rm{U}}^2({\beta ^4}\mu _{{H_{{\rm{iid}}}}}^4} \right.\\
 &+ 3{\beta ^2}\mu _{{H_{{\rm{iid}}}}}^2\sigma _N^2) + {\beta ^6}\mu _{{H_{{\rm{iid}}}}}^4\sigma _N^2]/\left( {{\beta ^6}\mu _{{H_{{\rm{iid}}}}}^4\sigma _{{\rm{RC}}}^2} \right),
 \end{aligned}
 \end{small}
\end{equation}
where $\sigma _{{\rm{R}}{{\rm{C}}_{{\rm{iid}}}}}^2 = \sigma _{{\rm{R}}{{\rm{C}}_{{\rm{det}}}}}^2$ and $\alpha _{{\rm{iid}}}^2 = \alpha _{{\rm{det}}}^2$ hold. These quantities can be denoted as $\sigma _{{\rm{RC}}}^2$ and ${\alpha ^2}$, respectively.

By substituting the intra-class and inter-class distances from (\ref{eq41}) into (\ref{eq5}) and utilizing (\ref{eq6}), the expected silhouette score $S_{\mathrm{RC}}^\mathrm{iid}$ can be obtained as shown in (\ref{eq42}) at the top of the following page.
\begin{figure*}[ht]
 	\centering
 \begin{small}
 	\begin{equation}	
    \label{eq42}
S_{\mathrm{RC}}^\mathrm{iid}\approx \frac{\alpha ^2f_\mathrm{RA}^{2}\sigma _\mathrm{U}^{2}\left( \beta ^2\mu _{H}^{2}+\sigma _{N}^{2} \right) ^2}{\alpha ^2f_\mathrm{RA}^{2}\left[ \mu _\mathrm{U}^{2}\sigma _{N}^{2}\left( \beta ^2\mu _{H}^{2}-\sigma _{N}^{2} \right) +\sigma _\mathrm{U}^{2}\left( \beta ^4\mu _{H}^{4}+3\beta ^2\mu _{H}^{2}\sigma _{N}^{2} \right) \right] +\beta ^6\mu _{H}^{4}\sigma _{N}^{2}}.
 	\end{equation}
 \end{small}
\vspace{-0.7cm}
\end{figure*}

\subsubsection{Non-i.i.d. Stochastic Channel Scenario}
Under the non-i.i.d. stochastic channel scenario, where $c = \rm{non} $ holds in (\ref{eq38}), the expected intra-class and inter-class distances for each training sample of device $i$ can be obtained by substituting the mean and variance of $r_{{\rm{RC}},i,d,k}^{\rm{non}}$ from (\ref{eq38}) into (\ref{eq2}) and (\ref{eq4}). These distances are shown in (\ref{eq70}) in Appendix \ref{appendix:H}, where ${\mu _{{\rm{R}}{{\rm{C}}_\mathrm{non}}}}$ and ${\sigma _{{\rm{R}}{{\rm{C}}_\mathrm{non}}}}$ represent the mean and standard deviation of the extracted RFF feature under the non-i.i.d. stochastic channel scenario. Subsequently, the expected silhouette score under the non-i.i.d. stochastic channel scenario can be obtained by substituting the intra-class and inter-class distances from (\ref{eq70}) in Appendix \ref{appendix:H} into (\ref{eq5}) and then applying (\ref{eq6}). The expression for the expected silhouette score is provided in (\ref{eq71}) in Appendix \ref{appendix:I}.

\vspace{-0.3cm}
\section{Silhouette Score Performance Comparison for RFF Feature Extraction}
\label{section:5}
\vspace{-0.1cm}
\subsection{Deterministic Channel Scenario}
First, we compare the expected silhouette scores of the existing SL and CR methods under the deterministic channel scenario, i.e., $S_{\mathrm{SL}}^\mathrm{det}$ and $S_{\mathrm{CR}}^\mathrm{det}$ as shown in (\ref{eq26}). For the convenience of comparison, $f_\mathrm{RA}$, $f_\mathrm{T{U_{L}}}$, $f_\mathrm{TA}$, $f_\mathrm{R{U}}$, ${\mu _\mathrm{{S}}}$, and ${\mu _\mathrm{{U}}}$ are set to the unit value. Since $\beta \triangleq f_{\mathrm{RU}}\cdot f_{\mathrm{TA}}\cdot x$ and $\gamma \triangleq f_{\mathrm{RA}}\cdot f_{\mathrm{TU}_{\mathrm{L}}}\cdot x$ hold, it follows that $\beta  = \gamma $. The difference between ${F_{\mathrm{T}{\mathrm{U}_{\mathrm{S},i}}}}$ and ${F_{\mathrm{T}{\mathrm{U}_{\mathrm{L},i}}}}$ is smaller than the difference between ${F_{\mathrm{T}{\mathrm{U}_i}}}$ and ${F_{\mathrm{R}{\mathrm{U}_i}}}$, because the former represents device $i$'s RFF within adjacent signals, i.e., the L-STF part and L-LTF part, while the later refers to device $i$'s RFF of transmitting and receiving front ends. Therefore, it can be assumed that $\sigma _{\rm{S}}^2 < \sigma _{\rm{U}}^2$ holds when $f_\mathrm{T{U_{L}}}$ and $f_\mathrm{R{U}}$ are normalized to the unit value. Consequently, it follows that $S_{\mathrm{CR}}^\mathrm{det}>S_{\mathrm{SL}}^\mathrm{det}$ according to (\ref{eq26}).

Second, we compare the expected silhouette scores of the existing CR and PC methods. In addition to $f_\mathrm{RA}$, $f_\mathrm{T{U_{L}}}$, $f_\mathrm{TA}$ and $f_\mathrm{R{U}}$, ${\mu _{H}}$ and $x$ are also normalized to the unit value for the convenience of comparison. Then, it can be derived that
 \begin{equation}	
 \label{eq43}
 \begin{array}{l}
S_{\mathrm{CR}}^\mathrm{det}\approx \frac{\sigma _\mathrm{U}^{2}\left( 1+2\sigma _{N}^{2} \right)}{\left[ \mu _\mathrm{U}^{2}\sigma _{N}^{2}+\sigma _\mathrm{U}^{2}\left( 1+3\sigma _{N}^{2} \right) \right] +\sigma _{N}^{2}+3\sigma _{H}^{2}\sigma _{N}^{2}+3\sigma _{N}^{4}},
\\
S_{\mathrm{PC}}^\mathrm{det}\approx \frac{\sigma _\mathrm{U}^{2}\left( 1+2\sigma _{N}^{2} \right)}{\left[ \mu _\mathrm{U}^{2}\sigma _{N}^{2}+\sigma _\mathrm{U}^{2}\left( 1+3\sigma _{N}^{2} \right) \right] +\sigma _{N}^{2}}.
\end{array}
 \end{equation}
As the denominator of $S_{\mathrm{CR}}^\mathrm{det}$ is greater than that of $S_{\mathrm{PC}}^\mathrm{det}$ in (\ref{eq43}), while both have the same numerator, it can be concluded that $S_{\mathrm{PC}}^\mathrm{det}>S_{\mathrm{CR}}^\mathrm{det}$ under these conditions.

Third, we focus on comparing the expected silhouette scores of the existing PC method and the proposed RC method. In this comparison, only $x$ is normalized to the unit value. Then, the expected silhouette scores for the PC method and the proposed RC method are
 \begin{equation}	
 \label{eq44}
 \begin{array}{l}
S_{\mathrm{PC}}^\mathrm{det}\approx \frac{f_\mathrm{RA}^{2}\sigma _\mathrm{U}^{2}\left( \beta ^2\mu _{H}^{2}+2\sigma _{N}^{2} \right)}{f_\mathrm{RA}^{2}\left[ \mu _\mathrm{U}^{2}\sigma _{N}^{2}+\sigma _\mathrm{U}^{2}\left( \beta ^2\mu _{H}^{2}+3\sigma _{N}^{2} \right) \right] +\beta ^4\mu _{H}^{2}\sigma _{N}^{2}},
\\
S_{\mathrm{RC}}^\mathrm{det}\approx \frac{f_\mathrm{RA}^{2}\sigma _\mathrm{U}^{2}\left( \beta ^2\mu _{H}^{2}+2\sigma _{N}^{2} \right)}{f_\mathrm{RA}^{2}\left[ \mu _\mathrm{U}^{2}\sigma _{N}^{2}+\sigma _\mathrm{U}^{2}\left( \beta ^2\mu _{H}^{2}+3\sigma _{N}^{2} \right) \right] +\frac{\beta ^4\mu _{H}^{2}\sigma _{N}^{2}}{\alpha ^2}}.
\end{array}
 \end{equation}
It should be noted that $\alpha =\sqrt{\frac{\eta}{P_\mathrm{det}}}$. Therefore, it follows that $S_{\mathrm{RC}}^\mathrm{det}>S_{\mathrm{PC}}^\mathrm{det}$ when $\eta  > P_\mathrm{det}$.

Fourth, we compare the expected silhouette scores of the RAW method and the proposed RC method. Using the same settings on $f_\mathrm{RA}$, $f_\mathrm{T{U_{L}}}$, $f_\mathrm{TA}$, $f_\mathrm{R{U}}$, ${\mu _{H}}$ and $x$ as in the comparison between the CR and PC methods, we can obtain

 \begin{equation}	
 \label{eq45}
 \begin{array}{l}
S_{\mathrm{RAW}}^\mathrm{det}=\frac{1}{1+\frac{\sigma _{N}^{2}}{\sigma _\mathrm{U}^{2}\left( 1+\sigma _{H}^{2} \right)}},
\\
S_{\mathrm{RC}}^\mathrm{det}\approx \frac{1}{1+\frac{(\alpha ^2+\alpha ^2\sigma _\mathrm{U}^{2}+1)\sigma _{N}^{2}}{\alpha ^2\sigma _\mathrm{U}^{2}(1+2\sigma _{N}^{2})}}.
\end{array}
 \end{equation}
It can then be observed from (\ref{eq45}) that $S_{\mathrm{RAW}}^\mathrm{det}>S_{\mathrm{RC}}^\mathrm{det}$ holds in the high SNR region.

In the low SNR region, by substituting $f_{\rm{RA}}$, ${\mu _{H}}$, and $x$ with the unit value, the expected silhouette score of the RAW method can be expressed as
 \begin{equation}	
 \begin{array}{l}
 \label{eq46}
S_{\mathrm{RAW}}^\mathrm{det}\approx \frac{\sigma _\mathrm{U}^{2}\left( 1+\sigma _{H}^{2} \right)}{\sigma _{N}^{2}}\rightarrow 0.
\end{array}
 \end{equation}

In the high SNR region, it can be observed from (\ref{eq26}) and (\ref{eq40}) that $S_{{\rm{RAW}}}^\mathrm{det}$, $S_{{\rm{SL}}}^\mathrm{det}$, $S_{{\rm{CR}}}^\mathrm{det}$, $S_{{\rm{PC}}}^\mathrm{det}$, and $S_{{\rm{RC}}}^\mathrm{det}$ all approximate one. This indicates that all of these methods can achieve high expected silhouette scores in the high SNR region under the deterministic channel scenario.
\vspace{-0.3cm}
\subsection{I.i.d. Stochastic Channel Scenario}
First, the expected silhouette scores of the existing SL and CR methods under the i.i.d. stochastic channel scenario are compared. $f_\mathrm{RA}$, $f_\mathrm{T{U_{L}}}$, $f_\mathrm{TA}$, $f_\mathrm{R{U}}$, ${\mu _\mathrm{{S}}}$, and ${\mu _\mathrm{{U}}}$ are normalized to the unit value. Additionally, it is also assumed that $\sigma _{\rm{S}}^2 < \sigma _{\rm{U}}^2$ holds. Under these conditions, it follows that $S_{\mathrm{CR}}^\mathrm{iid}>S_{\mathrm{SL}}^\mathrm{iid}$ according to (\ref{eq66}) in Appendix \ref{appendix:D}.

Second, the expected silhouette scores of the existing CR and PC methods are considered. In addition to $f_\mathrm{RA}$, $f_\mathrm{T{U_{L}}}$, $f_\mathrm{TA}$, and $f_\mathrm{R{U}}$, ${\mu _H}$ and $x$ are also normalized to the unit value. By substituting $f_\mathrm{RA}$, $f_\mathrm{T{U_{L}}}$, $f_\mathrm{TA}$, $f_\mathrm{R{U}}$, ${\mu _H}$, and $x$ with the unit value in (\ref{eq66}) in Appendix \ref{appendix:D}, it can be derived that

 \begin{equation}	
 \label{eq47}
 \begin{array}{l}
S_{\mathrm{CR}}^\mathrm{iid}\approx \frac{f_\mathrm{RA}^{2}\sigma _\mathrm{U}^{2}\left( 1+\sigma _{N}^{2} \right) ^2}{f_\mathrm{RA}^{2}x^2\left[ \mu _\mathrm{U}^{2}\sigma _{N}^{2}\left( 1-\sigma _{N}^{2} \right) +\sigma _\mathrm{U}^{2}\left( 1+3\sigma _{N}^{2} \right) \right] +\sigma _{N}^{2}+3\sigma _{H}^{2}\sigma _{N}^{2}+3\sigma _{N}^{4}},
\\
S_{\mathrm{PC}}^\mathrm{iid}\approx \frac{f_\mathrm{RA}^{2}\sigma _\mathrm{U}^{2}\left( 1+\sigma _{N}^{2} \right) ^2}{f_\mathrm{RA}^{2}\left[ \mu _\mathrm{U}^{2}\sigma _{N}^{2}\left( 1-\sigma _{N}^{2} \right) +\sigma _\mathrm{U}^{2}\left( 1+3\sigma _{N}^{2} \right) \right] +\sigma _{N}^{2}}.
\end{array}
 \end{equation}
It follows that $S_{\mathrm{PC}}^\mathrm{iid}>S_{\mathrm{CR}}^\mathrm{iid}$ since both have the same numerator, while $S_{\mathrm{CR}}^\mathrm{iid}$ has a greater denominator.

Third, we compare the expected silhouette scores of the existing PC method and the proposed RC method. In this comparison, only $x$ is normalized to the unit value. The expected silhouette scores of the PC method and the proposed RC method can be expressed as
 \begin{equation}	
 \begin{small}
 \label{eq48}
 \begin{array}{l}
S_{\mathrm{PC}}^\mathrm{iid}\approx \frac{f_\mathrm{RA}^{2}\sigma _\mathrm{U}^{2}\left( \beta ^2\mu _{H}^{2}+\sigma _{N}^{2} \right) ^2}{f_\mathrm{RA}^{2}\left[ \mu _\mathrm{U}^{2}\sigma _{N}^{2}\left( \beta ^2\mu _{H}^{2}-\sigma _{N}^{2} \right) +\sigma _\mathrm{U}^{2}\beta ^2\mu _{H}^{2}\left( \beta ^2\mu _{H}^{2}+3\sigma _{N}^{2} \right) \right] +\beta ^6\mu _{H}^{4}\sigma _{N}^{2}},
\\
S_{\mathrm{RC}}^\mathrm{iid}\approx \frac{f_\mathrm{RA}^{2}\sigma _\mathrm{U}^{2}\left( \beta ^2\mu _{H}^{2}+\sigma _{N}^{2} \right) ^2}{f_\mathrm{RA}^{2}\left[ \mu _\mathrm{U}^{2}\sigma _{N}^{2}\left( \beta ^2\mu _{H}^{2}-\sigma _{N}^{2} \right) +\sigma _\mathrm{U}^{2}\left( \beta ^4\mu _{H}^{4}+3\beta ^2\mu _{H}^{2}\sigma _{N}^{2} \right) \right] +\frac{\beta ^6\mu _{H}^{4}\sigma _{N}^{2}}{\alpha ^2}}.
\end{array}
\end{small}
 \end{equation}
Therefore, when the maximum power of the transmitted baseband signal, constrained by the transmitter's linear amplification region, exceeds the power of the processed baseband signal, i.e., $\eta  > P_\mathrm{{{det}}}$, it follows that ${\alpha} > 1$ and $S_{\mathrm{PC}}^\mathrm{iid}$ has a greater denominator than that of $S_{\mathrm{RC}}^\mathrm{iid}$ in (\ref{eq48}). Since both the numerators of $S_{\mathrm{PC}}^\mathrm{iid}$ and $S_{\mathrm{RC}}^\mathrm{iid}$ have the same value, it follows that $S_{{\rm{RC}}}^\mathrm{iid} > S_{{\rm{PC}}}^\mathrm{iid}$.

In the low SNR region, by substituting $f_{\rm{RA}}$, ${\mu _{H}}$, and $x$ with the unit value in (\ref{eq66}) in Appendix \ref{appendix:D}, the expected silhouette score of the RAW method can be expressed as
\begin{equation}	
\begin{array}{l}
 \label{eq49}
S_{{\rm{RAW}}}^\mathrm{iid} \approx \frac{{\sigma _\mathrm{{U}}^2}}{{\sigma _N^2}} \to 0.
\end{array}
\end{equation}
Thus, the expected silhouette score of the RAW method approximates zero in the low SNR region.

In the high SNR region, by substituting $f_{\rm{RA}}$, ${\mu _{H}}$, and $x$ with the unit value in (\ref{eq66}) in Appendix \ref{appendix:D}, all of $S_{{\rm{SL}}}^\mathrm{iid}$, $S_{{\rm{CR}}}^\mathrm{iid}$, $S_{{\rm{PC}}}^\mathrm{iid}$, and $S_{{\rm{RC}}}^\mathrm{iid}$ approach one. In contrast, according to (\ref{eq66}) in Appendix \ref{appendix:D}, $S_{\mathrm{RAW}}^\mathrm{iid}\approx \frac{\sigma _\mathrm{U}^{2}\mu _{H}^{2}}{\mu _\mathrm{U}^{2}\sigma _{H}^{2}+\sigma _\mathrm{U}^{2}\mu _{H}^{2}+\sigma _\mathrm{U}^{2}\sigma _{H}^{2}}<1$ holds, even under the high SNR region. This indicates that the RAW method exhibits a low expected silhouette score under the i.i.d. stochastic channel scenario, even when SNR is relatively high. As shown in Section \ref{section:6_c} and Section \ref{section:6_d}, its classification accuracy also fails to approach 100\%, even under the high SNR scenario.
\vspace{-0.3cm}
\subsection{Non-i.i.d. Stochastic Channel Scenario}
First, the expected silhouette scores of the existing SL and CR methods are compared. Similar to the previous setting, $f_\mathrm{RA}$, $f_\mathrm{T{U_{L}}}$, $f_\mathrm{TA}$, $f_\mathrm{R{U}}$, ${\mu _\mathrm{{S}}}$, and ${\mu _\mathrm{{U}}}$ are normalized to the unit value, i.e., $\beta = \gamma$. Assuming that $\sigma _{\rm{S}}^2 < \sigma _{\rm{U}}^2$ holds, it follows directly from (\ref{eq67}) in Appendix \ref{appendix:E} that $S_{\mathrm{CR}}^\mathrm{non}>S_{\mathrm{SL}}^\mathrm{non}$.

Second, the expected silhouette scores of the existing CR and PC methods are analyzed. In addition to $f_\mathrm{RA}$, $f_\mathrm{T{U_{L}}}$, $f_\mathrm{TA}$ and $f_\mathrm{R{U}}$, ${\mu _{H}}$, ${\mu _{H_\mathrm{non}}}$, and $x$ are also normalized to the unit value. Consequently, it is evident that $\mu _{\mathrm{CR}_\mathrm{det}}=\mu _{\mathrm{CR}_\mathrm{non}}$, $\sigma _{\mathrm{CR}_\mathrm{det}}^{2}=\sigma _{\mathrm{CR}_\mathrm{non}}^{2}$, $\mu _{\mathrm{PC}_\mathrm{det}}=\mu _{\mathrm{PC}_\mathrm{non}}$, and $\sigma _{\mathrm{PC}_\mathrm{det}}^{2}=\sigma _{\mathrm{PC}_\mathrm{non}}^{2}$ hold. Therefore, we obtain
 \begin{equation}
 \begin{small}	
 \label{eq50}
 \begin{aligned}
S_{\mathrm{CR}}^{\mathrm{non}}\approx& \sigma _{\mathrm{U}_i}^{2}\left( 1+\sigma _{N}^{2} \right) ^2/\left\{ \left( 1+3\sigma _{N}^{2} \right) \left( \mu _{\mathrm{U}}^{2}+\sigma _{\mathrm{U}}^{2} \right) -\mu _{\mathrm{U}}^{2}\left( 1+\sigma _{N}^{2} \right) ^2 \right.
\\
&\left. +\sigma _{N}^{2}+1.5\cdot \sigma _{N}^{2}\left[ \left( \sigma _{H}^{2}+\sigma _{H_{\mathrm{non}}}^{2} \right) +2\sigma _{N}^{2} \right] \right\} ,
\\
S_{\mathrm{PC}}^{\mathrm{non}}\approx& \sigma _{\mathrm{U}}^{2}\left( 1+\sigma _{N}^{2} \right) ^2/\left[ \left( 1+3\sigma _{N}^{2} \right) \left( \mu _{\mathrm{U}}^{2}+\sigma _{\mathrm{U}}^{2} \right) \right.
\\
&\left. -\mu _{\mathrm{U}}^{2}\left( 1+\sigma _{N}^{2} \right) ^2+\sigma _{N}^{2} \right] .
\end{aligned}
\end{small}
 \end{equation}
Under these conditions, it follows from (\ref{eq69}) in Appendix \ref{appendix:G} that $S_{\mathrm{PC}}^\mathrm{non}>S_{\mathrm{CR}}^\mathrm{non}$, since both $S_{\mathrm{CR}}^\mathrm{non}$ and $S_{\mathrm{PC}}^\mathrm{non}$ have the same value of numerator, while $S_{\mathrm{CR}}^\mathrm{non}$ has a greater denominator.

Third, the expected silhouette scores of the existing PC method and the proposed RC method are compared. For this comparison, $x$, ${f_\mathrm{TA}}$, ${f_\mathrm{RA}}$, ${f_\mathrm{R{U}}}$, ${\mu _\mathrm{{U}}}$, ${\mu _{{H}}}$, and ${\mu _{{H_\mathrm{non}}}}$ are normalized to the unit value. Then, we obtain $\mu _{\mathrm{RC}}=\alpha \left( 1+\sigma _{N}^{2} \right) $ and $\mu _{\mathrm{RC}_{\mathrm{non}}}=\alpha _{\mathrm{non}}\left( 1+\sigma _{N}^{2} \right)$. The expected silhouette score of the proposed method can be expressed as
 \begin{equation}	
 \begin{small}
 \label{eq51}
 \begin{aligned}
S_{\mathrm{RC}}^{\mathrm{non}}\approx& 2\alpha \alpha _{\mathrm{non}}\sigma _{\mathrm{RC}}\sigma _{\mathrm{RC}_{\mathrm{non}}}\sigma _{\mathrm{U}}^{2}\left( 1+\sigma _{N}^{2} \right) ^2/\left[ \left( \alpha _{\mathrm{non}}^{2}\sigma _{\mathrm{RC}}^{2}+\alpha ^2\sigma _{\mathrm{RC}_{\mathrm{non}}}^{2} \right) \right.
\\
&\cdot \left( 1+\sigma _{\mathrm{U}}^{2} \right) \left( 1+3\sigma _{N}^{2} \right) -\left( \alpha ^2\sigma _{\mathrm{RC}_{\mathrm{non}}}^{2}+\alpha _{\mathrm{non}}^{2}\sigma _{\mathrm{RC}}^{2} \right) \left( 1+\sigma _{N}^{2} \right) ^2
\\
&\left. +\sigma _{N}^{2}\left( \sigma _{\mathrm{RC}}^{2}+\sigma _{\mathrm{RC}_{\mathrm{non}}}^{2} \right) \right] .
\end{aligned}
\end{small}
 \end{equation}
To further simplify the expression of $S_{{\rm{RC}}}^\mathrm{non}$, we define ${\alpha _\mathrm{non}} = A{\alpha }$ and ${\sigma _{{\rm{R}}{{\rm{C}}_\mathrm{non}}}} = B{\sigma _{{\rm{RC}}}}$. Then, (\ref{eq51}) can be simplified as
 \begin{equation}	
 \begin{small}
 \label{eq52}
 \begin{aligned}
S_{\mathrm{RC}}^{\mathrm{non}}\approx& 2\sigma _{\mathrm{U}}^{2}(1+\sigma _{N}^{2})^2/\left[ \left( \frac{A^2+B^2}{AB} \right) \left( 1+\sigma _{\mathrm{U}}^{2} \right) \left( 1+3\sigma _{N}^{2} \right) \right.
\\
&\left. -\left( \frac{A^2+B^2}{AB} \right) (1+\sigma _{N}^{2})^2+\frac{2\sigma _{N}^{2}}{\frac{2AB\alpha ^2}{1+B^2}} \right] .
\end{aligned}
\end{small}
 \end{equation}

In the high SNR region, it follows that $A \approx B$ holds. Consequently, we have
 \begin{equation}
 \begin{small}
 \label{eq53}
 \begin{aligned}
S_{\mathrm{RC}}^{\mathrm{non}}\approx& 2\sigma _{\mathrm{U}}^{2}(1+\sigma _{N}^{2})^2/\left\{ 2(1+\sigma _{\mathrm{U}}^{2})(1+3\sigma _{N}^{2}) \right.
\\
&\left. -2(1+\sigma _{N}^{2})^2+2\sigma _{N}^{2}/\left( 2A^2\alpha ^2/\left( 1+A^2 \right) \right) \right\} .
\end{aligned}
\end{small}
 \end{equation}
Meanwhile, the expected silhouette score of the PC method can be expressed as
 \begin{equation}
 \begin{small}	
 \label{eq54}
 \begin{aligned}
S_{{\rm{PC}}}^\mathrm{non} \approx& 2\sigma _\mathrm{U}^2{(1 + \sigma _N^2)^2}/\{ 2(1 + \sigma _\mathrm{U}^2)(1 + 3\sigma _N^2)\\
 &- 2{(1 + \sigma _N^2)^2} + 2\sigma _N^2\} .
\end{aligned}
\end{small}
 \end{equation}
Therefore, $S_{{\rm{RC}}}^\mathrm{non} > S_{{\rm{PC}}}^\mathrm{non}$ holds when $\alpha >1$ and $\alpha _{\mathrm{non}}>1$. A more general analysis under broader parameter settings is provided in Appendix \ref{appendix:J}.

In the low SNR region, the expected silhouette score of the RAW method can be expressed as
\begin{equation}
\begin{array}{l}	
\label{eq55}
S_{{\rm{RAW}}}^\mathrm{non} \approx \frac{{\sigma _\mathrm{{U}}^2}}{{\sigma _N^2}} \to 0.
\end{array}
\end{equation}
The expected silhouette score of the RAW method also approximates zero in the low SNR region under the non-i.i.d. stochastic channel scenario.

In the high SNR region, it is evident from (\ref{eq69}) in Appendix \ref{appendix:G} and (\ref{eq71}) in Appendix \ref{appendix:I} that $S_{{\rm{SL}}}^\mathrm{non}$, $S_{{\rm{CR}}}^\mathrm{non}$, $S_{{\rm{PC}}}^\mathrm{non}$, and $S_{{\rm{RC}}}^\mathrm{non}$ all approximate one. For the RAW method, we have
\begin{equation}	
\begin{small}
\label{eq56}
\begin{aligned}
S_{{\rm{RAW}}}^\mathrm{non} \approx& 2{\sigma _{{\rm{RAW}}}}{\sigma _{{\rm{RA}}{{\rm{W}}_\mathrm{non}}}}\sigma _\mathrm{U}^2/\{ \sigma _{{\rm{RAW}}}^2\sigma _{{H_\mathrm{non}}}^2 + \sigma _{{\rm{RA}}{{\rm{W}}_\mathrm{non}}}^2\sigma _H^2\\
& + (\sigma _{{\rm{RAW}}}^2 + \sigma _{{\rm{RAW}}}^2\sigma _{{H_\mathrm{non}}}^2 + \sigma _{{\rm{RA}}{{\rm{W}}_\mathrm{non}}}^2 + \sigma _{{\rm{RA}}{{\rm{W}}_\mathrm{non}}}^2\sigma _H^2)\sigma _\mathrm{U}^2\} \\
 < &1.
\end{aligned}
\end{small}
\end{equation}
Therefore, the RAW method exhibits a low expected silhouette score under the non-i.i.d. stochastic channel scenario, even in the high SNR region.

\begin{table}[h!t]
\vspace{-0.1cm}
\scriptsize
   \centering
   \caption{Simulation Parameters in Frequency Domain}
    \begin{center}
     \begin{tabular}{||c c c c c c||}
     \hline
     Parameter & {\makecell[c] {Constant or \\ random variable?}} & {Value} & {Distribution} & {Mean} & {{\makecell[c] {Standard \\ deviation}}}\\
     \hline
     $x$ & Constant & 1 & \textbackslash & \textbackslash & \textbackslash \\
     \hline
     ${f_\mathrm{RA}}$ & Constant & 1 & \textbackslash & \textbackslash & \textbackslash \\
     \hline
     ${f_\mathrm{TA}}$ & Constant & 1 & \textbackslash & \textbackslash & \textbackslash \\
     \hline
     ${f_\mathrm{RU}}$ & Constant & 1 & \textbackslash & \textbackslash & \textbackslash \\
     \hline
     ${f_\mathrm{TU_L}}$ & Constant & 1 & \textbackslash & \textbackslash & \textbackslash \\
     \hline
     $\eta$ & Constant & 2 & \textbackslash & \textbackslash & \textbackslash \\
     \hline
     $R_\mathrm{L}$ & Constant & 52 & \textbackslash & \textbackslash & \textbackslash \\
     \hline
     $R_\mathrm{S}$ & Constant & 12 & \textbackslash & \textbackslash & \textbackslash \\
     \hline
     ${H_\mathrm{det}}$ & Random variable & \textbackslash & Gaussian & 1 & 0.15 \\
     \hline
     ${H_\mathrm{iid}}$ & Random variable & \textbackslash & Gaussian & 1 & 0.15 \\
     \hline
     ${H_\mathrm{non}}$ & Random variable & \textbackslash & Gaussian & 1 & 0.2 \\
     \hline
      ${F_\mathrm{TU}}$ & Random variable & \textbackslash & Gaussian & 1 & 0.1 \\
     \hline
      ${F_\mathrm{TU_S}}$ & Random variable & \textbackslash & Gaussian & 1 & 0.08 \\
     \hline
    \end{tabular}
    \end{center}
\label{tab2}
\vspace{-0.5cm}
\end{table}

\vspace{-0.3cm}
\section{Simulation Results}
\label{section:6}
In the simulation section, we evaluate the silhouette score and classification accuracy of different RFF extraction methods. Due to its simplicity, efficiency, and interpretability, linear discriminant analysis (LDA) \cite{xanthopoulos2013linear} is used as the classifier in the simulation.

\vspace{-0.1cm}
\subsection{Simulation Parameters}
The simulation parameters are set based on the parameters listed in Table \ref{tab2}. All parameters are defined in the frequency domain. Some parameters, such as $x$ and $f_\mathrm{RA}$, are constant across different subcarriers, while other parameters including $H_\mathrm{det}$ and $H_\mathrm{non}$, are set as Gaussian random variables across different subcarriers. Monte Carlo simulations are performed 1,000 times to obtain the average results.
\vspace{-0.3cm}
\subsection{Deterministic Channel Scenario}
\begin{figure*}[htbp]
    \centering
    \subfloat{
        \includegraphics[width=5.7cm]{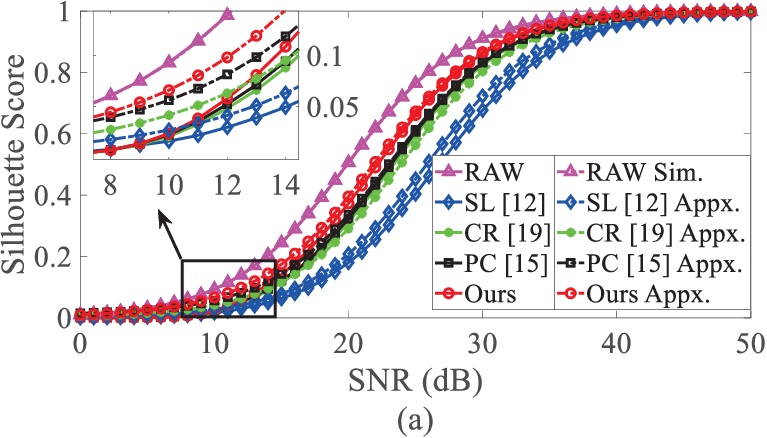}
        }\hfill
    \subfloat{
        \includegraphics[width=5.7cm]{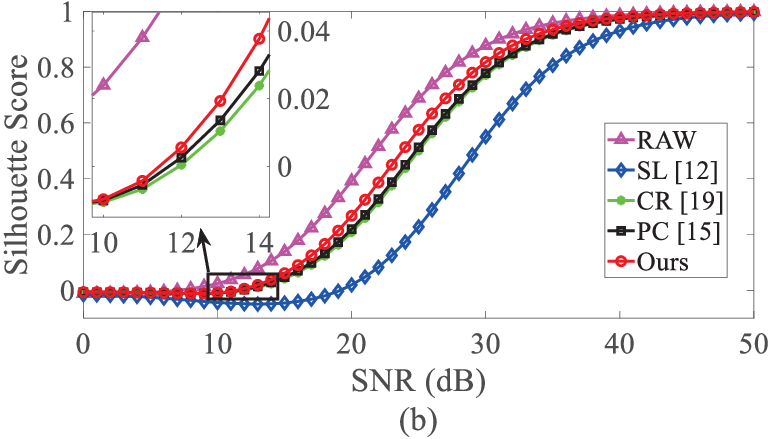}
        }\hfill
    \subfloat{
        \includegraphics[width=5.7cm]{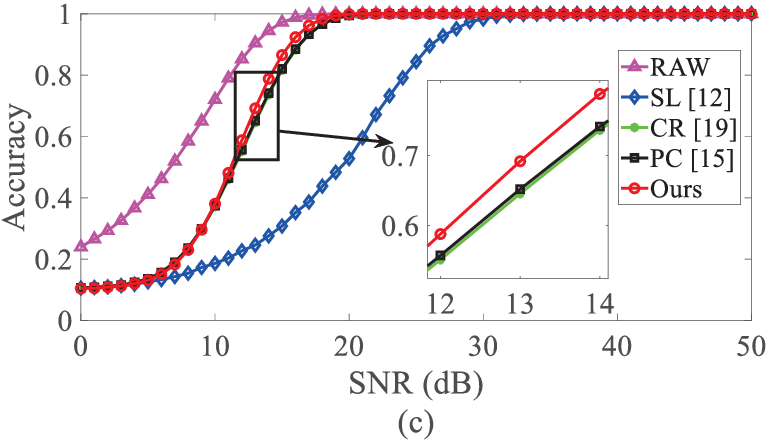}
        }
    \vspace{-0.1cm}
    \caption{Silhouette score and classification accuracy with varying SNR under the deterministic channel scenario: (a) silhouette score for two devices, (b) silhouette score for ten devices, and (c) classification accuracy for ten devices.}
    \label{fig2}
    \vspace{-0.5cm}
\end{figure*}

Fig. \ref{fig2}(a) shows the silhouette score of different methods with varying SNR under the deterministic channel scenario for two different devices. Comparison between the numerical and analytical results shows that the silhouette scores obtained using the Taylor series expansion approximation are accurate across different methods. For a given SNR, the RAW method and SL method achieve the highest and lowest silhouette scores, respectively. This is because the RAW method, without the division process, is more robust to noise effects under the deterministic channel scenario. In contrast, the SL method is more susceptible to noise due to the similar RFF effects between the L-STF and L-LTF parts. The CR method achieves a higher silhouette score than the SL method, as the RFF discrepancy between the device's transmitting and receiving front ends is greater than the RFF difference between the signals in the L-STF and L-LTF parts. Furthermore, the proposed method outperforms the PC method in terms of silhouette score. The rationale lies in that, in the proposed method, the baseband signal is amplified before transmission on the device side. This amplification process makes the RFF feature extracted by the proposed method more robust to noise effects compared to the PC method.

\begin{figure*}[htbp]
    \centering
    \subfloat{
        \includegraphics[width=5.7cm]{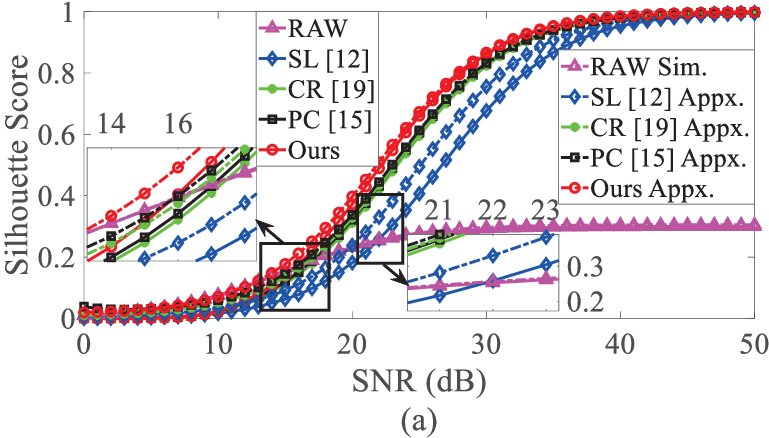}
        }\hfill
    \subfloat{
        \includegraphics[width=5.7cm]{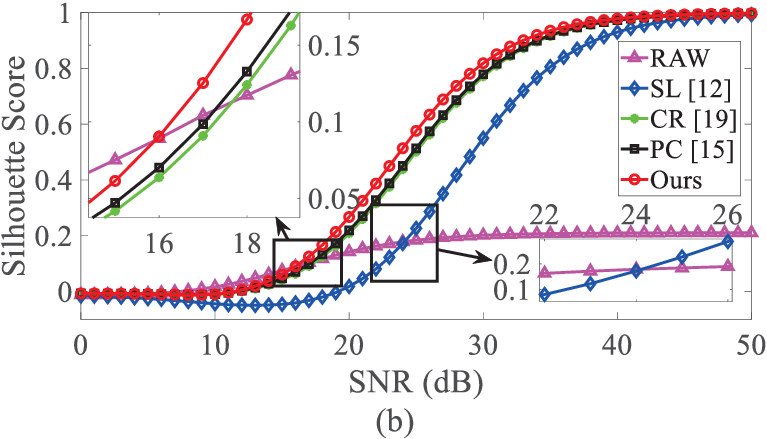}
        }\hfill
    \subfloat{
        \includegraphics[width=5.7cm]{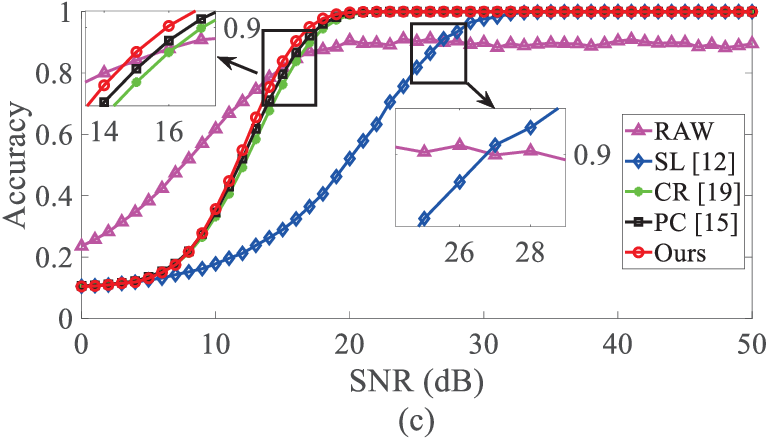}
        }
    \caption{Silhouette score and classification accuracy with varying SNR under the i.i.d. stochastic channel scenario: (a) silhouette score for two devices, (b) silhouette score for ten devices, and (c) classification accuracy for ten devices.}
    \label{fig3}
    \vspace{-0.5cm}
\end{figure*}

Fig. \ref{fig2}(b) illustrates the silhouette score of different methods with varying SNR under the deterministic channel scenario when the number of devices is ten. To calculate the inter-class distance for the $n$th training sample of device $i$, we first compute the average distance between this sample and the test samples of other devices. Subsequently, the smallest average distance is selected as the inter-class distance of the $n$th training sample from device $i$. As shown in Fig. \ref{fig2}(b), the silhouette scores across different methods remain similar to that observed in Fig. \ref{fig2}(a).

Fig. \ref{fig2}(c) shows the classification accuracy of different methods with varying SNR among ten devices. It can be observed from Fig. \ref{fig2}(c) that the RAW method consistently achieves the highest classification accuracy, as it has the highest silhouette score under the deterministic channel scenario. In contrast, the SL method consistently exhibits the lowest classification accuracy due to its lowest silhouette score. Across both low and high SNR regions, the CR method, the PC method, and the proposed method show comparable classification accuracy due to their similar silhouette scores. When SNR ranges from 10 dB to 20 dB, the proposed method achieves the highest classification accuracy among these three methods,
\begin{figure*}[htbp]
    \centering
    \subfloat{
        \includegraphics[width=5.7cm]{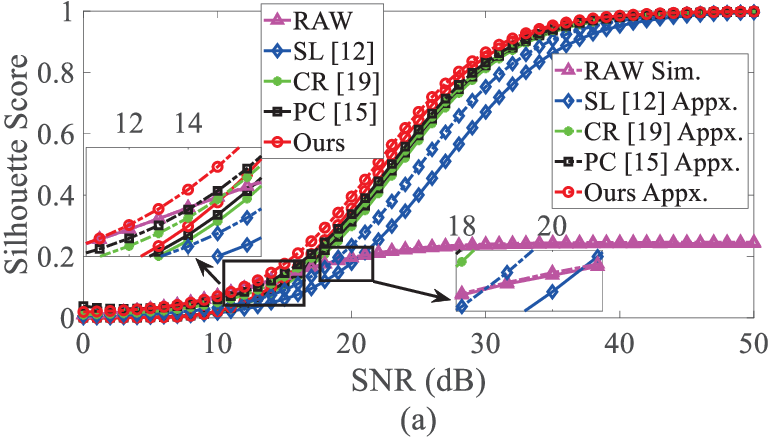}
        }\hfill
    \subfloat{
        \includegraphics[width=5.7cm]{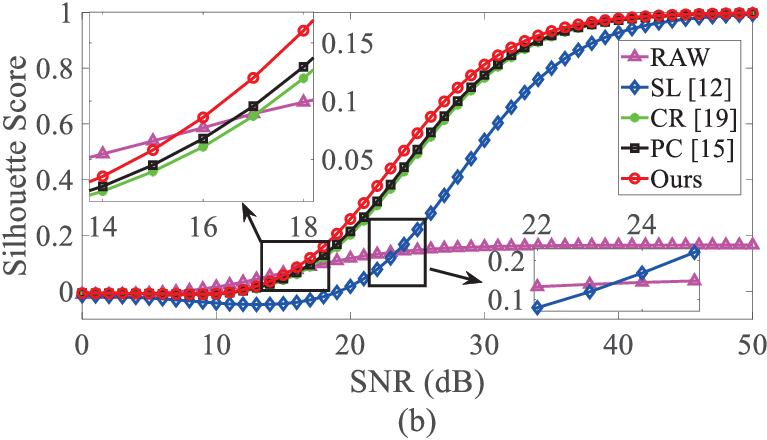}
        }\hfill
    \subfloat{
        \includegraphics[width=5.7cm]{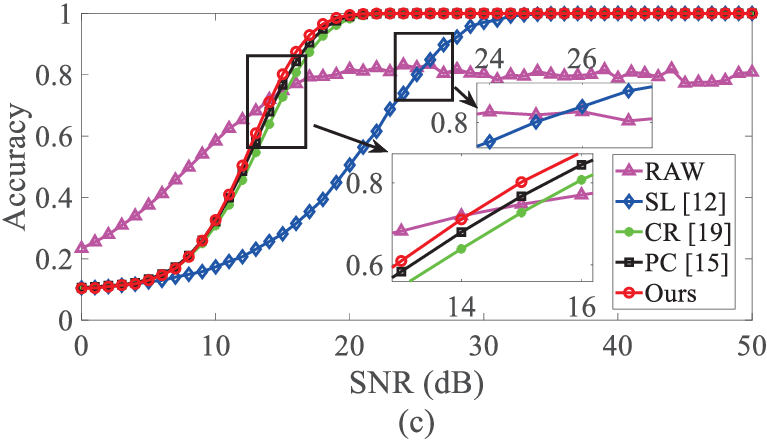}
        }
    \vspace{-0.1cm}
    \caption{Silhouette score and classification accuracy with varying SNR under the non-i.i.d. stochastic channel scenario: (a) silhouette score for two devices, (b) silhouette score for ten devices, and (c) classification accuracy for ten devices.}
    \label{fig4}
    \vspace{-0.7cm}
\end{figure*}
owing to its enhancement of the silhouette score. In the high SNR region, the classification accuracy of both methods can reach up to 100\%. By comparing Fig. \ref{fig2}(b) with Fig. \ref{fig2}(c), it can be concluded that the relationship of classification accuracy for different methods aligns with that of their silhouette scores. This correlation arises because a high silhouette score indicates a low intra-class distance and a high inter-class distance, which are beneficial for classification tasks.

\vspace{-0.3cm}
\subsection{I.i.d. Stochastic Channel Scenario}
\label{section:6_c}
Fig. \ref{fig3}(a) shows the silhouette score for different methods at varying SNR levels under the i.i.d. stochastic channel scenario, considering two different devices. It can be observed that the approximation is less accurate with the SL method. The reason is that the SL method is more susceptible to noise, making its Taylor series expansion less accurate compared to the CR, PC, and RC methods. In the low SNR region, the silhouette scores of all the methods are similar due to the dominant noise effect. When SNR ranges from 15 dB to 35 dB, the proposed method achieves the highest silhouette score, while the RAW method exhibits the lowest silhouette score. This is attributed to the absence of a channel removal process in the RAW method, resulting in higher intra-class distance and lower silhouette score due to channel variations. Similar to the deterministic channel scenario, the CR method shows a higher silhouette score than the SL method, while the PC method outperforms the CR method in terms of the silhouette score, consistent with the theoretical analysis.

Fig. \ref{fig3}(b) illustrates the silhouette score for different methods at varying SNR levels when there are ten devices. A comparative analysis of Fig. \ref{fig3}(a) and Fig. \ref{fig3}(b) reveals that the relative ordering of silhouette scores for all methods remains remarkably consistent between the two-device and ten-device scenarios.

Fig. \ref{fig3}(c) shows the classification accuracy of different methods with varying SNR under the i.i.d. stochastic channel scenario. As shown in Fig. \ref{fig3}(b) and Fig. \ref{fig3}(c), the relationship between classification accuracy and silhouette score is also similar for different methods. However, there are some exceptions. For instance, at SNR levels of 25 dB and 26 dB, the SL method yields higher silhouette scores than the RAW method, yet achieves lower classification accuracy. This indicates that, in addition to the silhouette score, factors like the quality of feature representations may also influence the classification accuracy of different methods. Therefore, while the silhouette score can serve as a useful indicator for classification accuracy, it is not a definitive indicator for classification accuracy. Besides the silhouette score, other metrics, such as the maximum Fisher's discriminant ratio, sum of the error distance by linear programming, and error rate of linear classifier \cite{lorena2019complex}, are potential candidates for indicating classification accuracy. The correlation between these metrics and classification accuracy, as well as the theoretical performance analysis of different RFF extraction methods based on these metrics, will be addressed in future work.

Moreover, by comparing Fig. \ref{fig3}(c) with Fig. \ref{fig2}(c), it is observed that the relationship between classification accuracy among different methods under the i.i.d. stochastic channel scenario is similar to that under the deterministic channel scenario in the low SNR region. However, in the high SNR region, the RAW method shows the lowest classification accuracy, while the relationship among the other methods in terms of classification accuracy remains unchanged. The RAW method's classification accuracy does not reach 100\% even at very high SNR levels, due to the absence of a channel removal process.

\begin{table}[h!t]
\scriptsize
   \centering
   \caption{The Experimental Devices}
   \vspace{-0.1cm}
   \label{tab3}
    \begin{center}
     \begin{tabular}{||c c||}
     \hline
     Notation & USRP module \\
     \hline
     Alice &  LW B210\\
     \hline
     Device 1 & LW B205 mini-i\\
     \hline
     Device 2 & LW B205 mini-i \\
     \hline
     Device 3 & LW B205 mini-i \\
     \hline
     Device 4 & USRP N210 \\
     \hline
     Device 5 & USRP N210 \\
     \hline
    \end{tabular}
    \end{center}
\vspace{-0.7cm}
\end{table}

\begin{figure}[htbp]
    \centering
    \subfloat{
        \includegraphics[width=4.18cm]{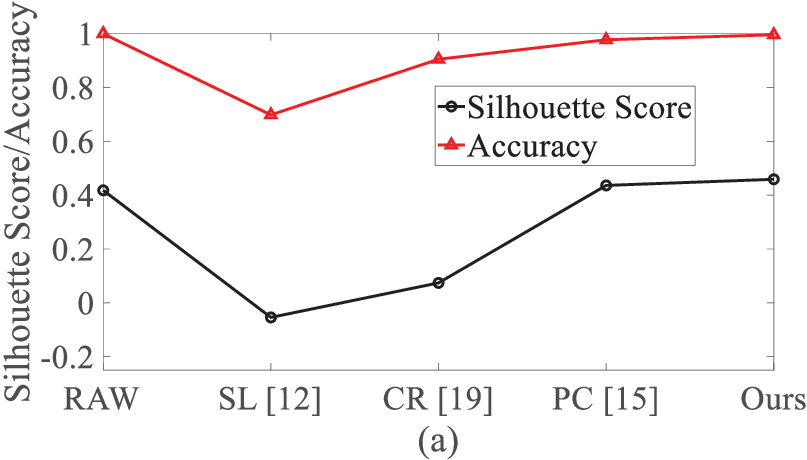}
        }\hfill
    \subfloat{
        \includegraphics[width=4.18cm]{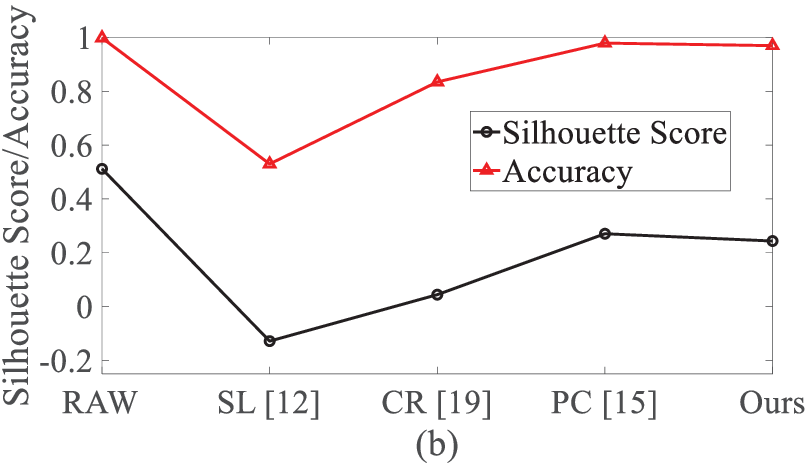}
        }
    \vspace{-0.1cm}
    \caption{Silhouette score and classification accuracy under the deterministic channel scenario: (a) LOS scenario and (b) NLOS scenario.}
    \label{fig5}
    \vspace{-0.5cm}
\end{figure}

\vspace{-0.3cm}
\subsection{Non-i.i.d. Stochastic Channel Scenario}
\label{section:6_d}
Fig. \ref{fig4}(a) illustrates the silhouette scores of different methods with varying SNR under the non-i.i.d. stochastic channel scenario for two different devices. By comparing Fig. \ref{fig4}(a) with Fig. \ref{fig3}(a), we observe that the relationship between the silhouette scores of different methods under the non-i.i.d. stochastic channel scenario is similar to that under the i.i.d. stochastic channel scenario.

Fig. \ref{fig4}(b) shows the silhouette scores of different methods with varying SNR under the non-i.i.d. stochastic channel scenario when the number of devices is ten. It can be observed from the comparison of Fig. \ref{fig4}(b) and Fig. \ref{fig4}(a) that the relationship between the silhouette scores of different methods for ten devices is similar to that for two devices.

Fig. \ref{fig4}(c) illustrates the classification accuracy of different methods with varying SNR. The proposed method achieves the highest classification accuracy when SNR ranges from 10 dB to 20 dB. The SL, CR, PC, and RC methods all achieve 100\% classification accuracy in the high SNR region. In contrast, the classification accuracy of the RAW method fluctuates around 80\% as SNR increases and is lower than that under the i.i.d. stochastic channel scenario.

This discrepancy arises because the channel differences are more pronounced in the non-i.i.d. stochastic channel scenario compared to the i.i.d. stochastic channel scenario.

\vspace{-0.3cm}
\section{Experimental Evaluation}
\label{section:7}
\vspace{-0.1cm}
\subsection{Experimental Setup}
The experiments are conducted in a typical indoor office environment characterized by rich multipath channel fading effects. Both line-of-sight (LOS) and non-line-of-sight (NLOS) scenarios are included. The LuoWave (LW) B210 is used as the legitimate node Alice, while the five devices to be classified are equipped with either LW B205 mini-i or USRP N210 modules. The configurations of different devices are detailed in Table \ref{tab3}. LDA is employed as the classifier in the experiments. The silhouette scores and classification accuracies of various methods are evaluated under both deterministic and stochastic channel scenarios.

\begin{figure}[htbp]
    \centering
    \subfloat{
        \includegraphics[width=4.18cm]{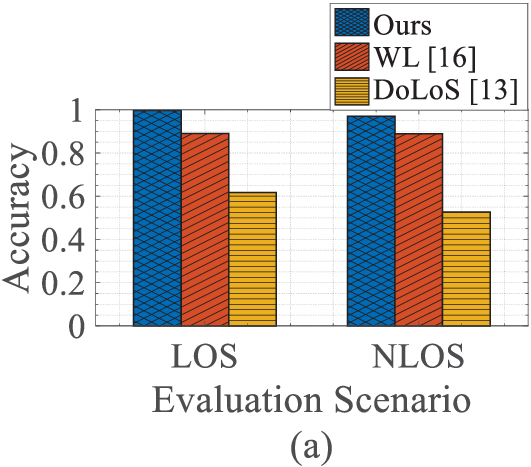}
        }\hfill
    \subfloat{
        \includegraphics[width=4.18cm]{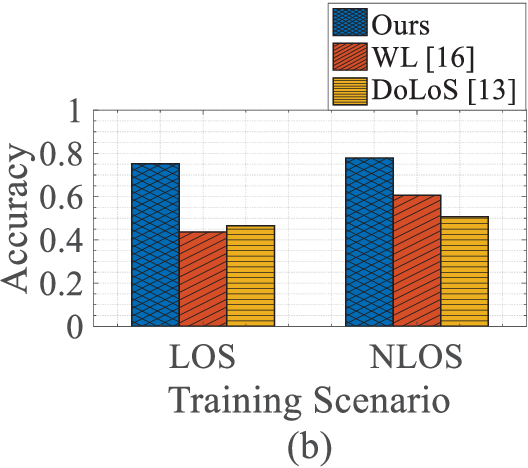}
        }
    \vspace{-0.1cm}
    \caption{Comparison with the deep learning based methods proposed in \cite{yang2023led} and \cite{xing2022design}. (a) Classification accuracy under the deterministic channel scenario and (b) classification accuracy under the varying channel scenario.}
    \label{fig6}
    \vspace{-0.5cm}
\end{figure}

\begin{figure}[htbp]
    \centering
    \subfloat{
        \includegraphics[width=4.18cm]{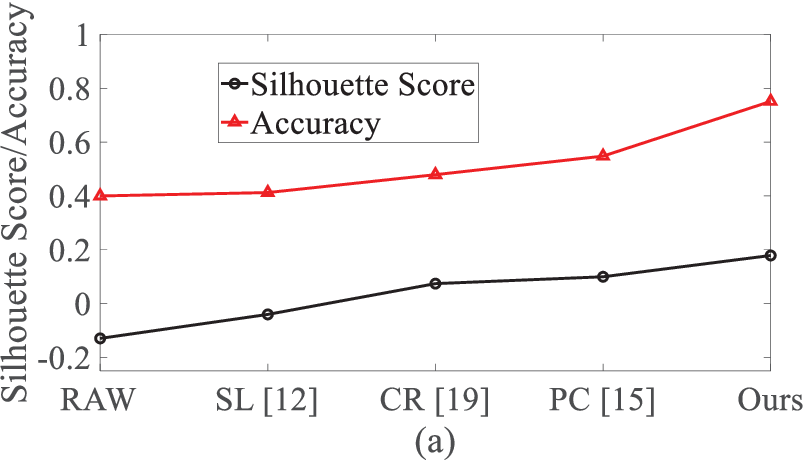}
        }\hfill
    \subfloat{
        \includegraphics[width=4.18cm]{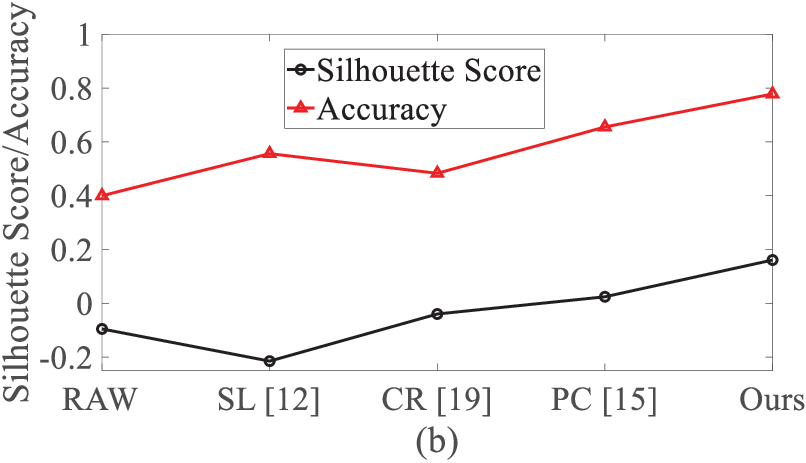}
        }
    \vspace{-0.1cm}
    \caption{Silhouette score and classification accuracy under varying channels: (a) training on LOS scenario, testing on NLOS scenario; (b) training on NLOS scenario, testing on LOS scenario.}
    \label{fig7}
    \vspace{-0.3cm}
\end{figure}
\vspace{-0.2cm}
\subsection{Deterministic Channel Scenario}
Fig. \ref{fig5}(a) and Fig. \ref{fig5}(b) present the silhouette scores and classification accuracies of different methods under LOS and NLOS scenarios, respectively. From these figures, it can be observed that the SL method consistently has the lowest silhouette score and classification accuracy in both scenarios. Additionally, in most cases, a higher silhouette score corresponds to a higher classification accuracy. The silhouette scores of the RAW, PC, and RC methods are similar, and their classification accuracies also show close performance under both LOS and NLOS scenarios.

To enable comparison with deep learning-based approaches, the performances of two existing methods proposed in \cite{yang2023led} and \cite{xing2022design} are evaluated, denoted as the WL and DoLoS methods, respectively. The WL method employs the InceptionTime neural network as its classifier, whereas the DoLoS method uses a convolutional neural network (CNN) for classification. The classification accuracy of the proposed RC method, the WL method, and the DoLoS method under the deterministic channel scenario is shown in Fig. \ref{fig6}(a).

It can be observed that both the WL and DoLoS methods achieve lower classification accuracy than the proposed RC method. This is because, although InceptionTime and CNN are more powerful than classical machine learning classifiers, the WL method fails to effectively extract representative RFF features from a limited number of WiFi subcarriers, while the DoLoS method requires a frame-stacking process for denoising, both of which lead to lower classification accuracy compared to the proposed RC method.
\begin{figure}[htbp]
    \centering
    \subfloat{
        \includegraphics[width=5cm]{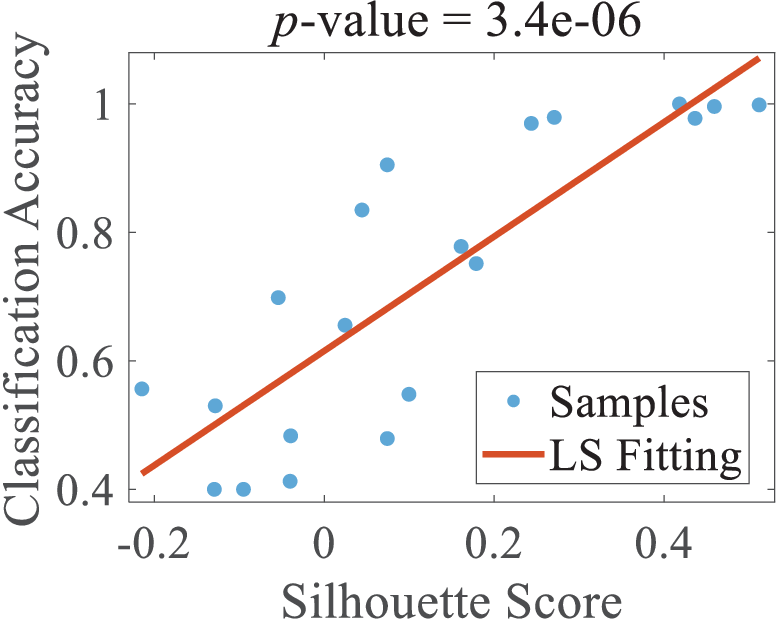}
        }
    \vspace{-0.1cm}
    \caption{The strong positive correlation between the silhouette score and classification accuracy.}
    \label{fig8}
    \vspace{-0.3cm}
\end{figure}

\vspace{-0.2cm}
\subsection{Stochastic Channel Scenario}
Under the stochastic channel scenario, we do not analyze the i.i.d. and non-i.i.d. stochastic channel scenarios separately, as done in the simulations. This is because the i.i.d. stochastic channel scenario cannot be reliably guaranteed in practice. Therefore, we primarily evaluate performance under the non-i.i.d. stochastic channel scenarios, i.e., by using the LOS scenario for training and the NLOS scenario for testing, or vice versa. Fig. \ref{fig7}(a) and Fig. \ref{fig7}(b) illustrate the silhouette scores and classification accuracies of different methods under channel variations. In Fig. \ref{fig7}(a), the training set is collected under the LOS scenario, while the test set is from the NLOS scenario. Conversely, in Fig. \ref{fig7}(b), the training set is obtained under the NLOS scenario, and the test set is from the LOS scenario. As shown in Fig. \ref{fig7}(a), the proposed method achieves the highest silhouette score and classification accuracy, whereas the RAW method has the lowest silhouette score and classification accuracy, due to the absence of a channel removing process. This reinforces the trend that a higher silhouette score generally corresponds to a higher classification accuracy. A similar result can be observed in Fig. \ref{fig7}(b) when comparing the RAW, CR, PC, and RC methods.

Note that although the SL method has the lowest silhouette score, its classification accuracy is higher than that of the RAW and CR methods. Fig. \ref{fig8} shows the silhouette score and classification accuracy of different methods under the deterministic and stochastic channel scenarios. The $p$-value for the correlation between the silhouette score and classification accuracy is $3.4\times 10^{-6}$, which is much smaller than 0.05. Furthermore, least-squares (LS) fitting indicates a positive correlation between the silhouette score and classification accuracy. These results demonstrate that the silhouette score and classification accuracy exhibit a strong positive correlation. Thus, rather than directly determining the classification accuracy, the silhouette score serves as a valuable metric for evaluating classification performance.

The classification accuracy of the WL, DoLoS, and the proposed RC methods under the stochastic channel scenario is shown in Fig. \ref{fig6}(b). Both the WL and DoLoS methods continue to achieve lower classification accuracy than the RC method, for reasons similar to those observed under the deterministic channel scenario.
\vspace{-0.3cm}
\section{Conclusion}
\label{section:8}
In this paper, we conducted theoretical analyses of various division-based and precoding-based RFF feature extraction methods using the silhouette score as the evaluation metric and proposed a precoding-based channel-robust RFF feature extraction method to enhance the silhouette score without requiring channel estimation. Simulation and experimental results showed that while the silhouette score cannot directly determine classification accuracy, it serves as an important metric for evaluating classification performance. Additionally, the expectation approximation using the Taylor series expansion was found to be accurate for calculating the silhouette scores. Moreover, the proposed method, which computes the reciprocal of the received signal and amplifies the processed signal in the baseband, achieves the highest silhouette score and classification accuracy under channel variations.
\vspace{-0.3cm}
\bibliographystyle{IEEEtran}
\bibliography{references}

\newpage

\maketitle
\vspace{-0cm}
\begin{appendices}
% Appendix A
\section{Proof of Claim 1}
\label{appendix:A}
Following the approach in \cite{van2000mean}, we employ the Taylor series expansion of $\frac{1}{{\rho G + W}}$ around $({\mu _G},{\mu _W})$, yielding
\begin{equation}
\begin{small}
\label{eq57}
\begin{aligned}
&\frac{G}{\rho G+W}=\frac{G}{\rho G+W}|_{\mu _G,\mu _W}+\left( G-\mu _G \right) \frac{\partial}{\partial G}\left( \frac{G}{\rho G+W} \right) |_{\mu _G,\mu _W}
\\
&+W\frac{\partial}{\partial W}\left( \frac{G}{\rho G+W} \right) |_{\mu _G,\mu _W}+\frac{\left( G-\mu _G \right) ^2}{2}\frac{\partial ^2}{\partial G^2}
\\
&\left( \frac{G}{\rho G+W} \right) |_{\mu _G,\mu _W}+\frac{W^2}{2}\frac{\partial ^2}{\partial W^2}\left( \frac{G}{\rho G+W} \right) |_{\mu _G,\mu _W}
\\
&+\left( G-\mu _G \right) W\frac{\partial ^2}{\partial G\partial W}\left( \frac{G}{\rho G+W} \right) |_{\mu _G,\mu _W}+o\left( G,W \right)
\\
&\approx \frac{1}{\rho}-\frac{W}{\rho ^2\mu _G}+\frac{W^2}{\rho ^3\mu _{G}^{3}}+\frac{\left( G-\mu _G \right) W}{\rho ^2\mu _{G}^{2}},
\end{aligned}
\end{small}
\end{equation}
where ${o}\left( G,W \right)$ represents the approximation error. Under the condition that ${\mu _W} = 0$, we obtain
\begin{equation}
\begin{small}
\label{eq58}
\mathbb{E} [\frac{G}{\rho G+W}]\approx \frac{\rho ^2\mu _{G}^{2}+\sigma _{W}^{2}}{\rho ^3\mu _{G}^{2}}.
\end{small}
\end{equation}
After applying the Taylor series expansion to $\frac{{{G^2}}}{{{{(\rho G + W)}^2}}}$ around $(\mu_G,\mu_W)$, where $\mu_W = 0$, we can get
\begin{equation}
\begin{small}
\label{eq59}
\begin{aligned}
&\frac{G^2}{\left( \rho G+W \right) ^2}=\frac{G^2}{\left( \rho G+W \right) ^2}|_{\mu _G,\mu _W}+\left( G-\mu _G \right) \frac{\partial}{\partial G}
\\
&\left( \frac{G^2}{\left( \rho G+W \right) ^2} \right) |_{\mu _G,\mu _W}+W\frac{\partial}{\partial W}\left( \frac{G^2}{\left( \rho G+W \right) ^2} \right) |_{\mu _G,\mu _W}
\\
&+\frac{\left( G-\mu _G \right) ^2}{2}\frac{\partial ^2}{\partial G^2}\left( \frac{G^2}{\left( \rho G+W \right) ^2} \right) |_{\mu _G,\mu _W}+\frac{W^2}{2}
\\
&\frac{\partial ^2}{\partial W^2}\left( \frac{G^2}{\left( \rho G+W \right) ^2} \right) |_{\mu _G,\mu _W}+\left( G-\mu _G \right) W
\\
&\frac{\partial ^2}{\partial G\partial W}\left( \frac{G^2}{\left( \rho G+W \right) ^2} \right) |_{\mu _G,\mu _W}+o\left( G,W \right)
\\
&\approx \frac{1}{\rho ^2}-\frac{2W}{\rho ^3\mu _G}+\frac{3W^2}{\rho ^4\mu _{G}^{2}}.
\end{aligned}
\end{small}
\end{equation}
Consequently, we obtain
\begin{equation}
\begin{small}
\label{eq60}
\mathbb{E} [\frac{G^2}{(\rho G+W)^2}]\approx \frac{\rho ^2\mu _{G}^{2}+3\sigma _{W}^{2}}{\rho ^4\mu _{G}^{2}}.
\end{small}
\end{equation}
$\hfill\blacksquare$

% Appendix B
\section{Proof of Claim 3}
\label{appendix:B}
Similar to the approach in \cite{van2000mean}, we employ the Taylor series expansion of $\frac{{{G_1}{W_2} - {G_2}{W_1}}}{{\left( {\rho {G_1} + {W_1}} \right)\left( {\rho {G_2} + {W_2}} \right)}}$ around $\left(\mu _{G_1}, \mu _{G_2},\mu _{W_1},\mu _{W_2} \right)$. The result is shown in (\ref{eq61}) on the top of the following page, where $o\left(G_1,G_2,W_1,W_2 \right)$ denotes the approximation error.

\begin{figure*}[ht]
\centering
\begin{equation}
\begin{small}
\label{eq61}
\begin{array}{l}
\frac{G_1W_2-G_2W_1}{\left( \rho G_1+W_1 \right) \left( \rho G_2+W_2 \right)}\approx \frac{G_1W_2-G_2W_1}{\left( \rho G_1+W_1 \right) \left( \rho G_2+W_2 \right)}|_{\mu _{G_1},\mu _{G_2},\mu _{W_1},\mu _{W_2}}+\left( G_1-\mu _{G_1} \right) \frac{\partial}{\partial G_1}\left( \frac{G_1W_2-G_2W_1}{\left( \rho G_1+W_1 \right) \left( \rho G_2+W_2 \right)} \right) |_{\mu _{G_1},\mu _{G_2},\mu _{W_1},\mu _{W_2}}
\\
+\left( G_2-\mu _{G_2} \right) \frac{\partial}{\partial G_2}\left( \frac{G_1W_2-G_2W_1}{\left( \rho G_1+W_1 \right) \left( \rho G_2+W_2 \right)} \right) |_{\mu _{G_1},\mu _{G_2},\mu _{W_1},\mu _{W_2}}+W_1\frac{\partial}{\partial W_1}\left( \frac{G_1W_2-G_2W_1}{\left( \rho G_1+W_1 \right) \left( \rho G_2+W_2 \right)} \right) |_{\mu _{G_1},\mu _{G_2},\mu _{W_1},\mu _{W_2}}
\\
+W_2\frac{\partial}{\partial W_2}\left( \frac{G_1W_2-G_2W_1}{\left( \rho G_1+W_1 \right) \left( \rho G_2+W_2 \right)} \right) |_{\mu _{G_1},\mu _{G_2},\mu _{W_1},\mu _{W_2}}+\frac{\left( G_1-\mu _{G_1} \right) ^2}{2}\frac{\partial ^2}{{\partial G_1}^2}\left( \frac{G_1W_2-G_2W_1}{\left( \rho G_1+W_1 \right) \left( \rho G_2+W_2 \right)} \right) |_{\mu _{G_1},\mu _{G_2},\mu _{W_1},\mu _{W_2}}
\\
+\frac{\left( G_2-\mu _{G_2} \right) ^2}{2}\frac{\partial ^2}{{\partial G_2}^2}\left( \frac{G_1W_2-G_2W_1}{\left( \rho G_1+W_1 \right) \left( \rho G_2+W_2 \right)} \right) |_{\mu _{G_1},\mu _{G_2},\mu _{W_1},\mu _{W_2}}+\frac{{W_1}^2}{2}\frac{\partial ^2}{{\partial W_1}^2}\left( \frac{G_1W_2-G_2W_1}{\left( \rho G_1+W_1 \right) \left( \rho G_2+W_2 \right)} \right) |_{\mu _{G_1},\mu _{G_2},\mu _{W_1},\mu _{W_2}}
\\
+\frac{{W_2}^2}{2}\frac{\partial ^2}{{\partial W_2}^2}\left( \frac{G_1W_2-G_2W_1}{\left( \rho G_1+W_1 \right) \left( \rho G_2+W_2 \right)} \right) |_{\mu _{G_1},\mu _{G_2},\mu _{W_1},\mu _{W_2}}+\left( G_1-\mu _{G_1} \right) \left( G_2-\mu _{G_2} \right) \frac{\partial ^2}{\partial G_1\partial G_2}\left( \frac{G_1W_2-G_2W_1}{\left( \rho G_1+W_1 \right) \left( \rho G_2+W_2 \right)} \right) |_{\mu _{G_1},\mu _{G_2},\mu _{W_1},\mu _{W_2}}
\\
+\left( G_1-\mu _{G_1} \right) W_1\frac{\partial ^2}{\partial G_1\partial W_1}\left( \frac{G_1W_2-G_2W_1}{\left( \rho G_1+W_1 \right) \left( \rho G_2+W_2 \right)} \right) |_{\mu _{G_1},\mu _{G_2},\mu _{W_1},\mu _{W_2}}+\left( G_1-\mu _{G_1} \right) W_2\frac{\partial ^2}{\partial G_1\partial W_2}\left( \frac{G_1W_2-G_2W_1}{\left( \rho G_1+W_1 \right) \left( \rho G_2+W_2 \right)} \right) |_{\mu _{G_1},\mu _{G_2},\mu _{W_1},\mu _{W_2}}
\\
+\left( G_2-\mu _{G_2} \right) W_1\frac{\partial ^2}{\partial G_2\partial W_1}\left( \frac{G_1W_2-G_2W_1}{\left( \rho G_1+W_1 \right) \left( \rho G_2+W_2 \right)} \right) |_{\mu _{G_1},\mu _{G_2},\mu _{W_1},\mu _{W_2}}+\left( G_2-\mu _{G_2} \right) W_2\frac{\partial ^2}{\partial G_2\partial W_2}\left( \frac{G_1W_2-G_2W_1}{\left( \rho G_1+W_1 \right) \left( \rho G_2+W_2 \right)} \right) |_{\mu _{G_1},\mu _{G_2},\mu _{W_1},\mu _{W_2}}
\\
+W_1W_2\frac{\partial ^2}{\partial W_1\partial W_2}\left( \frac{G_1W_2-G_2W_1}{\left( \rho G_1+W_1 \right) \left( \rho G_2+W_2 \right)} \right) |_{\mu _{G_1},\mu _{G_2},\mu _{W_1},\mu _{W_2}}+o\left( G_1,G_2,W_1,W_2 \right).
\end{array}
\end{small}
\end{equation}
\vspace{-0.7cm}
\end{figure*}
Under the condition that $\mu _{W_1}=0$ and $\mu _{W_2}=0$, we can get
\begin{equation}
\begin{small}
\begin{aligned}
\label{eq62}
\mathbb{E} \left[ \frac{G_1W_2-G_2W_1}{\left( \rho G_1+W_1 \right) \left( \rho G_2+W_2 \right)} \right] &\approx \frac{\sigma _{W}^{2}}{2}\frac{2}{\rho ^3{\mu _G}^2}-\frac{\sigma _{W}^{2}}{2}\frac{2}{\rho ^3{\mu _G}^2}
\\
&=0.
\end{aligned}
\end{small}
\end{equation}
Then, the Taylor series expansion of $\left( \frac{G_1W_2-G_2W_1}{\left( \rho G_1+W_1 \right) \left( \rho G_2+W_2 \right)} \right) ^2$ is employed around $\left(\mu _{G_1}, \mu _{G_2},\mu _{W_1},\mu _{W_2} \right)$ and the result is shown in (\ref{eq63}) on the top of this page.
\begin{figure*}[ht]
\centering
\begin{equation}
\begin{small}
\label{eq63}
\begin{array}{l}
\left( \frac{G_1W_2-G_2W_1}{\left( \rho G_1+W_1 \right) \left( \rho G_2+W_2 \right)} \right) ^2\approx \left( \frac{G_1W_2-G_2W_1}{\left( \rho G_1+W_1 \right) \left( \rho G_2+W_2 \right)} \right) ^2|_{\mu _{G_1},\mu _{G_2},\mu _{W_1},\mu _{W_2}}+\left( G_1-\mu _{G_1} \right) \frac{\partial}{\partial G_1}\left( \frac{G_1W_2-G_2W_1}{\left( \rho G_1+W_1 \right) \left( \rho G_2+W_2 \right)} \right) ^2|_{\mu _{G_1},\mu _{G_2},\mu _{W_1},\mu _{W_2}}
\\
+\left( G_2-\mu _{G_2} \right) \frac{\partial}{\partial G_2}\left( \frac{G_1W_2-G_2W_1}{\left( \rho G_1+W_1 \right) \left( \rho G_2+W_2 \right)} \right) ^2|_{\mu _{G_1},\mu _{G_2},\mu _{W_1},\mu _{W_2}}+W_1\frac{\partial}{\partial W_1}\left( \frac{G_1W_2-G_2W_1}{\left( \rho G_1+W_1 \right) \left( \rho G_2+W_2 \right)} \right) ^2|_{\mu _{G_1},\mu _{G_2},\mu _{W_1},\mu _{W_2}}
\\
+W_2\frac{\partial}{\partial W_2}\left( \frac{G_1W_2-G_2W_1}{\left( \rho G_1+W_1 \right) \left( \rho G_2+W_2 \right)} \right) ^2|_{\mu _{G_1},\mu _{G_2},\mu _{W_1},\mu _{W_2}}+\frac{\left( G_1-\mu _{G_1} \right) ^2}{2}\frac{\partial ^2}{{\partial G_1}^2}\left( \frac{G_1W_2-G_2W_1}{\left( \rho G_1+W_1 \right) \left( \rho G_2+W_2 \right)} \right) ^2|_{\mu _{G_1},\mu _{G_2},\mu _{W_1},\mu _{W_2}}
\\
+\frac{\left( G_2-\mu _{G_2} \right) ^2}{2}\frac{\partial ^2}{{\partial G_2}^2}\left( \frac{G_1W_2-G_2W_1}{\left( \rho G_1+W_1 \right) \left( \rho G_2+W_2 \right)} \right) ^2|_{\mu _{G_1},\mu _{G_2},\mu _{W_1},\mu _{W_2}}+\frac{{W_1}^2}{2}\frac{\partial ^2}{{\partial W_1}^2}\left( \frac{G_1W_2-G_2W_1}{\left( \rho G_1+W_1 \right) \left( \rho G_2+W_2 \right)} \right) ^2|_{\mu _{G_1},\mu _{G_2},\mu _{W_1},\mu _{W_2}}
\\
+\frac{{W_2}^2}{2}\frac{\partial ^2}{{\partial W_2}^2}\left( \frac{G_1W_2-G_2W_1}{\left( \rho G_1+W_1 \right) \left( \rho G_2+W_2 \right)} \right) ^2|_{\mu _{G_1},\mu _{G_2},\mu _{W_1},\mu _{W_2}}+\left( G_1-\mu _{G_1} \right) \left( G_2-\mu _{G_2} \right) \frac{\partial ^2}{\partial G_1\partial G_2}\left( \frac{G_1W_2-G_2W_1}{\left( \rho G_1+W_1 \right) \left( \rho G_2+W_2 \right)} \right) ^2|_{\mu _{G_1},\mu _{G_2},\mu _{W_1},\mu _{W_2}}
\\
+\left( G_1-\mu _{G_1} \right) W_1\frac{\partial ^2}{\partial G_1\partial W_1}\left( \frac{G_1W_2-G_2W_1}{\left( \rho G_1+W_1 \right) \left( \rho G_2+W_2 \right)} \right) ^2|_{\mu _{G_1},\mu _{G_2},\mu _{W_1},\mu _{W_2}}+\left( G_1-\mu _{G_1} \right) W_2\frac{\partial ^2}{\partial G_1\partial W_2}\left( \frac{G_1W_2-G_2W_1}{\left( \rho G_1+W_1 \right) \left( \rho G_2+W_2 \right)} \right) ^2|_{\mu _{G_1},\mu _{G_2},\mu _{W_1},\mu _{W_2}}
\\
+\left( G_2-\mu _{G_2} \right) W_1\frac{\partial ^2}{\partial G_2\partial W_1}\left( \frac{G_1W_2-G_2W_1}{\left( \rho G_1+W_1 \right) \left( \rho G_2+W_2 \right)} \right) ^2|_{\mu _{G_1},\mu _{G_2},\mu _{W_1},\mu _{W_2}}+\left( G_2-\mu _{G_2} \right) W_2\frac{\partial ^2}{\partial G_2\partial W_2}\left( \frac{G_1W_2-G_2W_1}{\left( \rho G_1+W_1 \right) \left( \rho G_2+W_2 \right)} \right) ^2|_{\mu _{G_1},\mu _{G_2},\mu _{W_1},\mu _{W_2}}
\\
+W_1W_2\frac{\partial ^2}{\partial W_1\partial W_2}\left( \frac{G_1W_2-G_2W_1}{\left( \rho G_1+W_1 \right) \left( \rho G_2+W_2 \right)} \right) ^2|_{\mu _{G_1},\mu _{G_2},\mu _{W_1},\mu _{W_2}}+o\left( G_1,G_2,W_1,W_2 \right).
\end{array}
\end{small}
\end{equation}
\vspace{-0.7cm}
\end{figure*}

Since $\mu _{G_1}=\mu _{G_2}=\mu _G$ and $\mu _{W_1}=\mu _{W_2}=0$, we have
\begin{equation}
\begin{small}
\begin{aligned}
\label{eq64}
\mathbb{E} \left[ \left( \frac{G_1W_2-G_2W_1}{\left( \rho G_1+W_1 \right) \left( \rho G_2+W_2 \right)} \right) ^2 \right] &\approx \frac{\sigma _{W}^{2}}{2}\frac{2}{\rho ^4{\mu _G}^2}+\frac{\sigma _{W}^{2}}{2}\frac{2}{\rho ^4{\mu _G}^2}
\\
&=\frac{2\sigma _{W}^{2}}{\rho ^4{\mu _G}^2}.
\end{aligned}
\end{small}
\end{equation}
This completes the proof.
$\hfill\blacksquare$

% Appendix C
\section{Inter-Class Distances under the i.i.d. Stochastic Channel Scenario}
\label{appendix:C}
The expected inter-class distances for different methods under the independent and identically distributed stochastic channel scenario are presented in (\ref{eq65}) on the following page.
\begin{figure*}[ht]
 \centering
 \begin{equation}	
 \label{eq65}
\begin{array}{l}
D_{\mathrm{RAW},i,j}^{\mathrm{Inter},\mathrm{iid}}=\frac{2R_{\mathrm{L}}\left[ f_{\mathrm{RA}}^{2}x^2\left( \mu _{\mathrm{U}}^{2}\sigma _{H}^{2}+\sigma _{\mathrm{U}}^{2}\mu _{H}^{2}+\sigma _{\mathrm{U}}^{2}\sigma _{H}^{2} \right) +\sigma _{N}^{2} \right]}{\sigma _{\mathrm{RAW}_{\mathrm{iid}}}^{2}},
\\
D_{\mathrm{SL},i,j}^{\mathrm{Inter},\mathrm{iid}}\approx \frac{2R_{\mathrm{S}}\left\{ f_{\mathrm{RA}}^{2}x^2\left[ \mu _{\mathrm{S}}^{2}\sigma _{N}^{2}\left( \gamma ^2\mu _{H}^{2}-\sigma _{N}^{2} \right) +\gamma ^2\sigma _{\mathrm{S}}^{2}\mu _{H}^{2}\left( \gamma ^2\mu _{H}^{2}+3\sigma _{N}^{2} \right) \right] +\gamma ^2\sigma _{N}^{2}\left( \gamma ^2\mu _{H}^{2}+3\gamma ^2\sigma _{H}^{2}+3\sigma _{N}^{2} \right) \right\}}{\sigma _{\mathrm{SL}_{\mathrm{iid}}}^{2}\gamma ^6\mu _{H}^{4}},
\\
D_{\mathrm{CR},i,j}^{\mathrm{Inter},\mathrm{iid}}\approx \frac{2R_{\mathrm{L}}\left\{ f_{\mathrm{RA}}^{2}x^2\left[ \mu _{\mathrm{U}}^{2}\sigma _{N}^{2}\left( \beta ^2\mu _{H}^{2}-\sigma _{N}^{2} \right) +\beta ^2\sigma _{\mathrm{U}}^{2}\mu _{H}^{2}\left( \beta ^2\mu _{H}^{2}+3\sigma _{N}^{2} \right) \right] +\beta ^2\sigma _{N}^{2}\left( \beta ^2\mu _{H}^{2}+3\beta ^2\sigma _{H}^{2}+3\sigma _{N}^{2} \right) \right\}}{\sigma _{\mathrm{CR}_{\mathrm{iid}}}^{2}\beta ^6\mu _{H}^{4}},
\\
D_{\mathrm{PC},i,j}^{\mathrm{Inter},\mathrm{iid}}\approx \frac{2R_{\mathrm{L}}}{\sigma _{\mathrm{PC}_{\mathrm{iid}}}^{2}}[\frac{f_{\mathrm{RA}}^{2}x^4[\mu _{\mathrm{U}}^{2}\sigma _{N}^{2}(\beta ^2\mu _{H}^{2}-\sigma _{N}^{2})+\sigma _{\mathrm{U}}^{2}\beta ^2\mu _{H}^{2}(\beta ^2\mu _{H}^{2}+3\sigma _{N}^{2})]}{\beta ^6\mu _{H}^{4}}+\sigma _{N}^{2}].
\end{array}
 \end{equation}
\vspace{-0.7cm}
\end{figure*}

% Appendix D
\section{Silhouette Scores under the i.i.d. Stochastic Channel Scenario}
\label{appendix:D}
The expected silhouette scores for different methods under the i.i.d. stochastic channel scenario are provided in (\ref{eq66}) on the following page.
\begin{figure*}[ht]
 	\centering
 	\begin{equation}	
    \label{eq66}
    \begin{array}{l}
S_{{\rm{RAW}}}^\mathrm{iid} = \frac{{f_\mathrm{RA}^2{x^2}\sigma _\mathrm{U}^2\mu _H^2}}{{f_\mathrm{RA}^2{x^2}\left( {\mu _\mathrm{U}^2\sigma _H^2 + \sigma _\mathrm{U}^2\mu _H^2 + \sigma _\mathrm{U}^2\sigma _H^2} \right) + \sigma _N^2}},\\
S_{{\rm{SL}}}^\mathrm{iid} \approx \frac{{f_\mathrm{RA}^2{x^2}\sigma _\mathrm{S}^2\left( {{\gamma ^4}\mu _H^4 + 2{\gamma ^2}\mu _H^2\sigma _N^2 + \sigma _N^4} \right)}}{{f_\mathrm{RA}^2{x^2}\left[ {\mu _\mathrm{S}^2\sigma _N^2\left( {{\gamma ^2}\mu _H^2 - \sigma _N^2} \right) + {\gamma ^2}\sigma _\mathrm{S}^2\mu _H^2\left( {{\gamma ^2}\mu _H^2 + 3\sigma _N^2} \right)} \right] + {\gamma ^2}\sigma _N^2\left( {{\gamma ^2}\mu _H^2 + 3{\gamma ^2}\sigma _H^2 + 3\sigma _N^2} \right)}},\\
S_{{\rm{CR}}}^\mathrm{iid} \approx \frac{{f_\mathrm{RA}^2{x^2}\sigma _\mathrm{U}^2\left( {{\beta ^4}\mu _H^4 + 2{\beta ^2}\mu _H^2\sigma _N^2 + \sigma _N^4} \right)}}{{f_\mathrm{RA}^2{x^2}\left[ {\mu _\mathrm{U}^2\sigma _N^2\left( {{\beta ^2}\mu _H^2 - \sigma _N^2} \right) + {\beta ^2}\sigma _\mathrm{U}^2\mu _H^2\left( {{\beta ^2}\mu _H^2 + 3\sigma _N^2} \right)} \right] + {\beta ^2}\sigma _N^2\left( {{\beta ^2}\mu _H^2 + 3{\beta ^2}\sigma _H^2 + 3\sigma _N^2} \right)}},\\
S_{{\rm{PC}}}^\mathrm{iid} \approx \frac{{f_\mathrm{RA}^2{x^4}\sigma _\mathrm{U}^2{{\left( {{\beta ^2}\mu _H^2 + \sigma _N^2} \right)}^2}}}{{f_\mathrm{RA}^2{x^4}\left[ {\mu _\mathrm{U}^2\sigma _N^2\left( {{\beta ^2}\mu _H^2 - \sigma _N^2} \right) + \sigma _\mathrm{U}^2{\beta ^2}\mu _H^2\left( {{\beta ^2}\mu _H^2 + 3\sigma _N^2} \right)} \right] + {\beta ^6}\mu _H^4\sigma _N^2}}.
    \end{array}
 	\end{equation}
\vspace{-0.8cm}
\end{figure*}

% Appendix E
\section{Intra-Class Distances under the Non-i.i.d. Stochastic Channel Scenario}
\label{appendix:E}
The expected intra-class distances for different methods under the non-i.i.d. stochastic channel scenario are presented in (\ref{eq67}) on the following page.
\begin{figure*}[ht]
 	\centering
 	\begin{equation}	
    \label{eq67}
    \begin{small}
    \begin{aligned}
D_{\mathrm{RAW},i}^{\mathrm{Intra},\mathrm{non}}=&R_{\mathrm{L}}\left\{ f_{\mathrm{RA}}^{2}x^2\left( \mu _{\mathrm{U}}^{2}+\sigma _{\mathrm{U}}^{2} \right) \left[ \sigma _{\mathrm{RAW}}^{2}\left( \mu _{H_{\mathrm{non}}}^{2}+\sigma _{H_{\mathrm{non}}}^{2} \right) +\sigma _{\mathrm{RAW}_{\mathrm{non}}}^{2}\left( \mu _{H}^{2}+\sigma _{H}^{2} \right) \right] -2f_{RA}x\mu _{\mathrm{U}} \right.
\\
&\cdot \left( \mu _H\mu _{\mathrm{RAW}}\sigma _{\mathrm{RAW}_{\mathrm{non}}}^{2}+\mu _{H_{\mathrm{non}}}\mu _{\mathrm{RAW}_{\mathrm{non}}}\sigma _{\mathrm{RAW}}^{2} \right) +\sigma _{N}^{2}\left( \sigma _{\mathrm{RAW}}^{2}+\sigma _{\mathrm{RAW}_{\mathrm{non}}}^{2} \right) +\mu _{\mathrm{RAW}}^{2}\sigma _{\mathrm{RAW}_{\mathrm{non}}}^{2}
\\
&+\mu _{\mathrm{RAW}_{\mathrm{non}}}^{2}\sigma _{\mathrm{RAW}}^{2}-2\sigma _{\mathrm{RAW}}\sigma _{\mathrm{RAW}_{\mathrm{non}}}\left[ \mu _H\mu _{H_{\mathrm{non}}}f_{\mathrm{RA}}^{2}x^2\left( \mu _{U}^{2}+\sigma _{U}^{2} \right) \right.
\\
&\left. \left. -f_{RA}x\mu _{\mathrm{U}}\left( \mu _{\mathrm{RAW}}\mu _{H_{\mathrm{non}}}+\mu _{\mathrm{RAW}_{\mathrm{non}}}\mu _H \right) +\mu _{\mathrm{RAW}}\mu _{\mathrm{RAW}_{\mathrm{non}}} \right] \right\} /\left( \sigma _{\mathrm{RAW}}^{2}\sigma _{\mathrm{RAW}_{\mathrm{non}}}^{2} \right) ,
\\
D_{\mathrm{SL},i}^{\mathrm{Intra},\mathrm{non}}\approx& R_{\mathrm{S}}\left\{ f_{\mathrm{RA}}^{2}x^2\gamma ^2\mu _{H}^{2}\mu _{H_{\mathrm{non}}}^{2}\left( \mu _{\mathrm{S}}^{2}+\sigma _{\mathrm{S}}^{2} \right) \left[ \sigma _{\mathrm{SL}}^{2}\mu _{H}^{2}\left( \gamma ^2\mu _{H_{\mathrm{non}}}^{2}+3\sigma _{N}^{2} \right) +\sigma _{\mathrm{SL}_{\mathrm{non}}}^{2}\mu _{H_{\mathrm{non}}}^{2}\left( \gamma ^2\mu _{H}^{2}+3\sigma _{N}^{2} \right) \right] \right.
\\
&+\sigma _{N}^{2}\gamma ^2\left[ \sigma _{\mathrm{SL}}^{2}\left( \gamma ^2\mu _{H_{\mathrm{non}}}^{2}+3\gamma ^2\sigma _{H_{\mathrm{non}}}^{2}+3\sigma _{N}^{2} \right) +\sigma _{\mathrm{SL}_{\mathrm{non}}}^{2}\left( \gamma ^2\mu _{H}^{2}+3\gamma ^2\sigma _{H}^{2}+3\sigma _{N}^{2} \right) \right]
\\
&-2f_{\mathrm{RA}}x\mu _{\mathrm{S}}\gamma ^3\mu _{H}^{2}\mu _{H_{\mathrm{non}}}^{2}\left[ \sigma _{\mathrm{SL}}^{2}\mu _{H}^{2}\left( \gamma ^2\mu _{H_{\mathrm{non}}}^{2}+\sigma _{N}^{2} \right) +\sigma _{\mathrm{SL}_{\mathrm{non}}}^{2}\mu _{H_{\mathrm{non}}}^{2}\left( \gamma ^2\mu _{H}^{2}+\sigma _{N}^{2} \right) \right]
\\
&-2\sigma _{\mathrm{SL}}\sigma _{\mathrm{SL}_{\mathrm{non}}}f_{\mathrm{RA}}^{2}x^2\mu _{H}^{2}\mu _{H_{\mathrm{non}}}^{2}\left( \mu _{\mathrm{S}}^{2}+\sigma _{\mathrm{S}}^{2} \right) \left( \gamma ^2\mu _{H}^{2}+\sigma _{N}^{2} \right) \left( \gamma ^2\mu _{H_{\mathrm{non}}}^{2}+\sigma _{N}^{2} \right)
\\
&+2\sigma _{\mathrm{SL}}\sigma _{\mathrm{SL}_{\mathrm{non}}}f_{\mathrm{RA}}x\mu _S\gamma ^3\mu _{H}^{2}\mu _{H_{\mathrm{non}}}^{2}\left[ \mu _{\mathrm{SL}}\mu _{H}^{2}\left( \gamma ^2\mu _{H_{\mathrm{non}}}^{2}+\sigma _{N}^{2} \right) +\mu _{\mathrm{SL}_{\mathrm{non}}}\mu _{H_{\mathrm{non}}}^{2}\left( \gamma ^2\mu _{H}^{2}+\sigma _{N}^{2} \right) \right]
\\
&\left. +\gamma ^6\mu _{H}^{4}\mu _{H_{\mathrm{non}}}^{4}\left( \sigma _{\mathrm{SL}}^{2}\mu _{\mathrm{SL}_{\mathrm{non}}}^{2}+\sigma _{\mathrm{SL}_{\mathrm{non}}}^{2}\mu _{\mathrm{SL}}^{2}-2\sigma _{\mathrm{SL}}\sigma _{\mathrm{SL}_{\mathrm{non}}}\mu _{\mathrm{SL}}\mu _{\mathrm{SL}_{\mathrm{non}}} \right) \right\} /\left( \gamma ^6\mu _{H}^{4}\mu _{H_{\mathrm{non}}}^{4}\sigma _{\mathrm{SL}}^{2}\sigma _{\mathrm{SL}_{\mathrm{non}}}^{2} \right) ,
\\
D_{\mathrm{CR},i}^{\mathrm{Intra},\mathrm{non}}\approx& R_{\mathrm{L}}\left\{ f_{\mathrm{RA}}^{2}x^2\beta ^2\mu _{H}^{2}\mu _{H_{\mathrm{non}}}^{2}\left( \mu _{U}^{2}+\sigma _{\mathrm{U}}^{2} \right) \left[ \sigma _{\mathrm{CR}}^{2}\mu _{H}^{2}\left( \beta ^2\mu _{H_{\mathrm{non}}}^{2}+3\sigma _{N}^{2} \right) +\sigma _{\mathrm{CR}_{\mathrm{non}}}^{2}\mu _{H_{\mathrm{non}}}^{2}\left( \beta ^2\mu _{H}^{2}+3\sigma _{N}^{2} \right) \right] \right.
\\
&+\sigma _{N}^{2}\beta ^2\left[ \sigma _{\mathrm{CR}}^{2}\left( \beta ^2\mu _{H_{\mathrm{non}}}^{2}+3\beta ^2\sigma _{H_{\mathrm{non}}}^{2}+3\sigma _{N}^{2} \right) +\sigma _{\mathrm{CR}_{\mathrm{non}}}^{2}\left( \beta ^2\mu _{H}^{2}+3\beta ^2\sigma _{H}^{2}+3\sigma _{N}^{2} \right) \right]
\\
&-2f_{RA}x\mu _{\mathrm{U}}\beta ^3\mu _{H}^{2}\mu _{H_{\mathrm{non}}}^{2}\left[ \sigma _{\mathrm{CR}}^{2}\mu _{H}^{2}\left( \beta ^2\mu _{H_{\mathrm{non}}}^{2}+\sigma _{N}^{2} \right) +\sigma _{\mathrm{CR}_{\mathrm{non}}}^{2}\mu _{H_{\mathrm{non}}}^{2}\left( \beta ^2\mu _{H}^{2}+\sigma _{N}^{2} \right) \right]
\\
&-2\sigma _{\mathrm{CR}}\sigma _{\mathrm{CR}_{\mathrm{non}}}f_{RA}^{2}x^2\mu _{H}^{2}\mu _{H_{\mathrm{non}}}^{2}\left( \mu _{\mathrm{U}}^{2}+\sigma _{\mathrm{U}}^{2} \right) \left( \beta ^2\mu _{H}^{2}+\sigma _{N}^{2} \right) \left( \beta ^2\mu _{H_{\mathrm{non}}}^{2}+\sigma _{N}^{2} \right)
\\
&+2\sigma _{\mathrm{CR}}\sigma _{\mathrm{CR}_{\mathrm{non}}}f_{RA}x\mu _{\mathrm{U}}\beta ^3\mu _{H}^{2}\mu _{H_{\mathrm{non}}}^{2}\left[ \mu _{\mathrm{CR}}\mu _{H}^{2}\left( \beta ^2\mu _{H_{\mathrm{non}}}^{2}+\sigma _{N}^{2} \right) +\mu _{\mathrm{CR}_{\mathrm{non}}}\mu _{H_{\mathrm{non}}}^{2}\left( \beta ^2\mu _{H}^{2}+\sigma _{N}^{2} \right) \right]
\\
&\left. +\beta ^6\mu _{H}^{4}\mu _{H_{\mathrm{non}}}^{4}\left( \sigma _{\mathrm{CR}}^{2}\mu _{\mathrm{CR}_{\mathrm{non}}}^{2}+\sigma _{\mathrm{CR}_{\mathrm{non}}}^{2}\mu _{\mathrm{CR}}^{2}-2\sigma _{\mathrm{CR}}\sigma _{\mathrm{CR}_{\mathrm{non}}}\mu _{\mathrm{CR}}\mu _{\mathrm{CR}_{\mathrm{non}}} \right) \right\} /\left( \beta ^6\mu _{H}^{4}\mu _{H_{\mathrm{non}}}^{4}\sigma _{\mathrm{CR}}^{2}\sigma _{\mathrm{CR}_{\mathrm{non}}}^{2} \right) ,\\
D_{\mathrm{PC},i}^{\mathrm{Intra},\mathrm{non}}\approx& R_{\mathrm{L}}\left\{ f_{\mathrm{RA}}^{2}x^4\left( \mu _{U}^{2}+\sigma _{U}^{2} \right) \beta ^2\left[ \sigma _{\mathrm{PC}}^{2}\mu _{H}^{2}\left( \beta ^2\mu _{H_{\mathrm{non}}}^{2}+3\sigma _{N}^{2} \right) +\sigma _{\mathrm{PC}_{\mathrm{non}}}^{2}\mu _{H_{\mathrm{non}}}^{2}\left( \beta ^2\mu _{H}^{2}+3\sigma _{N}^{2} \right) \right] \right.
\\
&-2f_{RA}x^2\mu _U\beta ^3\left[ \mu _{\mathrm{PC}}\sigma _{\mathrm{PC}_{\mathrm{non}}}^{2}\mu _{H_{\mathrm{non}}}^{2}\left( \beta ^2\mu _{H}^{2}+\sigma _{N}^{2} \right) +\mu _{\mathrm{PC}_{\mathrm{non}}}\sigma _{\mathrm{PC}}^{2}\mu _{H}^{2}\left( \beta ^2\mu _{H_{\mathrm{non}}}^{2}+\sigma _{N}^{2} \right) \right]
\\
&-2\sigma _{\mathrm{PC}}\sigma _{\mathrm{PC}_{\mathrm{non}}}f_{RA}^{2}x^4\left( \mu _{U}^{2}+\sigma _{U}^{2} \right) \left( \beta ^2\mu _{H}^{2}+\sigma _{N}^{2} \right) \left( \beta ^2\mu _{H_{\mathrm{non}}}^{2}+\sigma _{N}^{2} \right) +2\sigma _{\mathrm{PC}}\sigma _{\mathrm{PC}_{\mathrm{non}}}f_{RA}x^2\mu _U\beta ^3
\\
&\cdot \left[ \mu _{\mathrm{PC}}\mu _{H}^{2}\left( \beta ^2\mu _{H_{\mathrm{non}}}^{2}+\sigma _{N}^{2} \right) +\mu _{\mathrm{PC}_{\mathrm{non}}}\mu _{H_{\mathrm{non}}}^{2}\left( \beta ^2\mu _{H}^{2}+\sigma _{N}^{2} \right) \right] +\beta ^6\mu _{H}^{2}\mu _{H_{\mathrm{non}}}^{2}\left[ \sigma _{\mathrm{PC}}^{2}\mu _{\mathrm{PC}_{\mathrm{non}}}^{2}+\sigma _{\mathrm{PC}_{\mathrm{non}}}^{2}\mu _{\mathrm{PC}}^{2} \right.
\\
&\left. \left. +\sigma _{N}^{2}\left( \sigma _{\mathrm{PC}}^{2}+\sigma _{\mathrm{PC}_{\mathrm{non}}}^{2} \right) -2\sigma _{\mathrm{PC}}\sigma _{\mathrm{PC}_{\mathrm{non}}}\mu _{\mathrm{PC}}\mu _{\mathrm{PC}_{\mathrm{non}}} \right] \right\} /\left( \beta ^6\mu _{H}^{2}\mu _{H_{\mathrm{non}}}^{2}\sigma _{\mathrm{PC}}^{2}\sigma _{\mathrm{PC}_{\mathrm{non}}}^{2} \right) .
    \end{aligned}
    \end{small}
 	\end{equation}
\vspace{-0.7cm}
\end{figure*}

% Appendix F
\section{Inter-Class Distances under the Non-i.i.d. Stochastic Channel Scenario}
\label{appendix:F}
The expected inter-class distances for different methods under the non-i.i.d. stochastic channel scenario are shown in (\ref{eq68}) on the page following the next.
\begin{figure*}[ht]
 	\centering
 	\begin{equation}	
    \begin{small}
    \label{eq68}
    \begin{aligned}
D_{\mathrm{RAW},i,j}^{\mathrm{Inter},\mathrm{non}}=&R_{\mathrm{L}}\left\{ f_{RA}^{2}x^2\left( \mu _{\mathrm{U}}^{2}+\sigma _{\mathrm{U}}^{2} \right) \left[ \sigma _{\mathrm{RAW}}^{2}\left( \mu _{H_{\mathrm{non}}}^{2}+\sigma _{H_{\mathrm{non}}}^{2} \right) +\sigma _{\mathrm{RAW}_{\mathrm{non}}}^{2}\left( \mu _{H}^{2}+\sigma _{H}^{2} \right) \right] \right.
\\
&-2f_{\mathrm{RA}}x\mu _{\mathrm{U}}\left( \mu _H\mu _{\mathrm{RAW}}\sigma _{\mathrm{RAW}_{\mathrm{non}}}^{2}+\mu _{H_{\mathrm{non}}}\mu _{\mathrm{RAW}_{\mathrm{non}}}\sigma _{\mathrm{RAW}}^{2} \right) +\sigma _{N}^{2}\left( \sigma _{\mathrm{RAW}}^{2}+\sigma _{\mathrm{RAW}_{\mathrm{non}}}^{2} \right)
\\
&+\mu _{\mathrm{RAW}}^{2}\sigma _{\mathrm{RAW}_{\mathrm{non}}}^{2}+\mu _{\mathrm{RAW}_{\mathrm{non}}}^{2}\sigma _{\mathrm{RAW}}^{2}-2\sigma _{\mathrm{RAW}}\sigma _{\mathrm{RAW}_{\mathrm{non}}}\left[ \mu _H\mu _{H_{\mathrm{non}}}f_{RA}^{2}x^2\mu _{\mathrm{U}}^{2}-f_{RA}x\mu _{\mathrm{U}} \right.
\\
&\left. \left. \cdot \left( \mu _{\mathrm{RAW}}\mu _{H_{\mathrm{non}}}+\mu _{\mathrm{RAW}_{\mathrm{non}}}\mu _H \right) +\mu _{\mathrm{RAW}}\mu _{\mathrm{RAW}_{\mathrm{non}}} \right] \right\} /(\sigma _{\mathrm{RAW}}^{2}\sigma _{\mathrm{RAW}_{\mathrm{non}}}^{2}),
\\
D_{\mathrm{SL},i,j}^{\mathrm{Inter},\mathrm{non}}\approx& R_{\mathrm{S}}\left\{ f_{\mathrm{RA}}^{2}x^2\gamma ^2\mu _{H}^{2}\mu _{H_{\mathrm{non}}}^{2}\left( \mu _{\mathrm{S}}^{2}+\sigma _{\mathrm{S}}^{2} \right) \left[ \sigma _{\mathrm{SL}}^{2}\mu _{H}^{2}\left( \gamma ^2\mu _{H_{\mathrm{non}}}^{2}+3\sigma _{N}^{2} \right) +\sigma _{\mathrm{SL}_{\mathrm{non}}}^{2}\mu _{H_{\mathrm{non}}}^{2}\left( \gamma ^2\mu _{H}^{2}+3\sigma _{N}^{2} \right) \right] \right.
\\
&+\sigma _{N}^{2}\gamma ^2\left[ \sigma _{\mathrm{SL}}^{2}\left( \gamma ^2\mu _{H_{\mathrm{non}}}^{2}+3\gamma ^2\sigma _{H_{\mathrm{non}}}^{2}+3\sigma _{N}^{2} \right) +\sigma _{\mathrm{SL}_{\mathrm{non}}}^{2}\left( \gamma ^2\mu _{H}^{2}+3\gamma ^2\sigma _{H}^{2}+3\sigma _{N}^{2} \right) \right]
\\
&-2f_{\mathrm{RA}}x\mu _{\mathrm{S}}\gamma ^3\mu _{H}^{2}\mu _{H_{\mathrm{non}}}^{2}\left[ \sigma _{\mathrm{SL}}^{2}\mu _{H}^{2}\left( \gamma ^2\mu _{H_{\mathrm{non}}}^{2}+\sigma _{N}^{2} \right) +\sigma _{R\mathrm{SL}_{\mathrm{non}}}^{2}\mu _{H_{\mathrm{non}}}^{2}\left( \gamma ^2\mu _{H}^{2}+\sigma _{N}^{2} \right) \right]
\\
&-2\sigma _{\mathrm{SL}}\sigma _{\mathrm{SL}_{\mathrm{non}}}f_{\mathrm{RA}}^{2}x^2\mu _{H}^{2}\mu _{H_{\mathrm{non}}}^{2}\mu _{\mathrm{S}}^{2}\left( \gamma ^2\mu _{H}^{2}+\sigma _{N}^{2} \right) \left( \gamma ^2\mu _{H_{\mathrm{non}}}^{2}+\sigma _{N}^{2} \right)
\\
&+2\sigma _{\mathrm{SL}}\sigma _{\mathrm{SL}_{\mathrm{non}}}f_{\mathrm{RA}}x\mu _{\mathrm{S}}\gamma ^3\mu _{H}^{2}\mu _{H_{\mathrm{non}}}^{2}\left[ \mu _{\mathrm{SL}}\mu _{H}^{2}\left( \gamma ^2\mu _{H_{\mathrm{non}}}^{2}+\sigma _{N}^{2} \right) +\mu _{\mathrm{SL}_{\mathrm{non}}}\mu _{H_{\mathrm{non}}}^{2}\left( \gamma ^2\mu _{H}^{2}+\sigma _{N}^{2} \right) \right]
\\
&\left. +\gamma ^6\mu _{H}^{4}\mu _{H_{\mathrm{non}}}^{4}\left( \sigma _{\mathrm{SL}}^{2}\mu _{\mathrm{SL}_{\mathrm{non}}}^{2}+\sigma _{\mathrm{SL}_{\mathrm{non}}}^{2}\mu _{\mathrm{SL}}^{2}-2\sigma _{\mathrm{SL}}\sigma _{\mathrm{SL}_{\mathrm{non}}}\mu _{\mathrm{SL}}\mu _{\mathrm{SL}_{\mathrm{non}}} \right) \right\} /\left( \gamma ^6\mu _{H}^{4}\mu _{H_{\mathrm{non}}}^{4}\sigma _{\mathrm{SL}}^{2}\sigma _{\mathrm{SL}_{\mathrm{non}}}^{2} \right) ,
\\
D_{\mathrm{CR},i,j}^{\mathrm{Inter},\mathrm{non}}\approx& R_{\mathrm{L}}\left\{ f_{\mathrm{RA}}^{2}x^2\beta ^2\mu _{H}^{2}\mu _{H_{\mathrm{non}}}^{2}\left( \mu _{\mathrm{U}}^{2}+\sigma _{\mathrm{U}}^{2} \right) \left[ \sigma _{\mathrm{CR}}^{2}\mu _{H}^{2}\left( \beta ^2\mu _{H_{\mathrm{non}}}^{2}+3\sigma _{N}^{2} \right) +\sigma _{\mathrm{CR}_{\mathrm{non}}}^{2}\mu _{H_{\mathrm{non}}}^{2}\left( \beta ^2\mu _{H}^{2}+3\sigma _{N}^{2} \right) \right] \right.
\\
&+\sigma _{N}^{2}\beta ^2\left[ \sigma _{\mathrm{CR}}^{2}\left( \beta ^2\mu _{H_{\mathrm{non}}}^{2}+3\beta ^2\sigma _{H_{\mathrm{non}}}^{2}+3\sigma _{N}^{2} \right) +\sigma _{\mathrm{CR}_{\mathrm{non}}}^{2}\left( \beta ^2\mu _{H}^{2}+3\beta ^2\sigma _{H}^{2}+3\sigma _{N}^{2} \right) \right]
\\
&-2f_{\mathrm{RA}}x\mu _{\mathrm{U}}\beta ^3\mu _{H}^{2}\mu _{H_{\mathrm{non}}}^{2}\left[ \sigma _{\mathrm{CR}}^{2}\mu _{H}^{2}\left( \beta ^2\mu _{H_{\mathrm{non}}}^{2}+\sigma _{N}^{2} \right) +\sigma _{\mathrm{CR}_{\mathrm{non}}}^{2}\mu _{H_{\mathrm{non}}}^{2}\left( \beta ^2\mu _{H}^{2}+\sigma _{N}^{2} \right) \right]
\\
&-2\sigma _{\mathrm{CR}}\sigma _{\mathrm{CR}_{\mathrm{non}}}f_{\mathrm{RA}}^{2}x^2\mu _{H}^{2}\mu _{H_{\mathrm{non}}}^{2}\mu _{\mathrm{U}}^{2}\left( \beta ^2\mu _{H}^{2}+\sigma _{N}^{2} \right) \left( \beta ^2\mu _{H_{\mathrm{non}}}^{2}+\sigma _{N}^{2} \right)
\\
&+2\sigma _{\mathrm{CR}}\sigma _{\mathrm{CR}_{\mathrm{non}}}f_{\mathrm{RA}}x\mu _{\mathrm{U}}\beta ^3\mu _{H}^{2}\mu _{H_{\mathrm{non}}}^{2}\left[ \mu _{\mathrm{CR}}\mu _{H}^{2}\left( \beta ^2\mu _{H_{\mathrm{non}}}^{2}+\sigma _{N}^{2} \right) +\mu _{\mathrm{CR}_{\mathrm{non}}}\mu _{H_{\mathrm{non}}}^{2}\left( \beta ^2\mu _{H}^{2}+\sigma _{N}^{2} \right) \right]
\\
&\left. +\beta ^6\mu _{H}^{4}\mu _{H_{\mathrm{non}}}^{4}\left( \sigma _{\mathrm{CR}}^{2}\mu _{\mathrm{CR}_{\mathrm{non}}}^{2}+\sigma _{\mathrm{CR}_{\mathrm{non}}}^{2}\mu _{\mathrm{CR}}^{2}-2\sigma _{\mathrm{CR}}\sigma _{\mathrm{CR}_{\mathrm{non}}}\mu _{\mathrm{CR}}\mu _{\mathrm{CR}_{\mathrm{non}}} \right) \right\} /\left( \beta ^6\mu _{H}^{4}\mu _{H_{\mathrm{non}}}^{4}\sigma _{\mathrm{CR}}^{2}\sigma _{\mathrm{CR}_{\mathrm{non}}}^{2} \right) ,
\\
D_{\mathrm{PC},i,j}^{\mathrm{Inter},\mathrm{non}}\approx& R_{\mathrm{L}}\left\{ f_{\mathrm{RA}}^{2}x^4\left( \mu _{\mathrm{U}}^{2}+\sigma _{\mathrm{U}}^{2} \right) \beta ^2\left[ \sigma _{\mathrm{PC}}^{2}\mu _{H}^{2}\left( \beta ^2\mu _{H_{\mathrm{non}}}^{2}+3\sigma _{N}^{2} \right) +\sigma _{\mathrm{PC}_{\mathrm{non}}}^{2}\mu _{H_{\mathrm{non}}}^{2}\left( \beta ^2\mu _{H}^{2}+3\sigma _{N}^{2} \right) \right] \right.
\\
&-2f_{\mathrm{RA}}x^2\mu _{\mathrm{U}}\beta ^3\left[ \mu _{\mathrm{PC}}\sigma _{\mathrm{PC}_{\mathrm{non}}}^{2}\mu _{H_{\mathrm{non}}}^{2}\left( \beta ^2\mu _{H}^{2}+\sigma _{N}^{2} \right) +\mu _{\mathrm{PC}_{\mathrm{non}}}\sigma _{\mathrm{PC}}^{2}\mu _{H}^{2}\left( \beta ^2\mu _{H_{\mathrm{non}}}^{2}+\sigma _{N}^{2} \right) \right]
\\
&-2\sigma _{\mathrm{PC}}\sigma _{\mathrm{PC}_{\mathrm{non}}}f_{\mathrm{RA}}^{2}x^4\mu _{\mathrm{U}}^{2}\left( \beta ^2\mu _{H}^{2}+\sigma _{N}^{2} \right) \left( \beta ^2\mu _{H_{\mathrm{non}}}^{2}+\sigma _{N}^{2} \right) +2\sigma _{\mathrm{PC}}\sigma _{\mathrm{PC}_{\mathrm{non}}}f_{\mathrm{RA}}x^2\mu _{\mathrm{U}}\beta ^3
\\
&\cdot \left[ \mu _{\mathrm{PC}}\mu _{H}^{2}\left( \beta ^2\mu _{H_{\mathrm{non}}}^{2}+\sigma _{N}^{2} \right) +\mu _{\mathrm{PC}_{\mathrm{non}}}\mu _{H_{\mathrm{non}}}^{2}\left( \beta ^2\mu _{H}^{2}+\sigma _{N}^{2} \right) \right] +\beta ^6\mu _{H}^{2}\mu _{H_{\mathrm{non}}}^{2}[\sigma _{\mathrm{PC}}^{2}\mu _{\mathrm{PC}_{\mathrm{non}}}^{2}
\\
&\left. \left. +\sigma _{\mathrm{PC}_{\mathrm{non}}}^{2}\mu _{\mathrm{PC}}^{2}+\sigma _{N}^{2}\left( \sigma _{\mathrm{PC}}^{2}+\sigma _{\mathrm{PC}_{\mathrm{non}}}^{2} \right) -2\sigma _{\mathrm{PC}}\sigma _{\mathrm{PC}_{\mathrm{non}}}\mu _{\mathrm{PC}}\mu _{\mathrm{PC}_{\mathrm{non}}} \right] \right\} /\left( \beta ^6\mu _{H}^{2}\mu _{H_{\mathrm{non}}}^{2}\sigma _{\mathrm{PC}}^{2}\sigma _{\mathrm{PC}_{\mathrm{non}}}^{2} \right) .
    \end{aligned}
    \end{small}
 	\end{equation}
\vspace{-0.6cm}
\end{figure*}

% Appendix G
\section{Silhouette Scores under the Non-i.i.d. Stochastic Channel Scenario}
\label{appendix:G}
The expected silhouette scores for different methods under the non-i.i.d. stochastic channel scenario are presented in (\ref{eq69}) on the final page.
\begin{figure*}[ht]
 	\centering
 	\begin{equation}	
    \begin{small}
    \label{eq69}
    \begin{aligned}
S_{\mathrm{RAW}}^{\mathrm{non}}=&2\sigma _{\mathrm{RAW}}\sigma _{\mathrm{RAW}_{\mathrm{non}}}\mu _H\mu _{H_{\mathrm{non}}}f_{\mathrm{RA}}^{2}x^2\sigma _{\mathrm{U}}^{2}/\left\{ f_{\mathrm{RA}}^{2}x^2\left[ \sigma _{\mathrm{RAW}}^{2}\left( \mu _{H_{\mathrm{non}}}^{2}+\sigma _{H_{\mathrm{non}}}^{2} \right) +\sigma _{\mathrm{RAW}_{\mathrm{non}}}^{2}\left( \mu _{H}^{2}+\sigma _{H}^{2} \right) \right] \right.
\\
&\cdot \left( \mu _{\mathrm{U}}^{2}+\sigma _{\mathrm{U}}^{2} \right) -2f_{\mathrm{RA}}x\mu _{\mathrm{U}}\left( \mu _H\mu _{\mathrm{RAW}}\sigma _{\mathrm{RAW}_{\mathrm{non}}}^{2}+\mu _{H_{\mathrm{non}}}\mu _{\mathrm{RAW}_{\mathrm{non}}}\sigma _{\mathrm{RAW}}^{2} \right) +\sigma _{N}^{2}\left( \sigma _{\mathrm{RAW}}^{2}+\sigma _{\mathrm{RAW}_{\mathrm{non}}}^{2} \right)
\\
&+\mu _{\mathrm{RAW}}^{2}\sigma _{\mathrm{RAW}_{\mathrm{non}}}^{2}+\mu _{\mathrm{RAW}_{\mathrm{non}}}^{2}\sigma _{\mathrm{RAW}}^{2}-2\sigma _{\mathrm{RAW}}\sigma _{\mathrm{RAW}_{\mathrm{non}}}\left[ \mu _H\mu _{H_{\mathrm{non}}}f_{\mathrm{RA}}^{2}x^2\mu _{\mathrm{U}}^{2} \right.
\\
&\left. \left. -f_{\mathrm{RA}}x\mu _{\mathrm{U}}\left( \mu _{\mathrm{RAW}}\mu _{H_{\mathrm{non}}}+\mu _{\mathrm{RAW}_{\mathrm{non}}}\mu _H \right) +\mu _{\mathrm{RAW}}\mu _{\mathrm{RAW}_{\mathrm{non}}} \right] \right\} ,
\\
S_{\mathrm{SL}}^{\mathrm{non}}\approx& 2\sigma _{\mathrm{SL}}\sigma _{\mathrm{SL}_{\mathrm{non}}}f_{RA}^{2}x^2\mu _{H}^{2}\mu _{H_{\mathrm{non}}}^{2}\sigma _{\mathrm{S}}^{2}\left( \gamma ^2\mu _{H}^{2}+\sigma _{N}^{2} \right) \left( \gamma ^2\mu _{H_{\mathrm{non}}}^{2}+\sigma _{N}^{2} \right) /\left\{ f_{\mathrm{RA}}^{2}x^2\gamma ^2\mu _{H}^{2}\mu _{H_{\mathrm{non}}}^{2}\left( \mu _{\mathrm{S}}^{2}+\sigma _{\mathrm{S}}^{2} \right) \right.
\\
&\cdot \left[ \sigma _{\mathrm{SL}}^{2}\mu _{H}^{2}\left( \gamma ^2\mu _{H_{\mathrm{non}}}^{2}+3\sigma _{N}^{2} \right) +\sigma _{\mathrm{SL}_{\mathrm{non}}}^{2}\mu _{H_{\mathrm{non}}}^{2}\left( \gamma ^2\mu _{H}^{2}+3\sigma _{N}^{2} \right) \right] +\sigma _{N}^{2}\gamma ^2\left[ \sigma _{\mathrm{SL}}^{2}\left( \gamma ^2\mu _{H_{\mathrm{non}}}^{2}+3\gamma ^2\sigma _{H_{\mathrm{non}}}^{2}+3\sigma _{N}^{2} \right) \right.
\\
&\left. +\sigma _{\mathrm{SL}_{\mathrm{non}}}^{2}\left( \gamma ^2\mu _{H}^{2}+3\gamma ^2\sigma _{H}^{2}+3\sigma _{N}^{2} \right) \right] -2f_{\mathrm{RA}}x\mu _{\mathrm{S}}\gamma ^3\mu _{H}^{2}\mu _{H_{\mathrm{non}}}^{2}\left[ \mu _{\mathrm{SL}}\sigma _{\mathrm{SL}_{\mathrm{non}}}^{2}\mu _{H_{\mathrm{non}}}^{2}\left( \gamma ^2\mu _{H}^{2}+\sigma _{N}^{2} \right) \right.
\\
&\left. +\mu _{\mathrm{SL}_{\mathrm{non}}}\sigma _{\mathrm{SL}}^{2}\mu _{H}^{2}\left( \gamma ^2\mu _{H_{\mathrm{non}}}^{2}+\sigma _{N}^{2} \right) \right] -2\sigma _{\mathrm{SL}}\sigma _{\mathrm{SL}_{\mathrm{non}}}f_{\mathrm{RA}}^{2}x^2\mu _{H}^{2}\mu _{H_{\mathrm{non}}}^{2}\mu _{\mathrm{S}}^{2}\left( \gamma ^2\mu _{H}^{2}+\sigma _{N}^{2} \right) \left( \gamma ^2\mu _{H_{\mathrm{non}}}^{2}+\sigma _{N}^{2} \right)
\\
&+2\sigma _{\mathrm{SL}}\sigma _{\mathrm{SL}_{\mathrm{non}}}{f_{\mathrm{RA}}}_{\mathrm{S}}\mu _{\mathrm{S}}\gamma ^3\mu _{H}^{2}\mu _{H_{\mathrm{non}}}^{2}\left[ \mu _{\mathrm{SL}}\mu _{H}^{2}\left( \gamma ^2\mu _{H_{\mathrm{non}}}^{2}+\sigma _{N}^{2} \right) +\mu _{\mathrm{SL}_{\mathrm{non}}}\mu _{H_{\mathrm{non}}}^{2}\left( \gamma ^2\mu _{H}^{2}+\sigma _{N}^{2} \right) \right]
\\
&\left. +\gamma ^6\mu _{H}^{4}\mu _{H_{\mathrm{non}}}^{4}\left( \sigma _{\mathrm{SL}}^{2}\mu _{\mathrm{SL}_{\mathrm{non}}}^{2}+\sigma _{\mathrm{SL}_{\mathrm{non}}}^{2}\mu _{\mathrm{SL}}^{2}-2\sigma _{\mathrm{SL}}\sigma _{\mathrm{SL}_{\mathrm{non}}}\mu _{\mathrm{SL}}\mu _{\mathrm{SL}_{\mathrm{non}}} \right) \right\} ,
\\
S_{\mathrm{CR}}^{\mathrm{non}}\approx& 2\sigma _{\mathrm{CR}}\sigma _{\mathrm{CR}_{\mathrm{non}}}f_{\mathrm{RA}}^{2}x^2\mu _{H}^{2}\mu _{H_{\mathrm{non}}}^{2}\sigma _{\mathrm{U}}^{2}\left( \beta ^2\mu _{H}^{2}+\sigma _{N}^{2} \right) \left( \beta ^2\mu _{H_{\mathrm{non}}}^{2}+\sigma _{N}^{2} \right) /\left\{ f_{\mathrm{RA}}^{2}x^2\beta ^2\mu _{H}^{2}\mu _{H_{\mathrm{non}}}^{2}\left( \mu _{\mathrm{U}}^{2}+\sigma _{\mathrm{U}}^{2} \right) \right.
\\
&\cdot \left[ \sigma _{\mathrm{CR}}^{2}\mu _{H}^{2}\left( \beta ^2\mu _{H_{\mathrm{non}}}^{2}+3\sigma _{N}^{2} \right) +\sigma _{\mathrm{CR}_{\mathrm{non}}}^{2}\mu _{H_{\mathrm{non}}}^{2}\left( \beta ^2\mu _{H}^{2}+3\sigma _{N}^{2} \right) \right] +\sigma _{N}^{2}\beta ^2\left[ \sigma _{\mathrm{CR}}^{2}\left( \beta ^2\mu _{H_{\mathrm{non}}}^{2}+3\beta ^2\sigma _{H_{\mathrm{non}}}^{2}+3\sigma _{N}^{2} \right) \right.
\\
&\left. +\sigma _{\mathrm{CR}_{\mathrm{non}}}^{2}\left( \beta ^2\mu _{H}^{2}+3\beta ^2\sigma _{H}^{2}+3\sigma _{N}^{2} \right) \right] -2f_{\mathrm{RA}}x\mu _{\mathrm{U}}\beta ^3\mu _{H}^{2}\mu _{H_{\mathrm{non}}}^{2}\left[ \mu _{\mathrm{CR}}\sigma _{\mathrm{CR}_{\mathrm{non}}}^{2}\mu _{H_{\mathrm{non}}}^{2}\left( \beta ^2\mu _{H}^{2}+\sigma _{N}^{2} \right) \right.
\\
&\left. +\mu _{\mathrm{CR}_{\mathrm{non}}}\sigma _{\mathrm{CR}}^{2}\mu _{H}^{2}\left( \beta ^2\mu _{H_{\mathrm{non}}}^{2}+\sigma _{N}^{2} \right) \right] -2\sigma _{\mathrm{CR}}\sigma _{\mathrm{CR}_{\mathrm{non}}}f_{\mathrm{RA}}^{2}x^2\mu _{H}^{2}\mu _{H_{\mathrm{non}}}^{2}\mu _{\mathrm{U}}^{2}\left( \beta ^2\mu _{H}^{2}+\sigma _{N}^{2} \right) \left( \beta ^2\mu _{H_{\mathrm{non}}}^{2}+\sigma _{N}^{2} \right)
\\
&+2\sigma _{\mathrm{CR}}\sigma _{\mathrm{CR}_{\mathrm{non}}}f_{\mathrm{RA}}x\mu _{\mathrm{U}}\beta ^3\mu _{H}^{2}\mu _{H_{\mathrm{non}}}^{2}\left[ \mu _{\mathrm{CR}}\mu _{H}^{2}\left( \beta ^2\mu _{H_{\mathrm{non}}}^{2}+\sigma _{N}^{2} \right) +\mu _{\mathrm{CR}_{\mathrm{non}}}\mu _{H_{\mathrm{non}}}^{2}\left( \beta ^2\mu _{H}^{2}+\sigma _{N}^{2} \right) \right]
\\
&\left. +\beta ^6\mu _{H}^{4}\mu _{H_{\mathrm{non}}}^{4}\left( \sigma _{\mathrm{CR}}^{2}\mu _{\mathrm{CR}_{\mathrm{non}}}^{2}+\sigma _{\mathrm{CR}_{\mathrm{non}}}^{2}\mu _{\mathrm{CR}}^{2}-2\sigma _{\mathrm{CR}}\sigma _{\mathrm{CR}_{\mathrm{non}}}\mu _{\mathrm{CR}}\mu _{\mathrm{CR}_{\mathrm{non}}} \right) \right\} ,
\\
S_{\mathrm{PC}}^{\mathrm{non}}\approx& 2\sigma _{\mathrm{PC}}\sigma _{\mathrm{PC}_{\mathrm{non}}}f_{\mathrm{RA}}^{2}x^4\sigma _{\mathrm{U}}^{2}\left( \beta ^2\mu _{H}^{2}+\sigma _{N}^{2} \right) \left( \beta ^2\mu _{H_{\mathrm{non}}}^{2}+\sigma _{N}^{2} \right) /\left\{ f_{\mathrm{RA}}^{2}x^4\left( \mu _{\mathrm{U}}^{2}+\sigma _{\mathrm{U}}^{2} \right) \beta ^2\left[ \sigma _{\mathrm{PC}}^{2}\mu _{H}^{2} \right. \right.
\\
&\left. \cdot \left( \beta ^2\mu _{H_{\mathrm{non}}}^{2}+3\sigma _{N}^{2} \right) +\sigma _{\mathrm{PC}_{\mathrm{non}}}^{2}\mu _{H_{\mathrm{non}}}^{2}\left( \beta ^2\mu _{H}^{2}+3\sigma _{N}^{2} \right) \right] -2f_{\mathrm{RA}}x^2\mu _{\mathrm{U}}\beta ^3\left[ \mu _{\mathrm{PC}}\sigma _{\mathrm{PC}_{\mathrm{non}}}^{2}\mu _{H_{\mathrm{non}}}^{2}\left( \beta ^2\mu _{H}^{2}+\sigma _{N}^{2} \right) \right.
\\
&\left. +\mu _{\mathrm{PC}_{\mathrm{non}}}\sigma _{\mathrm{PC}}^{2}\mu _{H}^{2}\left( \beta ^2\mu _{H_{\mathrm{non}}}^{2}+\sigma _{N}^{2} \right) \right] -2\sigma _{\mathrm{PC}}\sigma _{\mathrm{PC}_{\mathrm{non}}}f_{\mathrm{RA}}^{2}x^4\mu _{\mathrm{U}}^{2}\left( \beta ^2\mu _{H}^{2}+\sigma _{N}^{2} \right) \left( \beta ^2\mu _{H_{\mathrm{non}}}^{2}+\sigma _{N}^{2} \right)
\\
&+2\sigma _{\mathrm{PC}}\sigma _{\mathrm{PC}_{\mathrm{non}}}f_{\mathrm{RA}}x^2\mu _{\mathrm{U}}\beta ^3\left[ \mu _{\mathrm{PC}}\mu _{H}^{2}\left( \beta ^2\mu _{H_{\mathrm{non}}}^{2}+\sigma _{N}^{2} \right) +\mu _{\mathrm{PC}_{\mathrm{non}}}\mu _{H_{\mathrm{non}}}^{2}\left( \beta ^2\mu _{H}^{2}+\sigma _{N}^{2} \right) \right]
\\
&\left. +\beta ^6\mu _{H}^{2}\mu _{H_{\mathrm{non}}}^{2}\left[ \sigma _{\mathrm{PC}}^{2}\mu _{\mathrm{PC}_{\mathrm{non}}}^{2}+\sigma _{\mathrm{PC}_{\mathrm{non}}}^{2}\mu _{\mathrm{PC}}^{2}+\sigma _{N}^{2}\left( \sigma _{\mathrm{PC}}^{2}+\sigma _{\mathrm{PC}_{\mathrm{non}}}^{2} \right) -2\sigma _{\mathrm{PC}}\sigma _{\mathrm{PC}_{\mathrm{non}}}\mu _{\mathrm{PC}}\mu _{\mathrm{PC}_{\mathrm{non}}} \right] \right\} .
    \end{aligned}
    \end{small}
 	\end{equation}
\vspace{-0cm}
\end{figure*}

% Appendix H
\section{Intra-Class and Inter-Class Distances of the RC Method under the Non-i.i.d. Stochastic Channel Scenario}
\label{appendix:H}
The expected intra-class and inter-class distances of the proposed RC method under the non-i.i.d. stochastic channel scenario are provided in (\ref{eq70}) on the final page.

\begin{figure*}[ht]
 	\centering
 	\begin{equation}	
    \begin{small}
    \label{eq70}
    \begin{aligned}
D_{{\rm{RC}},i}^\mathrm{Intra,non} \approx& {R_\mathrm{L}}\{ f_\mathrm{RA}^2\left( {\mu _{\mathrm{U}}^2 + \sigma _{\mathrm{U}}^2} \right){\beta ^2}\left[ {\alpha _\mathrm{non}^2\sigma _{{\rm{RC}}}^2\mu _H^2\left( {{\beta ^2}\mu _{{H_\mathrm{non}}}^2 + 3\sigma _N^2} \right) + {\alpha ^2}\sigma _{{\rm{R}}{{\rm{C}}_\mathrm{non}}}^2\mu _{{H_\mathrm{non}}}^2\left( {{\beta ^2}\mu _H^2 + 3\sigma _N^2} \right)} \right]\\
& - 2{f_\mathrm{RA}}{\mu _{{U}}}{\beta ^3}\left[ {\alpha {\mu _{{\rm{RC}}}}\sigma _{{\rm{R}}{{\rm{C}}_\mathrm{non}}}^2\mu _{{H_\mathrm{non}}}^2\left( {{\beta ^2}\mu _H^2 + \sigma _N^2} \right) + {\alpha _\mathrm{non}}{\mu _{{\rm{R}}{{\rm{C}}_\mathrm{non}}}}\sigma _{{\rm{RC}}}^2\mu _H^2\left( {{\beta ^2}\mu _{{H_\mathrm{non}}}^2 + \sigma _N^2} \right)} \right]\\
& - 2{\sigma _{{\rm{RC}}}}{\sigma _{{\rm{R}}{{\rm{C}}_\mathrm{non}}}}\alpha {\alpha _\mathrm{non}}f_\mathrm{RA}^2\left( {\mu _{\mathrm{U}}^2 + \sigma _{\mathrm{U}}^2} \right)\left( {{\beta ^2}\mu _H^2 + \sigma _N^2} \right)\left( {{\beta ^2}\mu _{{H_\mathrm{non}}}^2 + \sigma _N^2} \right) + 2{\sigma _{{\rm{RC}}}}{\sigma _{{\rm{R}}{{\rm{C}}_\mathrm{non}}}}{f_\mathrm{RA}}{\mu _{\mathrm{U}}}{\beta ^3}\\
& \cdot \left[ {{\alpha _\mathrm{non}}{\mu _{{\rm{RC}}}}\mu _H^2\left( {{\beta ^2}\mu _{{H_\mathrm{non}}}^2 + \sigma _N^2} \right) + \alpha {\mu _{{\rm{R}}{{\rm{C}}_\mathrm{non}}}}\mu _{{H_\mathrm{non}}}^2\left( {{\beta ^2}\mu _H^2 + \sigma _N^2} \right)} \right] + {\beta ^6}\mu _H^2\mu _{{H_\mathrm{non}}}^2[\sigma _{{\rm{RC}}}^2\mu _{{\rm{R}}{{\rm{C}}_\mathrm{non}}}^2\\
& + \sigma _{{\rm{R}}{{\rm{C}}_\mathrm{non}}}^2\mu _{{\rm{RC}}}^2 + \sigma _N^2\left( {\sigma _{{\rm{RC}}}^2 + \sigma _{{\rm{R}}{{\rm{C}}_\mathrm{non}}}^2} \right) - 2{\sigma _{{\rm{RC}}}}{\sigma _{{\rm{R}}{{\rm{C}}_\mathrm{non}}}}{\mu _{{\rm{RC}}}}{\mu _{{\rm{R}}{{\rm{C}}_\mathrm{non}}}}]\} /\left( {{\beta ^6}\mu _H^2\mu _{{H_\mathrm{non}}}^2\sigma _{{\rm{RC}}}^2\sigma _{{\rm{R}}{{\rm{C}}_\mathrm{non}}}^2} \right),\\
D_{\mathrm{RC},i,j}^{\mathrm{Inter},\mathrm{non}}\approx& R_{\mathrm{L}}\{f_{\mathrm{RA}}^{2}\left( \mu _{\mathrm{U}}^{2}+\sigma _{\mathrm{U}}^{2} \right) \beta ^2\left[ \alpha _{\mathrm{non}}^{2}\sigma _{\mathrm{RC}}^{2}\mu _{H}^{2}\left( \beta ^2\mu _{H_{\mathrm{non}}}^{2}+3\sigma _{N}^{2} \right) +\alpha ^2\sigma _{\mathrm{RC}_{\mathrm{non}}}^{2}\mu _{H_{\mathrm{non}}}^{2}\left( \beta ^2\mu _{H}^{2}+3\sigma _{N}^{2} \right) \right]
\\
&-2f_{\mathrm{RA}}\mu _{\mathrm{U}}\beta ^3\left[ \alpha \mu _{\mathrm{RC}}\sigma _{\mathrm{RC}_{\mathrm{non}}}^{2}\mu _{H_{\mathrm{non}}}^{2}\left( \beta ^2\mu _{H}^{2}+\sigma _{N}^{2} \right) +\alpha _{\mathrm{non}}\mu _{\mathrm{RC}_{\mathrm{non}}}\sigma _{\mathrm{RC}}^{2}\mu _{H}^{2}\left( \beta ^2\mu _{H_{\mathrm{non}}}^{2}+\sigma _{N}^{2} \right) \right]
\\
&-2\sigma _{\mathrm{RC}}\sigma _{\mathrm{RC}_{\mathrm{non}}}\alpha \alpha _{\mathrm{non}}f_{\mathrm{RA}}^{2}\mu _{\mathrm{U}}^{2}\left( \beta ^2\mu _{H}^{2}+\sigma _{N}^{2} \right) \left( \beta ^2\mu _{H_{\mathrm{non}}}^{2}+\sigma _{N}^{2} \right) +2\sigma _{\mathrm{RC}}\sigma _{\mathrm{RC}_{\mathrm{non}}}f_{\mathrm{RA}}\mu _{\mathrm{U}}\beta ^3
\\
&\cdot \left[ \alpha _{\mathrm{non}}\mu _{\mathrm{RC}}\mu _{H}^{2}\left( \beta ^2\mu _{H_{\mathrm{non}}}^{2}+\sigma _{N}^{2} \right) +\alpha \mu _{\mathrm{RC}_{\mathrm{non}}}\mu _{H_{\mathrm{non}}}^{2}\left( \beta ^2\mu _{H}^{2}+\sigma _{N}^{2} \right) \right] +\beta ^6\mu _{H}^{2}\mu _{H_{\mathrm{non}}}^{2}[\sigma _{\mathrm{RC}}^{2}\mu _{\mathrm{RC}_{\mathrm{non}}}^{2}
\\
&+\sigma _{\mathrm{RC}_{\mathrm{non}}}^{2}\mu _{\mathrm{RC}}^{2}+\sigma _{N}^{2}\left( \sigma _{\mathrm{RC}}^{2}+\sigma _{\mathrm{RC}_{\mathrm{non}}}^{2} \right) -2\sigma _{\mathrm{RC}}\sigma _{\mathrm{RC}_{\mathrm{non}}}\mu _{\mathrm{RC}}\mu _{\mathrm{RC}_{\mathrm{non}}}]\}/\left( \beta ^6\mu _{H}^{2}\mu _{H_{\mathrm{non}}}^{2}\sigma _{\mathrm{RC}}^{2}\sigma _{\mathrm{RC}_{\mathrm{non}}}^{2} \right) .
    \end{aligned}
    \end{small}
 	\end{equation}
\vspace{-0.7cm}
\end{figure*}

% Appendix I
\section{Silhouette Scores of the RC method under the Non-i.i.d. Stochastic Channel Scenario}
\label{appendix:I}
The expected silhouette score of the proposed RC method under the non-i.i.d. stochastic channel scenario is presented in (\ref{eq71}) on the final page.
\begin{figure*}[ht]
 	\centering
 	\begin{equation}	
    \begin{small}
    \label{eq71}
    \begin{aligned}
S_{\mathrm{RC}}^{\mathrm{non}}\approx& 2\alpha \alpha _{\mathrm{non}}\sigma _{\mathrm{RC}}\sigma _{\mathrm{RC}_{\mathrm{non}}}f_{\mathrm{RA}}^{2}\sigma _{\mathrm{U}}^{2}\left( \beta ^2\mu _{H}^{2}+\sigma _{N}^{2} \right) \left( \beta ^2\mu _{H_{\mathrm{non}}}^{2}+\sigma _{N}^{2} \right) /\left\{ f_{\mathrm{RA}}^{2}\left( \mu _{\mathrm{U}}^{2}+\sigma _{\mathrm{U}}^{2} \right) \beta ^2[\alpha _{\mathrm{non}}^{2}\sigma _{\mathrm{RC}}^{2}\mu _{H}^{2} \right.
\\
&\cdot \left( \beta ^2\mu _{H_{\mathrm{non}}}^{2}+3\sigma _{N}^{2} \right) +\alpha ^2\sigma _{\mathrm{RC}_{\mathrm{non}}}^{2}\mu _{H_{\mathrm{non}}}^{2}\left( \beta ^2\mu _{H}^{2}+3\sigma _{N}^{2} \right) ]-2f_{\mathrm{RA}}\mu _{\mathrm{U}}\beta ^3[\alpha \mu _{\mathrm{RC}}\sigma _{\mathrm{RC}_{\mathrm{non}}}^{2}\mu _{H_{\mathrm{non}}}^{2}\left( \beta ^2\mu _{H}^{2}+\sigma _{N}^{2} \right)
\\
&+\alpha _{\mathrm{non}}\mu _{\mathrm{RC}_{\mathrm{non}}}\sigma _{\mathrm{RC}}^{2}\mu _{H}^{2}\left( \beta ^2\mu _{H_{\mathrm{non}}}^{2}+\sigma _{N}^{2} \right) ]-2\sigma _{\mathrm{RC}}\sigma _{\mathrm{RC}_{\mathrm{non}}}\alpha \alpha _{\mathrm{non}}f_{\mathrm{RA}}^{2}\mu _{\mathrm{U}}^{2}\left( \beta ^2\mu _{H}^{2}+\sigma _{N}^{2} \right) \left( \beta ^2\mu _{H_{\mathrm{non}}}^{2}+\sigma _{N}^{2} \right)
\\
&+2\sigma _{\mathrm{RC}}\sigma _{\mathrm{RC}_{\mathrm{non}}}f_{\mathrm{RA}}\mu _{\mathrm{U}}\beta ^3\left[ \alpha _{\mathrm{non}}\mu _{\mathrm{RC}}\mu _{H}^{2}\left( \beta ^2\mu _{H_{\mathrm{non}}}^{2}+\sigma _{N}^{2} \right) +\alpha \mu _{\mathrm{RC}_{\mathrm{non}}}\mu _{H_{\mathrm{non}}}^{2}\left( \beta ^2\mu _{H}^{2}+\sigma _{N}^{2} \right) \right]
\\
&\left. +\beta ^6\mu _{H}^{2}\mu _{H_{\mathrm{non}}}^{2}\left[ \sigma _{\mathrm{RC}}^{2}\mu _{\mathrm{RC}_{\mathrm{non}}}^{2}+\sigma _{\mathrm{RC}_{\mathrm{non}}}^{2}\mu _{\mathrm{RC}}^{2}+\sigma _{N}^{2}\left( \sigma _{\mathrm{RC}}^{2}+\sigma _{\mathrm{RC}_{\mathrm{non}}}^{2} \right) -2\sigma _{\mathrm{RC}}\sigma _{\mathrm{RC}_{\mathrm{non}}}\mu _{\mathrm{RC}}\mu _{\mathrm{RC}_{\mathrm{non}}} \right] \right\} .
    \end{aligned}
    \end{small}
 	\end{equation}
 \vspace{-0.8cm}
\end{figure*}
\vspace{-0.4cm}

% Appendix J
\section{Comparison of Silhouette Scores between the PC and RC Methods under the Non-i.i.d. Stochastic Channel Scenario}
\label{appendix:J}
For a theoretical comparison under more general parameter settings, the PC and RC methods are analyzed under two primary assumptions: $\mu _H=\mu _{H_{\mathrm{non}}}$ and that the amplitude of the frequency-domain signal $X$ is normalized to 1. Under these assumptions, the expected silhouette score of the proposed RC method can be expressed as
 	\begin{equation}	
    \begin{small}
    \label{eq72}
    \begin{aligned}
S_{\mathrm{RC}}^{\mathrm{non}}\approx& 2f_{\mathrm{RA}}^{2}\sigma _{\mathrm{U}}^{2}\left( \beta ^2\mu _{H}^{2}+\sigma _{N}^{2} \right) ^2/\left\{ \beta ^2f_{\mathrm{RA}}^{2}\mu _{H}^{2}\left( \mu _{\mathrm{U}}^{2}+\sigma _{\mathrm{U}}^{2} \right) \right.
\\
&\cdot \left( \beta ^2\mu _{H}^{2}+3\sigma _{N}^{2} \right) \frac{A^2+B^2}{AB}-\frac{2\beta ^3}{\alpha}f_{\mathrm{RA}}\mu _{\mathrm{U}}\mu _{\mathrm{RC}}\mu _{H}^{2}
\\
&\cdot \left( \beta ^2\mu _{H}^{2}+\sigma _{N}^{2} \right) \frac{\left( A-B \right) ^2}{AB}-2f_{\mathrm{RA}}^{2}\mu _{\mathrm{U}}^{2}\left( \beta ^2\mu _{H}^{2}+\sigma _{N}^{2} \right) ^2
\\
&\left. +\frac{\beta ^6}{\alpha ^2}\mu _{H}^{4}\mu _{\mathrm{RC}}^{2}\frac{\left( A-B \right) ^2}{AB}+\frac{\beta ^6}{\alpha ^2}\mu _{H}^{4}\sigma _{N}^{2}\frac{\left( 1+B^2 \right)}{AB} \right\} .
    \end{aligned}
    \end{small}
 	\end{equation}
In the high SNR region, it follows that $A\approx B$. Therefore, we have
 	\begin{equation}	
    \begin{small}
    \label{eq73}
    \begin{aligned}
S_{\mathrm{RC}}^{\mathrm{non}}\approx& 2f_{\mathrm{RA}}^{2}\sigma _{\mathrm{U}}^{2}\left( \beta ^2\mu _{H}^{2}+\sigma _{N}^{2} \right) ^2/\left\{ 2\beta ^2f_{\mathrm{RA}}^{2}\mu _{H}^{2} \right. \left( \mu _{\mathrm{U}}^{2}+\sigma _{\mathrm{U}}^{2} \right)
\\
&\cdot \left( \beta ^2\mu _{H}^{2}+3\sigma _{N}^{2} \right) -2f_{\mathrm{RA}}^{2}\mu _{\mathrm{U}}^{2}\left( \beta ^2\mu _{H}^{2}+\sigma _{N}^{2} \right) ^2
\\
&\left. +2\beta ^6\mu _{H}^{4}\sigma _{N}^{2}/\left[ 2A^2\alpha ^2/\left( 1+A^2 \right) \right] \right\} .
    \end{aligned}
    \end{small}
 	\end{equation}
Meanwhile, the expected silhouette score of the PC method is given by
 	\begin{equation}	
    \begin{small}
    \label{eq74}
    \begin{aligned}
S_{\mathrm{PC}}^{\mathrm{non}}\approx& 2f_{\mathrm{RA}}^{2}\sigma _{\mathrm{U}}^{2}\left( \beta ^2\mu _{H}^{2}+\sigma _{N}^{2} \right) ^2/\left\{ 2\beta ^2f_{\mathrm{RA}}^{2}\mu _{H}^{2}\left( \mu _{\mathrm{U}}^{2}+\sigma _{\mathrm{U}}^{2} \right) \right.
\\
&\cdot \left( \beta ^2\mu _{H}^{2}+3\sigma _{N}^{2} \right) \left. -2f_{\mathrm{RA}}^{2}\mu _{\mathrm{U}}^{2}\left( \beta ^2\mu _{H}^{2}+\sigma _{N}^{2} \right) ^2+2\beta ^6\mu _{H}^{4}\sigma _{N}^{2} \right\} .
    \end{aligned}
    \end{small}
 	\end{equation}
When $\alpha >1$ and $\alpha _{\mathrm{non}}>1$ hold, it follows that $S_{{\rm{RC}}}^\mathrm{non} > S_{{\rm{PC}}}^\mathrm{non}$. Thus, under more general parameter settings, the proposed RC method can achieve a higher silhouette score than the PC method under the non-i.i.d. stochastic channel scenario.

\end{appendices}

\end{document}